\documentclass[useAMS,usenatbib,]{mn2e}
\usepackage{txfonts,graphicx} 
\usepackage{natbib,longtable,url} 
\usepackage[usenames,dvips]{color} 
\newcommand{\Fe}[5]{\mbox{$#1\,^#2{\rm #3}^{{\rm #4}}_{\rm #5}$}} 
\newcommand\ion[2]{#1$\;${\scshape{#2}}} 
\newcommand{\opd}{\log \tau_{\rm 5000}}    
\newcommand{\td}{\langle3\rm{D}\rangle}    
\definecolor{mygray}{gray}{.75}
\newcommand{\SH}{S\!_{\rm H}}             
\newcommand{\Te}{T_{\rm e}}               
\newcommand{\mA}{{\rm m\AA}}              
\newcommand{\Elow}{E_{\rm low}}           
\newcommand{\EW}{W_{\lambda}}
\newcommand{\Teff}{\ensuremath{T_{\mathrm{eff}}}}     
\newcommand{\logg}{\ensuremath{\log g}}               
\newcommand{\kms}{km s$^{-1}$}
\newcommand{\Vmic}{\xi_{\rm{t}}}          
\newcommand{\Vmac}{\xi_{\rm{RT}}}         

\newcommand{\absun}{\log \varepsilon_{\odot}}

\newcommand{\logeFesun}[1]{\log\varepsilon_{\rm Fe, \odot}}
\newcommand{\logeFeone}[1]{\log\varepsilon_{\rm Fe I}}
\newcommand{\logeFetwo}[1]{\log\varepsilon_{\rm Fe II}}
\newcommand{\logeFeones}[1]{\log\varepsilon_{\rm Fe I, \odot}}
\newcommand{\logeFetwos}[1]{\log\varepsilon_{\rm Fe II, \odot}}
\newcommand{\logemean}[1]{\log\varepsilon_{\rm mean}}

\title[Fe line formation for late-type stars]{NLTE line formation of Fe for
late-type stars. I. Standard stars with 1D and $\td$ model atmospheres} 
\author[M. Bergemann]{Maria Bergemann$^1$\thanks{E-mail: 
mbergema@mpa-garching.mpg.de}, K. Lind$^1$, R. Collet$^{2,3,1}$, Z. Magic$^1$,
M. Asplund$^{4,1}$\\
$^1$ Max Planck Institute for Astrophysics, Karl-Schwarzschild Str. 1, 85741,
Garching, Germany \\
$^2$ Natural History Museum of Denmark, Centre for Star and Planet Formation,
{\O}ster Voldgade 5-7, DK--1350 Copenhagen, Denmark \\
$^3$ Astronomical Observatory / Niels Bohr Institute, Juliane Maries Vej 30,
DK--2100 Copenhagen, Denmark \\ 
$^4$ Australian National University, Ellery Crescent  Acton ACT 2601,
Australia}

\begin{document}

\date{Accepted Date. Received Date; in original Date}

\pagerange{\pageref{firstpage}--\pageref{lastpage}} \pubyear{2012}

\maketitle

\label{firstpage}

\begin{abstract}
We investigate departures from LTE in the line formation of Fe for a
number of well-studied late-type stars in different evolutionary stages. A new
model of Fe atom was constructed from the most up-to-date theoretical and
experimental atomic data available so far. Non-local thermodynamic equilibrium
(NLTE) line formation calculations for Fe were performed using 1D hydrostatic
MARCS and MAFAGS-OS model atmospheres, as well as the spatial and temporal
average stratifications from full 3D hydrodynamical simulations of stellar
convection computed using the Stagger code. It is shown that the Fe I/Fe II
ionization balance can be well established with the 1D and mean 3D models under
NLTE including calibrated inelastic collisions with H I calculated from the
Drawin's (1969) formulae. Strong low-excitation Fe I lines are very sensitive to
the atmospheric structure; classical 1D models fail to provide consistent
excitation balance, particularly so for cool metal-poor stars. A better 
agreement between Fe I lines spanning a range of excitation potentials is
obtained with the mean 3D models. Mean NLTE metallicities determined for the
standard stars
using the 1D and mean 3D models are fully consistent. Also, the NLTE
spectroscopic effective temperatures and gravities from ionization balance
agree with that determined by other methods, e.g., infrared flux method
and
parallaxes, if one of the stellar parameters is constrained independently.
\end{abstract}

\begin{keywords} Atomic data -- Line: profiles -- Line: formation -- Stars:
abundances \end{keywords} 
%
%
\section{Introduction} 

Accurate determination of basic stellar parameters is fundamental for
calculations of chemical composition, ages, and evolutionary stages of stars.
One of the most commonly used methods to determine effective temperature
$\Teff$, surface gravity $\logg$ and metallicity [Fe/H], is to exploit
excitation-ionization equilibria of various chemical elements in stellar
atmospheres. Iron, with its partly filled $3d$ subshell, has, by far, the
largest number of lines all over the spectrum of a typical late-type star. This
atomic property coupled to a relatively large abundance makes it a reference
element for spectroscopic estimates of stellar parameters.

The goal of this work is to study systematic uncertainties in this method, which
are related to 1) the assumption of local thermodynamic equilibrium (LTE) in
spectral line formation calculations, and 2) the use of theoretical 1D
hydrostatic model atmospheres. These two approximations are inherent in most of
the line formation codes utilized in spectroscopic studies, since they strongly
reduce the complexity of the problem permitting the analysis of very large
stellar samples in short timescales. Yet, in conditions when the breakdown of
LTE/1D modelling occurs the inferred stellar parameters may suffer from large
systematic biases (e.g., Asplund 2005). To assess the extent of the latter, more
physically realistic modeling is necessary.

Non-local thermodynamic equilibrium (NLTE) effects on the
\ion{Fe}{i}/\ion{Fe}{ii} level populations for FGK stars have been extensively
discussed in the literature (\citealt*{1972ApJ...176..809A};
\citealt*{1980ApJ...241..374C};
\citealt*{1990SvAL...16...91B}; \citealt{1999ApJ...521..753T};
\citealt{2001A&A...366..981G}; \citealt*{2001A&A...380..645G}; Collet et al.
2005; Mashonkina et al. 2011). These and other studies showed that NLTE effects
in the ionization balance of \ion{Fe}{i}/\ion{Fe}{ii} are large for giants and
metal-poor stars. The effect on solar-metallicity stars is smaller, but it is
must be taken into account if one aims at the accuracy of few percent, as is the
case for the Sun. Despite major efforts aimed at understanding how
non-equilibrium thermodynamics affects the line formation of Fe, there have been
only few attempts to quantify these deviations in a systematic manner across the
Hertzsprung-Russell diagram. \citet{2011mast.conf..314M} provided a small grid
of NLTE corrections to five \ion{Fe}{i} lines for the solar-metallicity stars
with $\Teff > 6500$ K and $\logg >3$. \citet{1999ApJ...521..753T} explored a
larger range of stellar parameters including FGK stars down to [Fe/H] $\approx
-3$, however, they did not present a regular grid of NLTE corrections.

Furthermore, it has only recently become possible to perform full
time-dependent, 3D, hydrodynamical simulations of radiative and convective
energy transport in stellar atmospheres. Such simulations (Nordlund \& Drawins
1990, Asplund et al. 1999, Collet et al. 2006, Ludwig et al. 2006,
\citealt*{2009LRSP....6....2N}, Freytag et al. 2010) have evidenced important
shortcomings of 1D, stationary, hydrostatic models. Especially at low
metallicity, it has been realized that 1D models, by necessarily enforcing
radiative equilibrium, overestimate the
average temperatures of shallow atmospheric layers, with profound implications
for the spectral line formation (\citealt*[e.g., ][]{2007A&A...469..687C}).

In this work, we perform NLTE line formation calculations for
\ion{Fe}{i} and \ion{Fe}{ii} using 1D hydrostatic and mean 3D model atmospheres
obtained from temporal and spatial averaging of 3D hydrodynamical simulations
(hereafter, $\td$). A new model of Fe atom is constructed from the most
up-to-date theoretical and experimental atomic data available so far.  Non-local
thermodynamic equilibrium Fe abundances, effective temperatures, and surface
gravities are derived for the Sun, Procyon, and four metal-poor stars. The
efficiency of thermalization caused by inelastic \ion{H}{i} collisions is
calibrated as to satisfy ionization equilibrium by scaling the classical Drawin
(1968, 1969) formulae. In the next paper in the series \citep*[][hereafter Paper
II]{Paper2}, we discuss NLTE effects on \ion{Fe}{i} and \ion{Fe}{ii} over a wide
range of stellar parameters. That paper also presents a large grid of 1D NLTE
abundance corrections for a wealth of lines in metal-rich and metal-poor dwarf
and giant spectra.

Before proceeding with the description of the methods, we shall point
out a few important aspects of our study. First, due to a comparative nature of
the analysis (1D LTE vs. 1D NLTE and $\td$ NLTE) no attempt is made to fine-tune
various parameters in order to achieve full agreement with other results in the
literature. Second, although by the use of the averaged 3D models we roughly
account for hydrodynamic cooling associated with convective overshoot in the
simulations, the effect of horizontal inhomogeneities is not addressed because
such calculations with our new realistic extensive model atom are beyond current
computational capabilities. We note, however, that our results are expected to
closely resemble any future full 3D NLTE calculations once these become
feasible. The reason is that in NLTE the \ion{Fe}{i} line formation is largely
dictated by the radiation field (as explained in detail below) originating in
deep atmospheric layers, where the significance of the atmospheric
inhomogeneities is greatly reduced. Detailed tests of the $\td$ models, which
involve comparison with other observable quantities, will be presented elsewhere
(Collet et al., in prep.). Full 3D NLTE calculations with a simpler model atom
of Fe will be a subject of a forthcoming publication.

The paper is structured as follows. Section~\ref{sec:methods} gives a brief
description of the input model atmospheres, model atom, and programs used to
compute NLTE populations and line formation. The results of statistical
equilibrium calculations are presented in Sec.~\ref{sec:statec}. The
analysis of the solar spectrum along with the re-determination of the Fe
abundances for the Sun  and Procyon is described in detail in
Sec.~\ref{sec:solspec}.
Section \ref{sec:abstars} presents and discusses metallicities, temperatures,
and gravities obtained for the metal-poor stars. Finally, a comparison with
stellar evolution calculations is given in Sec.~\ref{sec:evolution}. A short
summary of the work and conclusions are given in Sec.~\ref{sec:conclusions}.

%
%
\section{Methods}{\label{sec:methods}}
\begin{figure*}
\begin{center}
\includegraphics[scale=0.8]{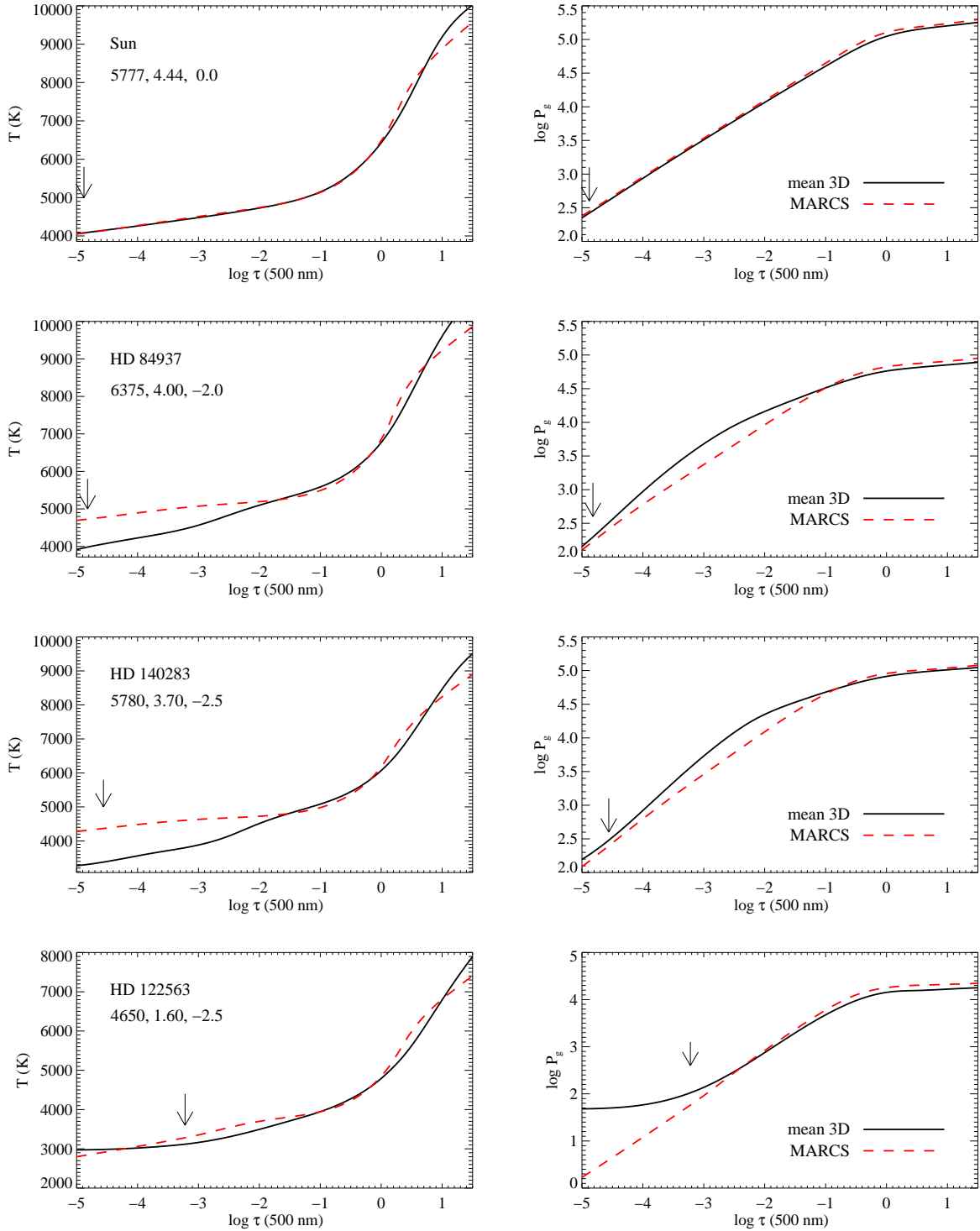}
\caption{Comparison of the temperature and gas pressure structures in the
\textsc{marcs} and $\td$ model atmospheres. Arrows indicate the optical depths,
above which the \textsc{marcs} models were extrapolated.} 
\label{atm_abs}
\end{center}
\end{figure*}
\subsection{Model atmospheres}{\label{sec:atmos}}
In the present study, we employ both \textsc{mafags-os}
(\citealt{2004A&A...420..289G}, \citealt*{2004A&A...426..309G},
\citealt*{2009A&A...503..177G}) and \textsc{marcs} \citep*{2008A&A...486..951G}
1D LTE hydrostatic model atmospheres, as well as 3D hydrodynamical models
computed with the Stagger code (Galsgaard \& Nordlund 1995, Collet et al. 2011).

The \textsc{mafags-os} models for all investigated stars were kindly provided
by F. Grupp. The models are plane-parallel with convective energy
transport based on the turbulent convection model of
\citet{1991ApJ...370..295C}. Compressible turbulence is accounted for using a
mixing length, $\alpha_{\rm cm} = l/\rm{H_p}$\footnote{$H_p$ is the pressure
scale height and $l$ the so-called 'mixing length'}$ = 0.82$. The reasons for
this choice are discussed in \citet{1998A&A...332..127B}. We note that this
$\alpha_{\rm cm}$ is also consistent with the results of
\citet*{1999ASPC..173..225F}, who attempted to calibrate the mixing length
parameter using the 2D radiation hydrodynamics simulations of stellar
convection. Line absorption is computed by the method of opacity sampling
including $\sim 86\,000$ wavelength points from extreme UV to far IR. Extensive
line lists were extracted from Kurucz and VALD databases. The reference solar
abundances were compiled from various literature sources, giving preference to
NLTE determinations by the Munich group \citep[e.g., ][]{2001A&A...366..981G}.
In particular, the solar Fe abundance is set to $\absun = 7.50$.

The \textsc{marcs} models are based on opacity sampling with $\sim 10^4 - 10^5$
wavelength points using Kurucz and VALD linelists. Convection is included in the
\citet*{1965ApJ...142..841H} version of the mixing length theory with the mixing
length parameter set to $\alpha_{\rm MLT} = 1.5$. A detailed description of the
models can be found on the MARCS
web-site\footnote{http://marcs.astro.uu.se/index.php}. For the majority of the
elements, the reference solar abundances are that of Asplund et al. (2005).

Three-dimensional radiation-hydrodynamic simulations of convection at the
surface of the Sun and reference stars were  computed using the Copenhagen
Stagger code (\citealt*{nordlund95}). The physical domains of the simulations
cover a representative portion of the stellar surface. They include the whole
photosphere as well as the upper part of the convective layers, typically
encompassing $12$ to $15$ pressure scale heights vertically. Horizontally, they
extend over an area sufficiently large to host about ten granules at the
surface. The simulations use realistic input physics, including state-of-the-art
equation-of-state, opacities, and treatment of non-grey radiative transfer.  The
adopted reference solar chemical composition for the simulations is that of
\citet*{2009ARA&A..47..481A}. For a more detailed description of the
simulations' setup, we refer to \citet*{2011JPhCS.328a2003C}.

For the purpose of the current study, we computed spatial and temporal averages
of these simulations over surfaces of equal optical depth $\opd$ at the chosen
reference wavelength at $5000$ \AA. The independent thermodynamic variables, gas
density $\rho$ and internal energy per unit mass $\varepsilon$, as well as the
temperature $T$ are first interpolated with cubic splines for all columns in the
full 3D structure to the reference optical depth scale. The reference scale was
constructed to cover the relevant range for line-formation calculations
(${-5}{\la}~{\opd}~{\la}{2}$) evenly in $\opd$. For density and internal energy,
a logarithmic interpolation is used. Other physical quantities, namely gas
pressure $P_{\rm gas}$ and electron number density $n_{\rm el}$, are looked up
from the simulations' equation-of-state tables as a function of density and
internal energy. Finally, mean $\td$ stratifications are constructed by
averaging the various physical quantities pertinent to line-formation
calculations, $\ln{\rho}$, $T$, $\ln{P_{\rm gas}}$, and $\ln{n_{\rm el}}$, on
surfaces of equal optical depth $\opd$ and over time (i.e., over all simulation
snapshots). We emphasize that no hydrostatic equilibrium was enforced after the
averaging process. In the present application of mean $\td$ stratifications to
line-formation calculations, averages of velocity fields were not considered,
and line broadening associated with bulk gas flows and turbulence was accounted
for by means of a classical, depth-independent micro-turbulence parameter
(Sect.~\ref{sec:methods}) as in 1D models.

The temperature and pressure stratifications from the 1D \textsc{marcs} model
atmospheres and the mean $\td$ hydrodynamical models of the Sun and three
metal-poor standard stars are shown in Fig.~\ref{atm_abs} as a function of
continuum optical depth at $5000$~{\AA}.The input parameters are the ones listed
in Table~\ref{tab:init_param}. The \textsc{mafags-os} models are not included in
the plots, because they adopt slightly different stellar parameters. Due to the
similar equation-of-state, \textsc{marcs} and \textsc{mafags-os} model
atmospheres are essentially identical in the outer layers. The differences
between 1D and $\td$ models are more pronounced, especially in the case of
metal-poor stellar atmospheres. At low metallicities, 3D stellar surface
convection simulations predict cooler upper photospheric stratifications
(\citealt*{1999A&A...346L..17A}, \citealt*{2007A&A...469..687C}), than
corresponding classical, hydrostatic, stationary 1D model atmospheres generated
for the same stellar parameters.
The temperature in the outer layers of time-dependent, 3D, hydrodynamical
simulations is mainly regulated by two mechanisms: radiative heating due to
reabsorption of continuum radiation by spectral lines and adiabatic cooling
associated with expanding gas above granules. At low metallicities, the coupling
between radiation and matter is weakened with respect to the solar-metallicity
case because of the decreased line opacities; the adiabatic cooling term
therefore prevails, causing the thermal balance to shift toward on average lower
temperatures. Stationary, 1D, hydrostatic models do not account for this cooling
term associated with diverging gas flows, and the thermal balance in the upper
photosphere is controlled by radiative heating and cooling only, ultimately
leading, at low metallicities, to higher equilibrium temperatures than predicted
by 3D models (Fig.~\ref{atm_abs}, left panels).
The lower average temperatures in the outer layers of metal-poor 3D models also
imply smaller values of the pressure scale height and, consequently, steeper
pressure stratifications on a geometrical scale with respect to corresponding 1D
models. However, the lower temperatures also result in lower opacities;
therefore, on an optical depth scale, the average gas pressure in the optically
thin layers of the metal-poor 3D models appears typically higher at a given
optical depth than in 1D models (Fig.~\ref{atm_abs}, right panels).
%
%
\subsection{Statistical equilibrium codes}{\label{sec:nlte_codes}}

The NLTE level populations of \ion{Fe}{i} and \ion{Fe}{ii} were computed with an
updated version of the \textsc{detail} code \citep*{Butler85} and
\textsc{multi} \citep*{1986UppOR..33.....C, 1992ASPC...26..499C}. In both
codes, the solution of the coupled statistical equilibrium and radiative
transfer equations is obtained using an approximate lambda iteration method. In
\textsc{detail}, the latter is implemented following a $\Psi$-operator approach
of the kind described by \citet*{1991A&A...245..171R, 1992A&A...262..209R},
which allows for self-consistent treatment of overlapping transitions and
continua. \textsc{multi} is based on the method described by
\citet*{1985JCoPh..59...56S} with the local operator by
\citet*{1986JQSRT..35..431O}. Scattering in the bound-bound (hereafter, b-b)
transitions included in the model atom follows complete frequency
redistribution. Upper boundary conditions differ in that \textsc{detail} assumes
no incoming radiation at the top, whereas \textsc{multi} estimates the minor
contribution from the optically thin gas and uses a second order Taylor
expansion of the Feautrier variables. Ng convergence acceleration
\citep{1974JChPh..61.2680N,1987nrt..book..101A} is implemented in both codes.

In both codes, the background line opacity is computed with a Planckian
source function. However, there is one difference. In \textsc{detail}, line
opacity is consistently added at the frequencies of all b-b and b-f transitions
in the NLTE atom. In \textsc{multi}, metal line opacities were added to the
continuous opacities for the calculation of photoionization rates only, as
described by \citet{2005A&A...442..643C}, while blends were neglected in the
calculation of the bound-bound radiative rates. This approximation is well
justified and saves computational time. Firstly, Fe itself dominates line
opacity in the UV. Secondly, according to our tests, the upper limit to the
differences in the NLTE equivalent widths computed with and without blends is
$\sim 0.2$ percent for the Sun and $5$ percent for HD 122563.

When solving for statistical equilibrium, Fe line profiles were computed with a
Gaussian in \textsc{detail} and with Voigt profiles in \textsc{multi}. Although
the latter is also possible with \textsc{detail}, it is un-necessary. Our tests
showed that NLTE effects in \ion{Fe}{i} are insensitive to the adopted profile
functions. Once the level populations were converged with \textsc{detail}, the
synthetic line profiles were computed with SIU \citep{reetz}.

Finally, we remark on the handling of the equation-of-state in the NLTE codes.
Whereas \textsc{multi} includes a subroutine to compute ionization fractions and
molecular equilibria internally, \textsc{detail} and \textsc{siu} require
partial pressures of all atoms and important molecules to be supplied with a
model atmosphere. These are included in the \textsc{mafags-os} models. For the
\textsc{marcs} and $\td$ models, we computed the atomic partial pressures using
the equation-of-state package built in \textsc{multi}.
%
%
\subsection{Model atom of Fe}{\label{sec:modelatom}} 
\subsubsection{Levels and radiative transitions}
\begin{figure*} 
\resizebox{\textwidth}{!}{\rotatebox{90}
{\includegraphics{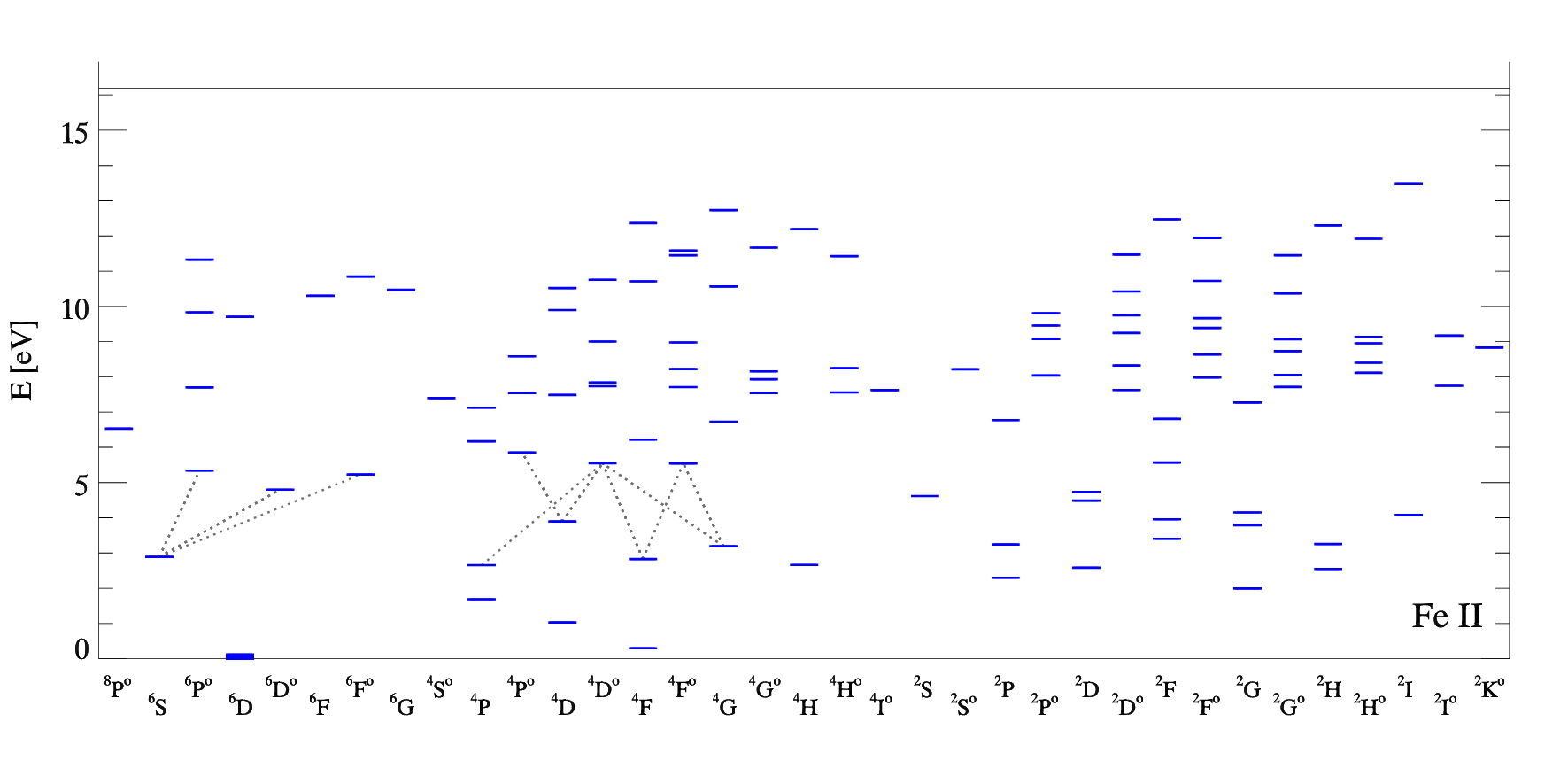}}}
\resizebox{\textwidth}{!}{\rotatebox{90}
{\includegraphics{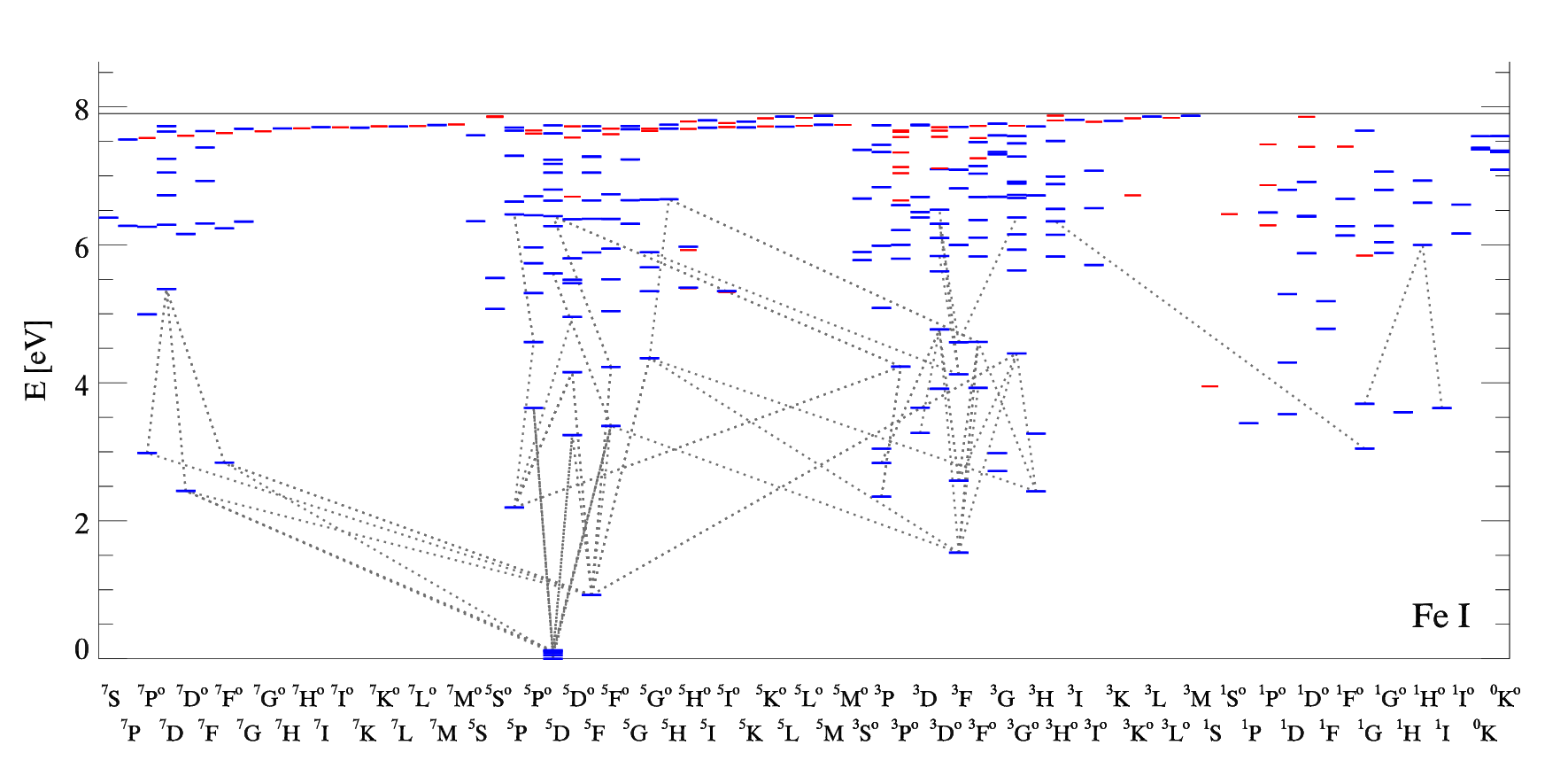}}} 
\caption[]{Grotrian diagram of the \ion{Fe}{ii} (top) and \ion{Fe}{i} atoms
(bottom). Predicted levels are shown in red. Only lines selected for the
abundances analysis are shown.}
\label{fe_grotrian} 
\end{figure*}
The number of levels and discrete radiative transitions in Fe is enormous.
Recent calculations of Kurucz\footnote{http://kurucz.harvard.edu/atoms.html}
predict $\sim 37\,000$ theoretical energy levels below and above the first 
ionization threshold of \ion{Fe}{i}, as well as $6\,025\,000$
radiatively-permitted transitions between them. Such atomic models are not
tractable with our NLTE codes; thus, we combined atomic levels and transitions
into super-levels and super-lines (see below). In our
final model, which also includes all experimental data from the NIST atomic
database \citep{2012AAS...21944301R}, the number of energy
levels is $296$ for \ion{Fe}{i} and $112$ for \ion{Fe}{ii}, with uppermost
excited levels located at $0.03$ eV and $2.72$ eV below the respective
ionization limits, $7.9$ eV and $16.19$ eV. The model is closed by the
\ion{Fe}{iii} ground state. The total number of radiatively-allowed transitions
is $16\,207$ ($13\,888$ \ion{Fe}{i} and $2\,316$ \ion{Fe}{ii}). Fine structure
was neglected for all levels, but the ground states of \ion{Fe}{i}
\Fe{a}{5}{D}{}{} (configuration $1s^2 2s^2 2p^6 3s^2 3p^6 3d^6 4s^2$) and
\ion{Fe}{ii} \Fe{a}{6}{D}{}{}. Excitation energy of each LS term is a weighted
mean of statistical weights and excitation energies of fine structure levels.
The mean wavelength for a multiplet is computed from the weighted energy levels.

All predicted energy levels of \ion{Fe}{i} with the same parity above $\Elow
\geq 5.1$ eV were grouped into superlevels. For the upper levels above $54000$
cm$^-1$ ($6.7$ eV), we combined all levels within $1000$ cm$^{-1}$ ($0.12$
eV)\footnote{For a typical F-type star ($\Teff = 6000$ K), the thermal energy is
$kT \sim 0.4$ eV.}, whereas below this energy limit only levels deviating by
less than $10$ cm$^{-1}$ ($0.001$ eV) were combined. Thus, not only predicted,
but also some experimental levels above $5.1$ eV were combined to superlevels.
None of the Fe lines selected for the subsequent abundance analysis has either
the lower or the upper level combined.

Transitions between the components of the superlevels were also grouped
preserving the parity and angular momentum conservation rules. The total
transition probability of a superline is a weighted average of $\log gf$'s of
individual transitions, and is computed in analogy to a $gf$-value for a
multiplet \citep*{1988atps.book.....M}:
\begin{equation} f_{\rm mean} = \frac{1}{\bar{\lambda}_{\rm mean} \sum_{l} g_l}
\sum_{l,u} g_l \times \lambda_{\rm lu} \times f_{\rm lu} 
\end{equation} 
where $l$ and $u$ are indices of lower and upper levels of the un-grouped
transition, $\bar{\lambda}_{\rm mean}$ is the mean Ritz wavelength of a
superline. Grotrian diagram of the Fe model atom is shown in Fig.
\ref{fe_grotrian}. Although the accuracy of individual $gf$-values for the
multitude of theoretically-computed transitions is hard to quantify, we expect
our model to be a good representation of the global atomic properties of
the \ion{Fe}{i}/\ion{Fe}{ii} and provide physically valid description of
statistical equilibrium of the atom in the atmospheres of late-type stars.

Accurate radiative bound-free cross-sections for \ion{Fe}{i} computed using the
close-coupling method were kindly provided by M. Bautista (private communication
2011, \citealt[see also][]{1997A&AS..122..167B}). These data are
computed on a more accurate energy mesh and provide better resolution of
photoionization resonances compared to the older data, e.g. provided in the
TOPbase. Thus, $136$ levels of \ion{Fe}{i}, including different multiplet
systems, from singlets to septets, are represented by quantum-mechanical data.
Hydrogenic approximation was used for the other levels.

\subsubsection{Collisional transitions}

The rates of transitions induced by inelastic collisions with free electrons
(e$^-$) and \ion{H}{i} atoms were computed using different recipes. For states
coupled by allowed b-b and b-f transitions, we used the formulas of
\citet{1962ApJ...136..906V} and \citet{1962amp..conf..375S}, respectively, in
the case of e$^-$ collisions and that of \citet{1984A&A...130..319S}, in the
case of \ion{H}{i} collisions. Also, all states are coupled by forbidden
transitions induced by $e^-$ collisions using the formula of
\citet{1973asqu.book.....A} and by \ion{H}{i} collisions with the formula of
\citet{1994PASJ...46...53T}.

We constrain the efficiency of collisions with \ion{H}{i} empirically (Sect.
\ref{sec:solspec}, \ref{sec:abstars}). We apply a scaling factor $\SH$ to the
cross-sections, which is iteratively derived from the constraint of
\ion{Fe}{i}/\ion{Fe}{ii} ionization equilibrium for the standard stars with
independently-fixed stellar parameters. As pointed out by
\citet{2011A&A...530A..94B}, such approach is an over-simplification of the
problem. However, at present there are no useful alternatives to this classical
recipe. Evidence for the necessity of including inelastic collisions with
\ion{H}{i} in statistical equilibrium calculations for Fe and other elements has
been demonstrated in many studies \citep[e.g., ][]{2011A&A...528A..87M}, which
is also confirmed by us (Sect. \ref{sec:solspec}, \ref{sec:abstars}).
%
%
\section{Statistical equilibrium of Fe}{\label{sec:statec}}
In the following section, we present the results of statistical equilibrium
calculations for Fe obtained with \textsc{detail} and \textsc{multi}.
We briefly describe the physical processes responsible for deviations from LTE
level populations in the \ion{Fe}{i} and \ion{Fe}{ii} atoms for atmospheric
conditions typical of FGK stars. A comprehensive description of NLTE effects as
a function of stellar parameters is deferred to the second paper of this series 
\citep{Paper2}. The main aim here is to understand the differences in terms of
NLTE \ion{Fe}{i} line formation with 1D and $\td$ model atmospheres, which is
necessary in order to explain the large differences between the spectroscopic
values of $\Teff$, $\log g$, and [Fe/H] obtained for the reference stars in the
two cases (Sect. \ref{sec:abstars}).

\ion{Fe}{i} is a minority ion in the atmospheres of late-type stars, which 
increases its sensitivity to NLTE over-ionization. The overall effect is that,
compared to LTE, the statistical equilibrium of Fe favors lower number
densities of the neutral ion, \ion{Fe}{i}, although the number densities of
relevant \ion{Fe}{ii} levels remain nearly thermalized. In general,
this leads to a weakening of \ion{Fe}{i} lines compared to LTE that, in turn,
requires larger Fe
abundance to fit a given observed spectral line. The actual magnitude of
departures and NLTE abundance corrections depends on stellar parameters (Sect.
\ref{sec:abstars}).
\begin{figure*}
\centering
\hbox{
\resizebox{\columnwidth}{!}{\includegraphics[scale=1.0]
{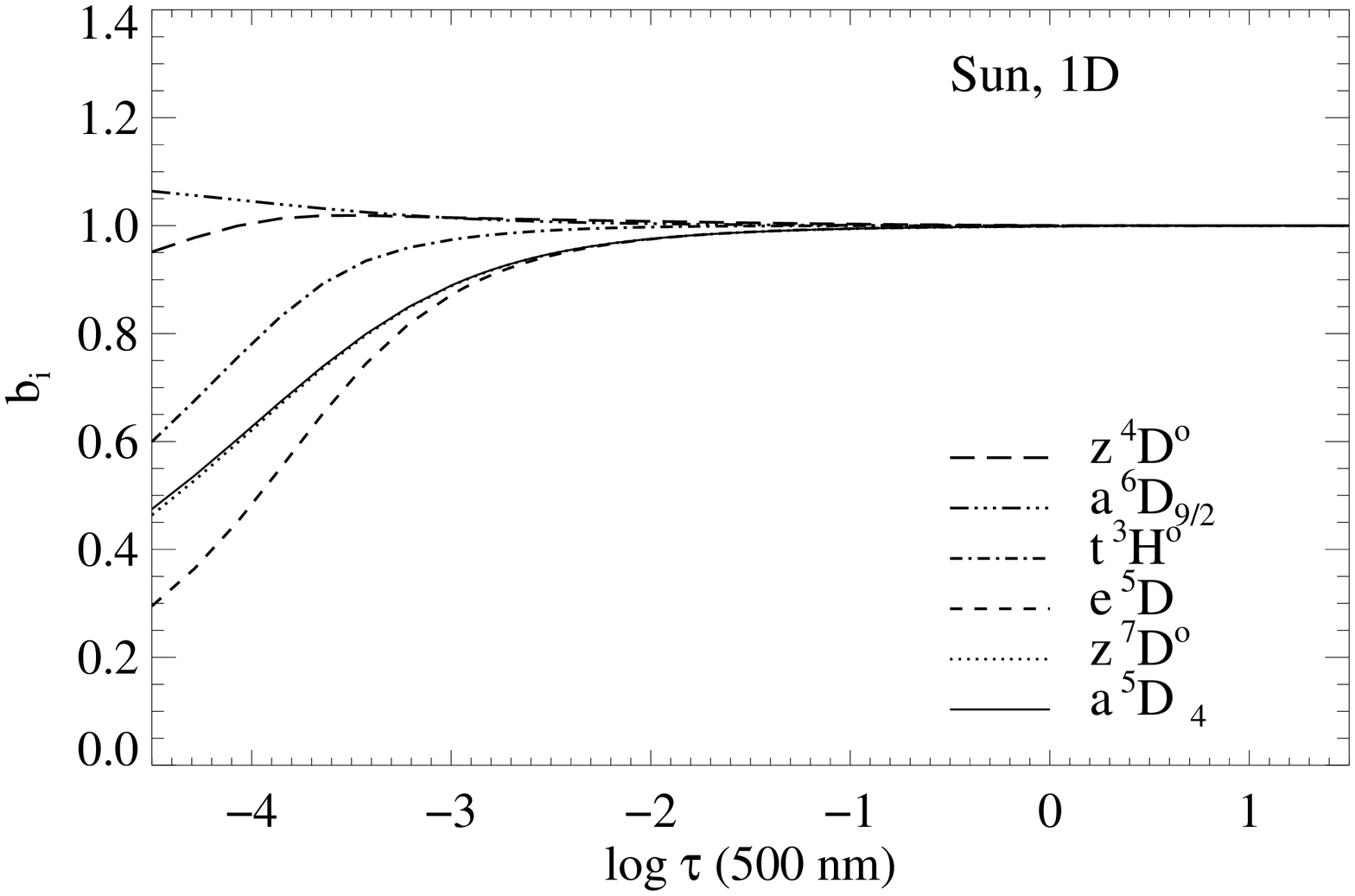}}\hfill
\resizebox{\columnwidth}{!}{\includegraphics[scale=1.0]
{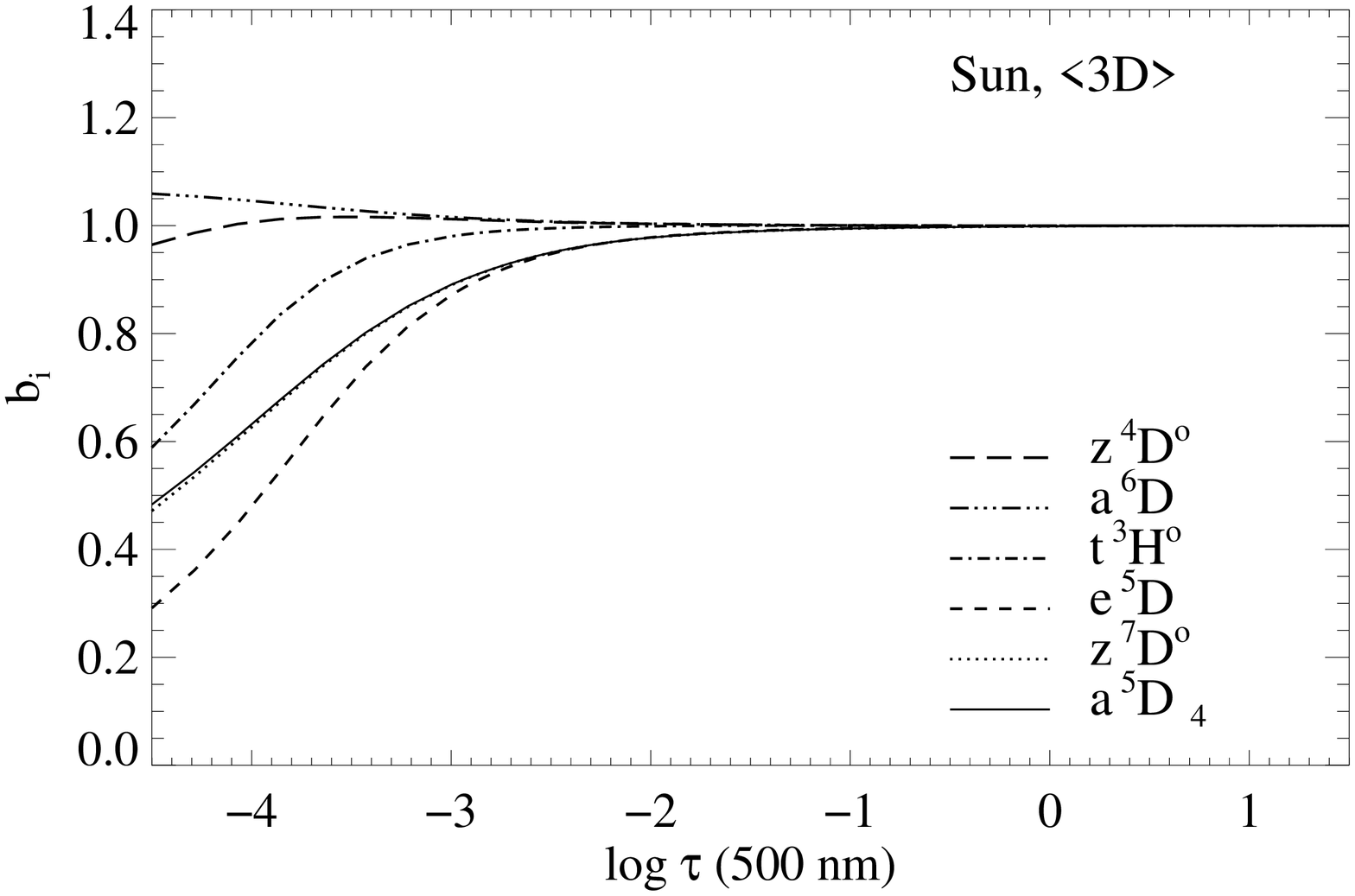}}} 
\hbox{
\resizebox{\columnwidth}{!}{\includegraphics[scale=1.0]
{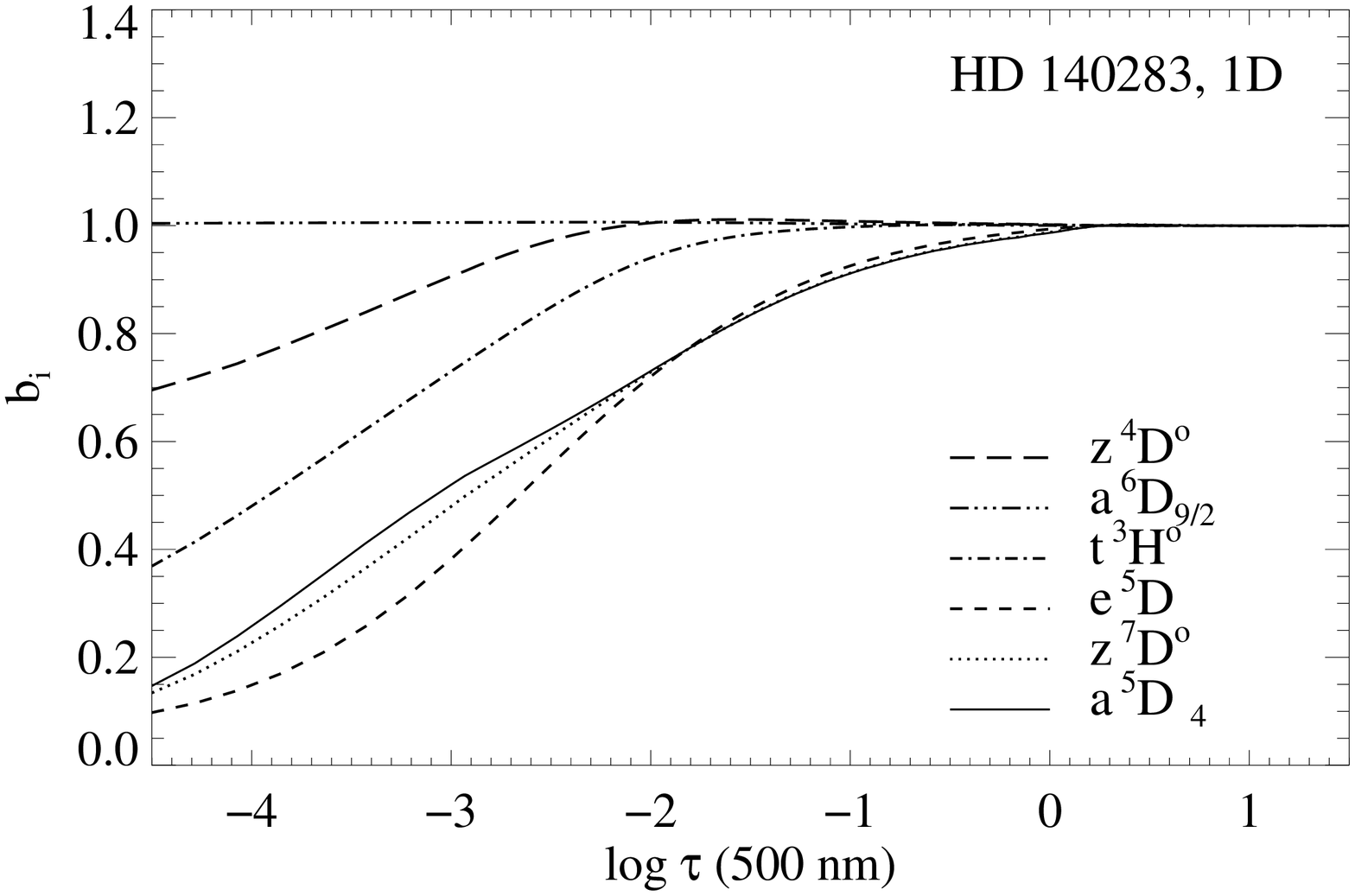}}
\hfill
\resizebox{\columnwidth}{!}{\includegraphics[scale=1.0]
{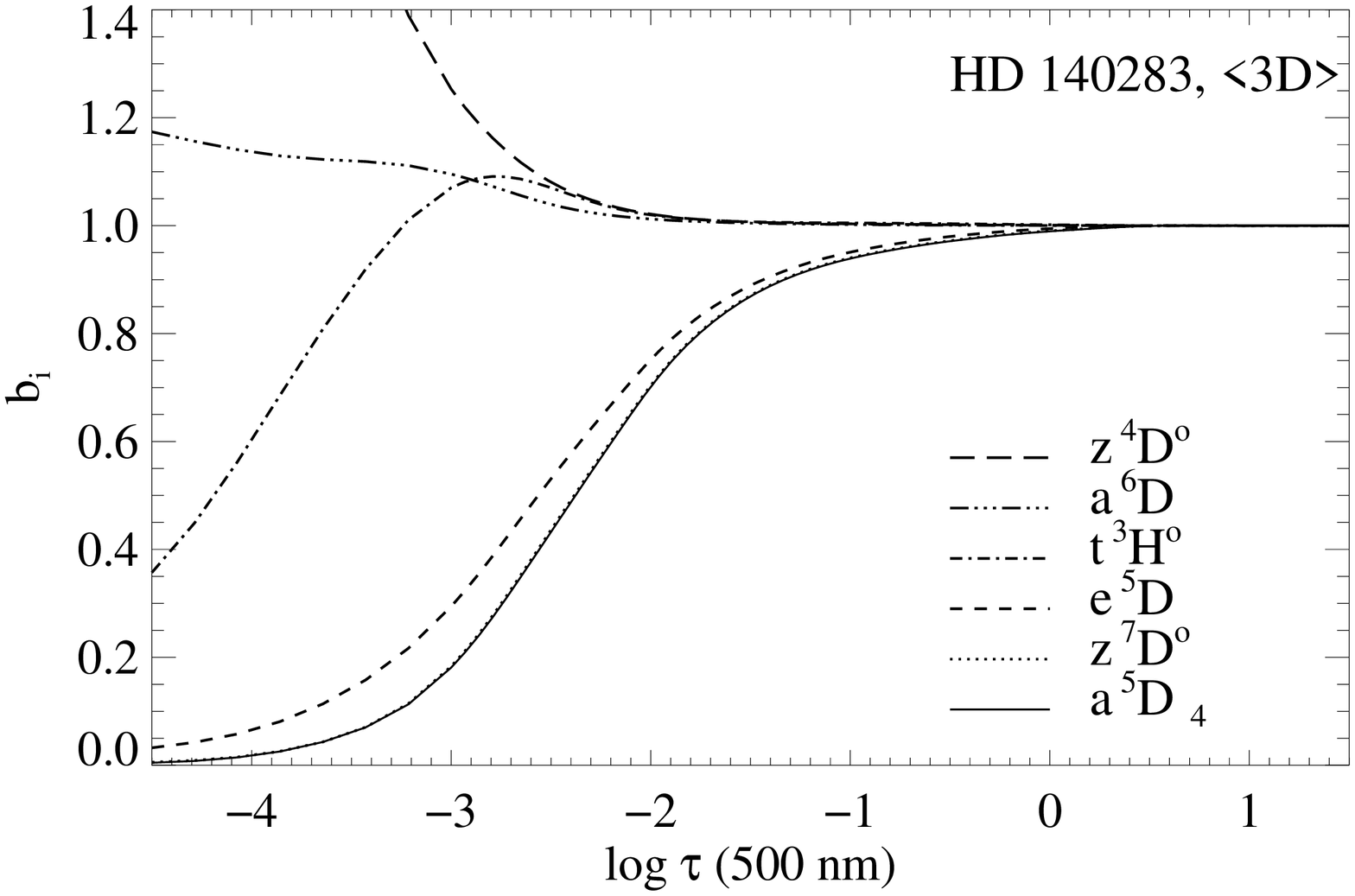}}}
\hbox{
\resizebox{\columnwidth}{!}{\includegraphics[scale=1.0]
{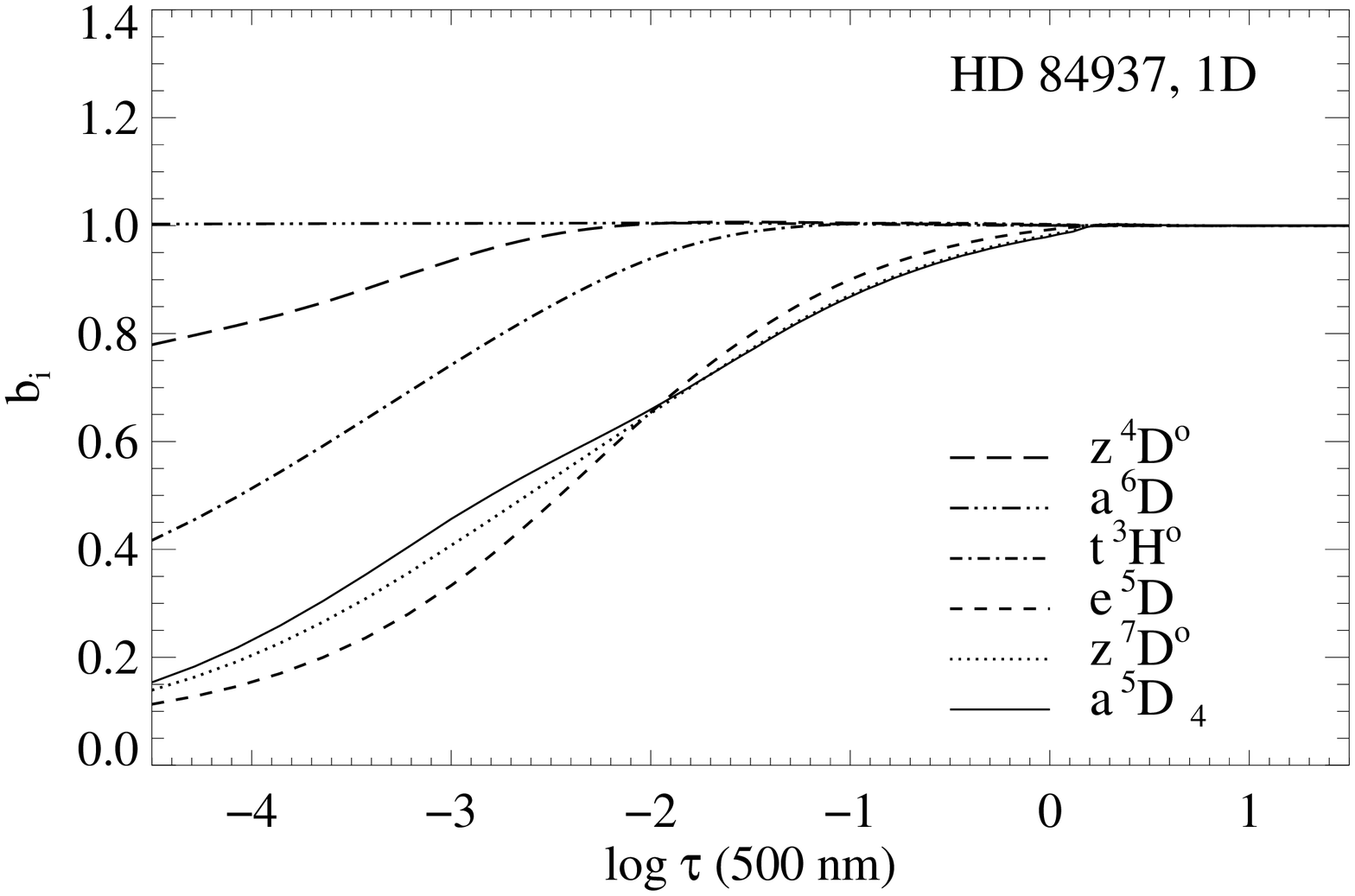}}\hfill
\resizebox{\columnwidth}{!}{\includegraphics[scale=1.0]
{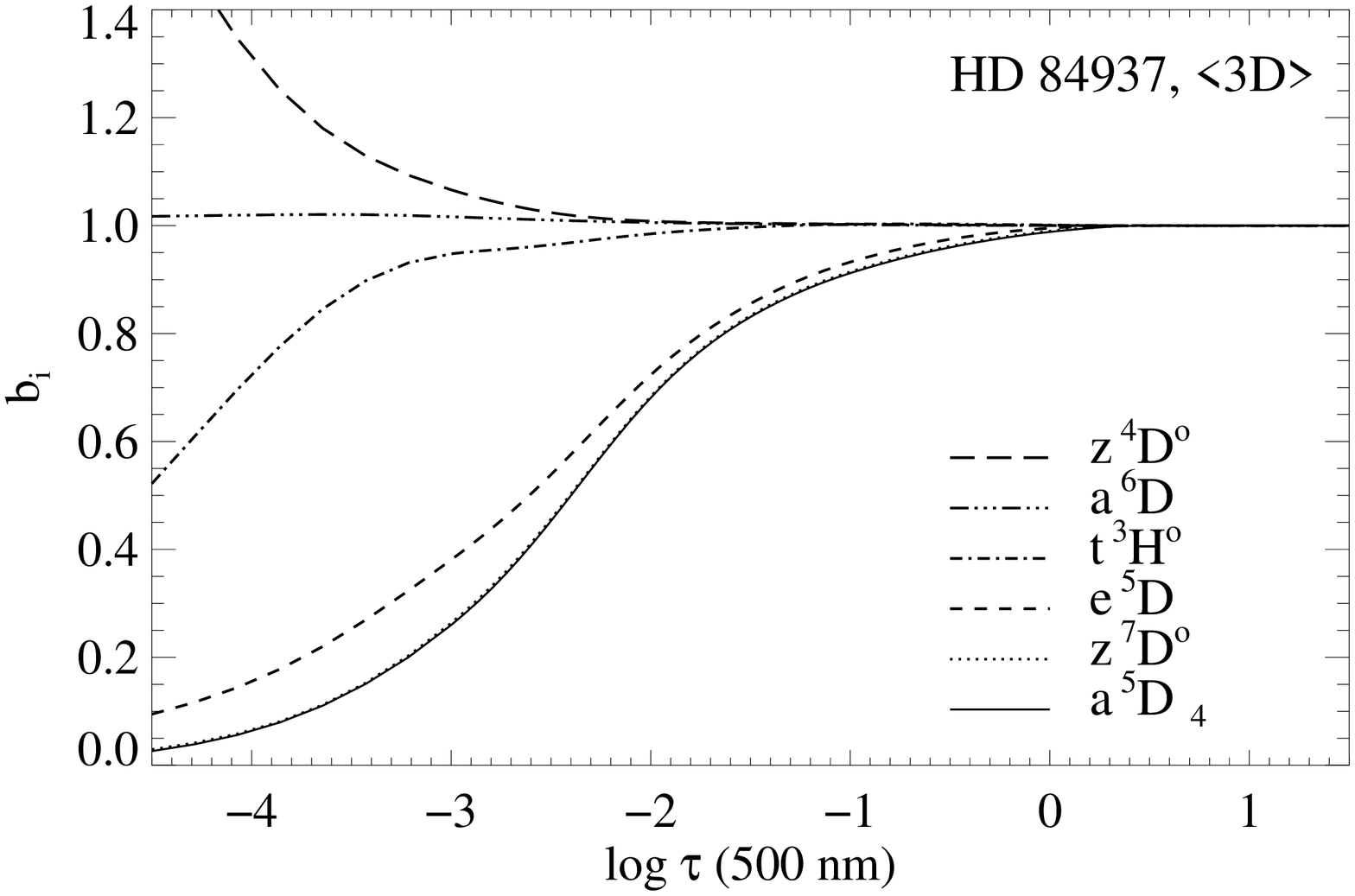}}}
\hbox{
\resizebox{\columnwidth}{!}{\includegraphics[scale=1.0]
{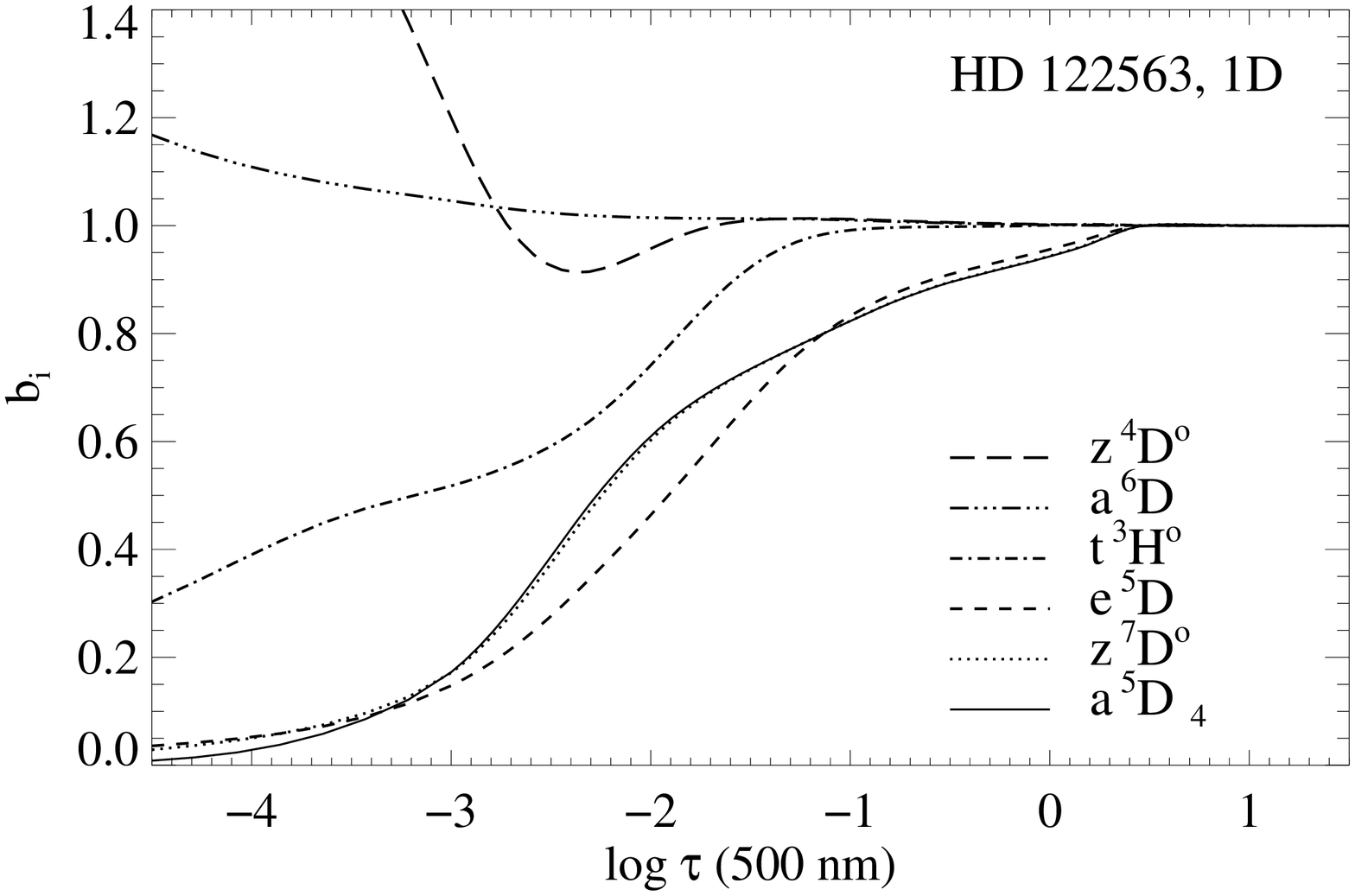}}\hfill
\resizebox{\columnwidth}{!}{\includegraphics[scale=1.0]
{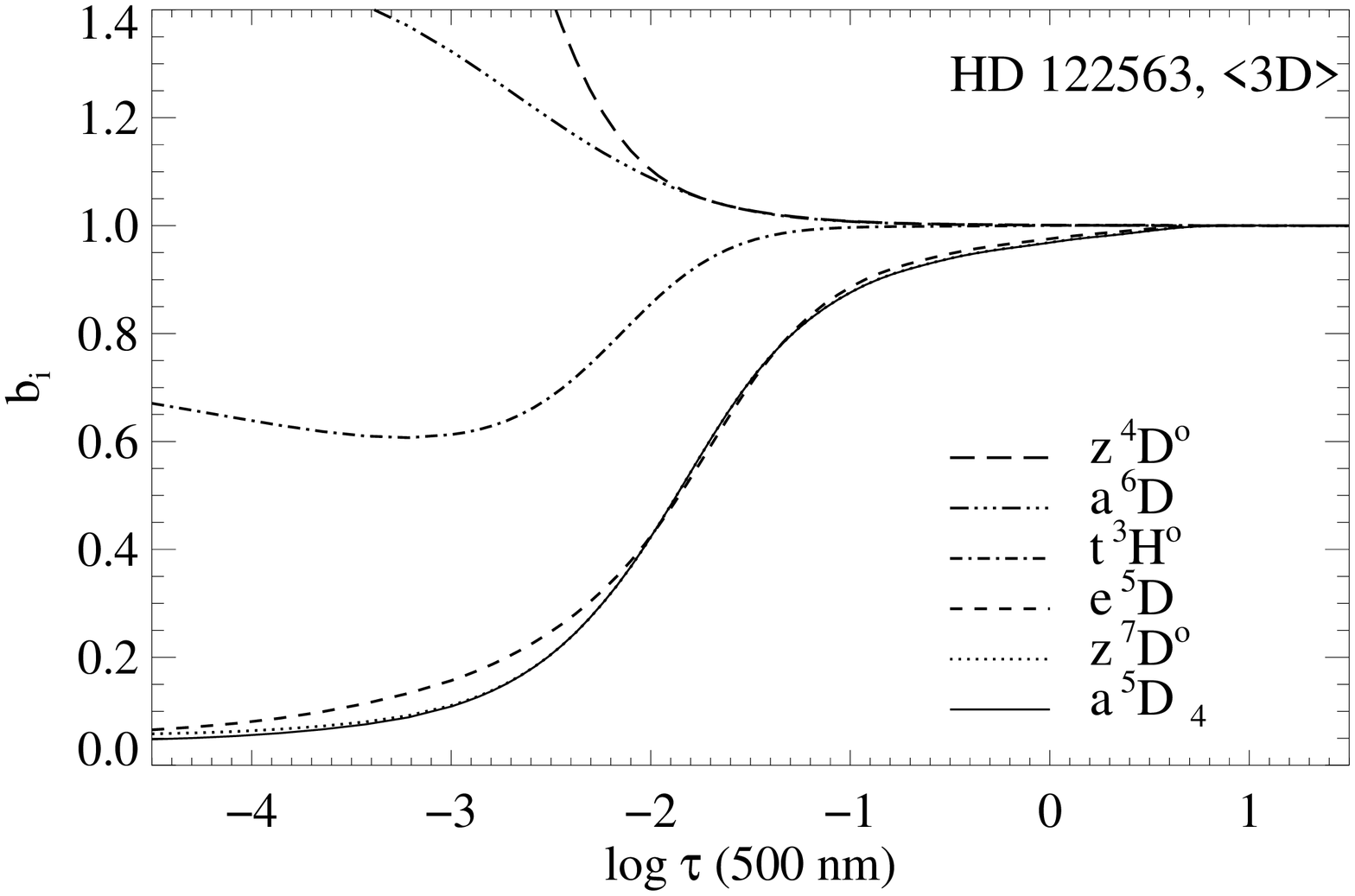}}}
\caption{Departure coefficients for the four reference stars computed with
the 1D \textsc{marcs} (left panel) and $\td$ (right panel) model atmospheres.}
\label{dep}
\end{figure*}

Figure~\ref{dep} presents the departure coefficients\footnote{We follow the
definition of \cite{1972SoPh...23..265W}, in which $b_i$ is a ratio of NLTE to
LTE atomic level populations, $b_i = n_i^{\rm NLTE}/n_i^{\rm LTE}$.} of selected
\ion{Fe}{i} and \ion{Fe}{ii} levels computed for the \textsc{marcs} and $\td$
model atmospheres with \textsc{multi}. The results are shown for the Sun, the
metal-poor subdwarf HD 84937, the metal-poor sub-giant HD 140283 and the
metal-poor giant HD 122563. Only a few, selected levels, typical for the
dependence of their associated departure coefficients with depth, are included
in the plot: \Fe{a}{5}{D}{}{4} (ground state of \ion{Fe}{i}),
\Fe{z}{7}{D}{\circ}{} ($2.4$ eV), \Fe{e}{5}{D}{}{} ($5.4$ eV),
\Fe{t}{3}{H}{}{\circ}, and two levels, which belong to the \ion{Fe}{ii} ion:
\Fe{a}{6}{D}{}{9/2} (ground state) and \Fe{z}{4}{D}{\circ}{} ($5.5$ eV). In the
optically thin atmospheric layers these and majority of other \ion{Fe}{i} levels
are underpopulated compared to LTE, $b_i < 1$. For all stars, but HD 122563, the
\ion{Fe}{ii} number densities remain close to LTE values throughout the
atmosphere, $b_i \approx 1$, and a minor overpopulation of the \ion{Fe}{ii}
ground state develops only close to the outer atmospheric boundary.
\begin{figure}
\resizebox{\columnwidth}{!}{\includegraphics[scale=1]
{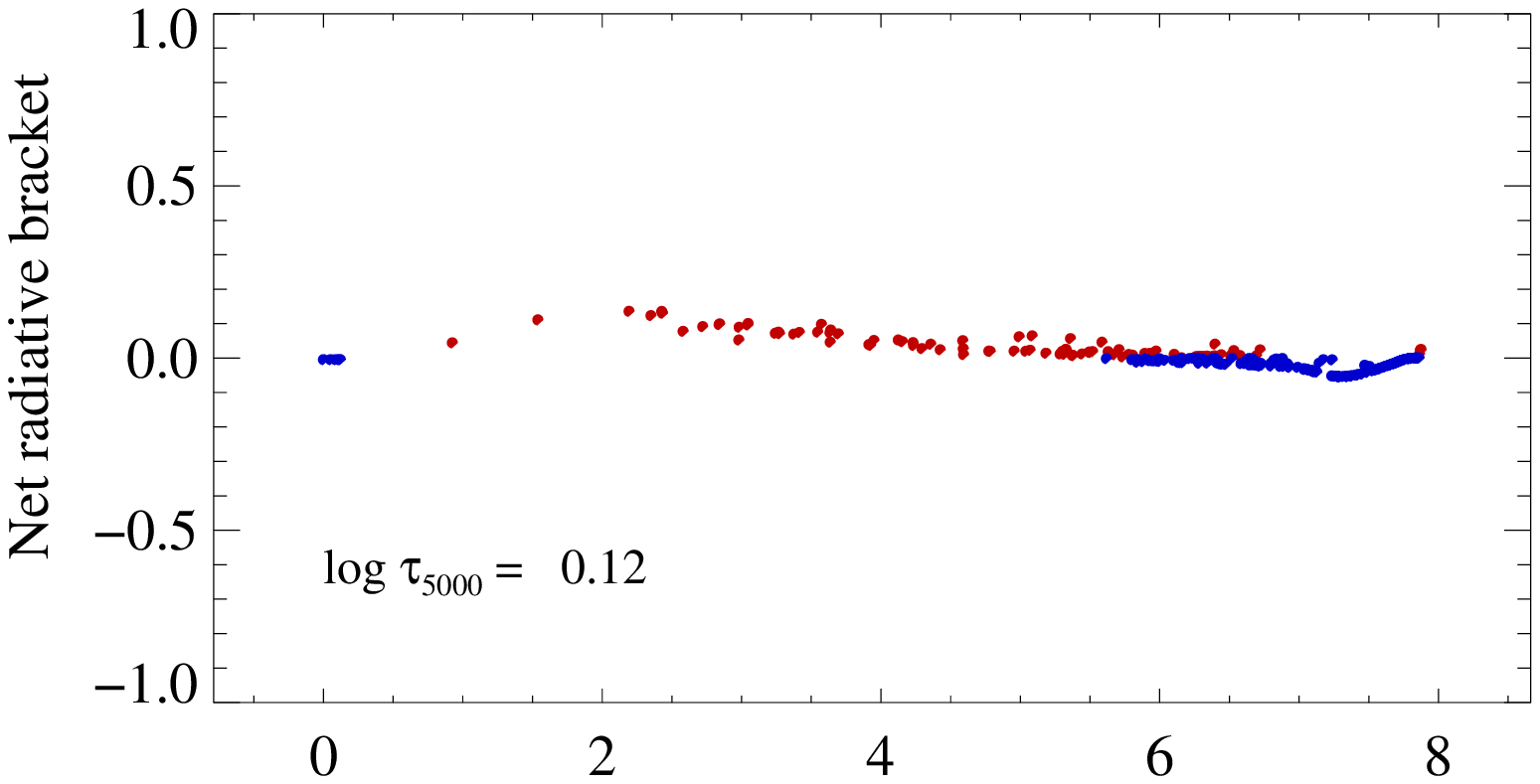}}
\resizebox{\columnwidth}{!}{\includegraphics[scale=1]
{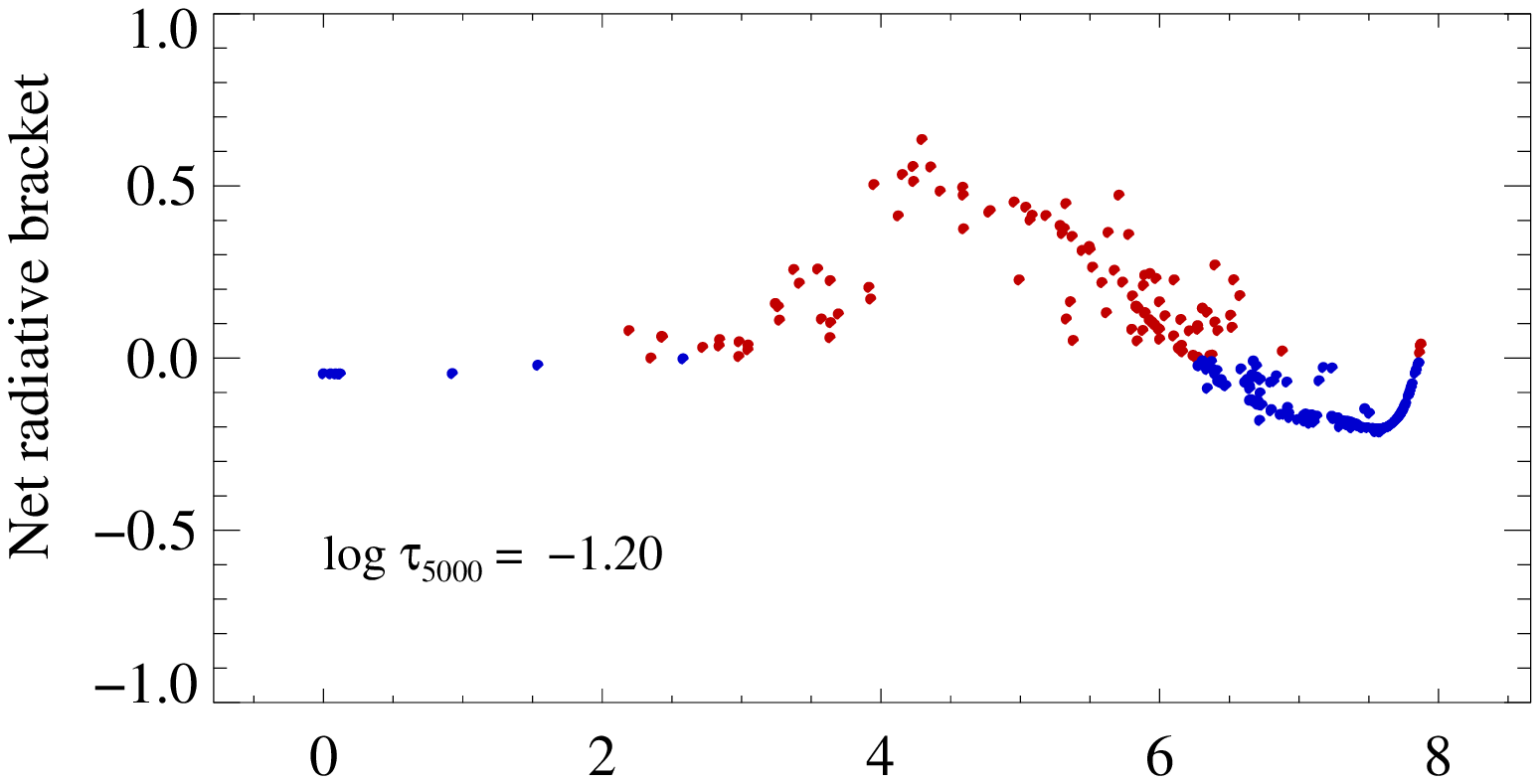}}
\resizebox{\columnwidth}{!}{\includegraphics[scale=1]
{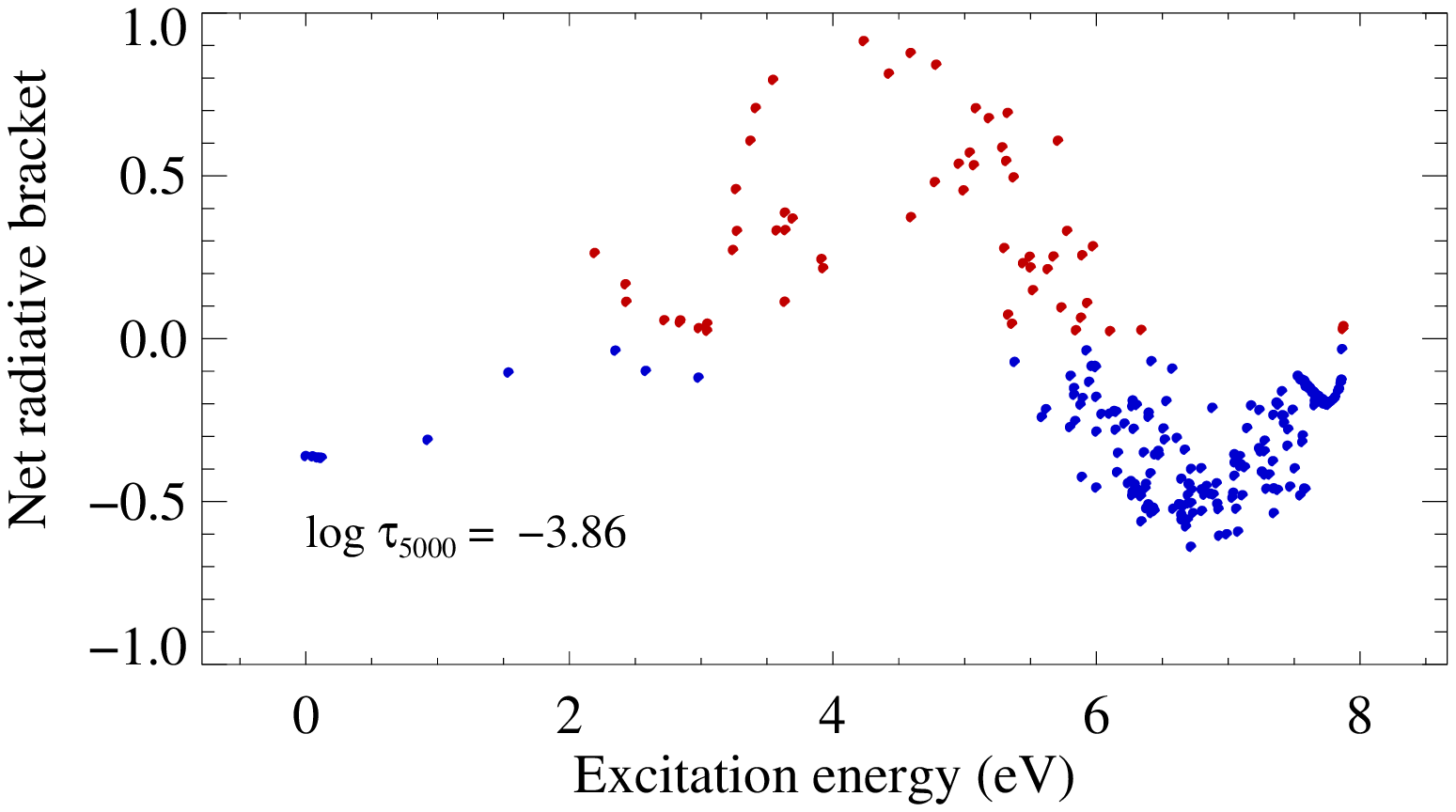}}
\caption{Net bound-free radiative brackets computed for the solar
\textsc{marcs} model atmosphere (top and middle panel). Optical depth is
indicated in each panel. Red and blue dots indicate over-ionization and
over-recombination, respectively.} 
\label{rates}
\end{figure}

Deviations from LTE in the distribution of atomic level populations arise
because the mean radiation field, $J_\nu$, at different depths and frequencies
is not equal to the Planck function, $B_\nu [T_{\rm e}(\tau)]$. For \ion{Fe}{i},
excess of the mean intensity over Planck function in the UV continua leads to
over-ionization, which sets in at $\opd \approx 0.2$, i.e., as soon as the
optical depth in the photoionization continuum of low-excitation \ion{Fe}{i}
levels with $E \approx 2$ eV falls below unity (Fig. \ref{dep}a).

In the layers with $\opd < -1$, the dominant mechanism is over-ionization from
the \ion{Fe}{i} levels with excitation energies at $2-5$ eV. The lower-lying
levels, including the ground state of \ion{Fe}{i}, maintain underpopulation due
to radiative and collisional coupling with the former. Excitation balance of
\ion{Fe}{i} is also mildly affected by radiative imbalances in line transitions.
These include radiative pumping by the non-local UV radiative field, as well as
photon suction driven by photon losses in large-probability \ion{Fe}{i}
transitions between highly-excited levels. These processes leave a
characteristic imprint on the behavior of $b_i$-factors in the outer atmospheric
layers, $\opd < -2$. In the infrared continuum, $J_\nu < B_\nu$ leads to
over-recombination, which is very efficient for our atomic model with only
$0.03$ eV energy gap of the upper \ion{Fe}{i} levels from the \ion{Fe}{ii}
ground state. This is illustrated in Fig. \ref{rates}, which shows net radiative
brackets\footnote{Net radiative brackets $\rho$ are defined as $\rho R_{ci}n_{c}
= R_{ci}n_{c} - R_{ic}n_{i}$, where $R_{ci}$ are rate coefficients for radiative
transitions from the continuum $c$ to a bound level $i$ of an atom and $n_{c}$
are atomic level populations \citep{1973ARA&A..11..187M}.} for all \ion{Fe}{i}
levels in the model atom at the depths $\opd = 0.12$, $-1.2$, and $-3.86$. All
\ion{Fe}{i} levels with excitation energy between $2$ and $6$ eV experience net
over-ionization, and the loop is closed by net over-recombination to the upper
levels. Note that only radiative rates are plotted. We also compared absolute
radiative rates from \textsc{multi} and \textsc{detail} and found that they
agree very well with each other.
\begin{figure*}
\includegraphics[scale=0.5]{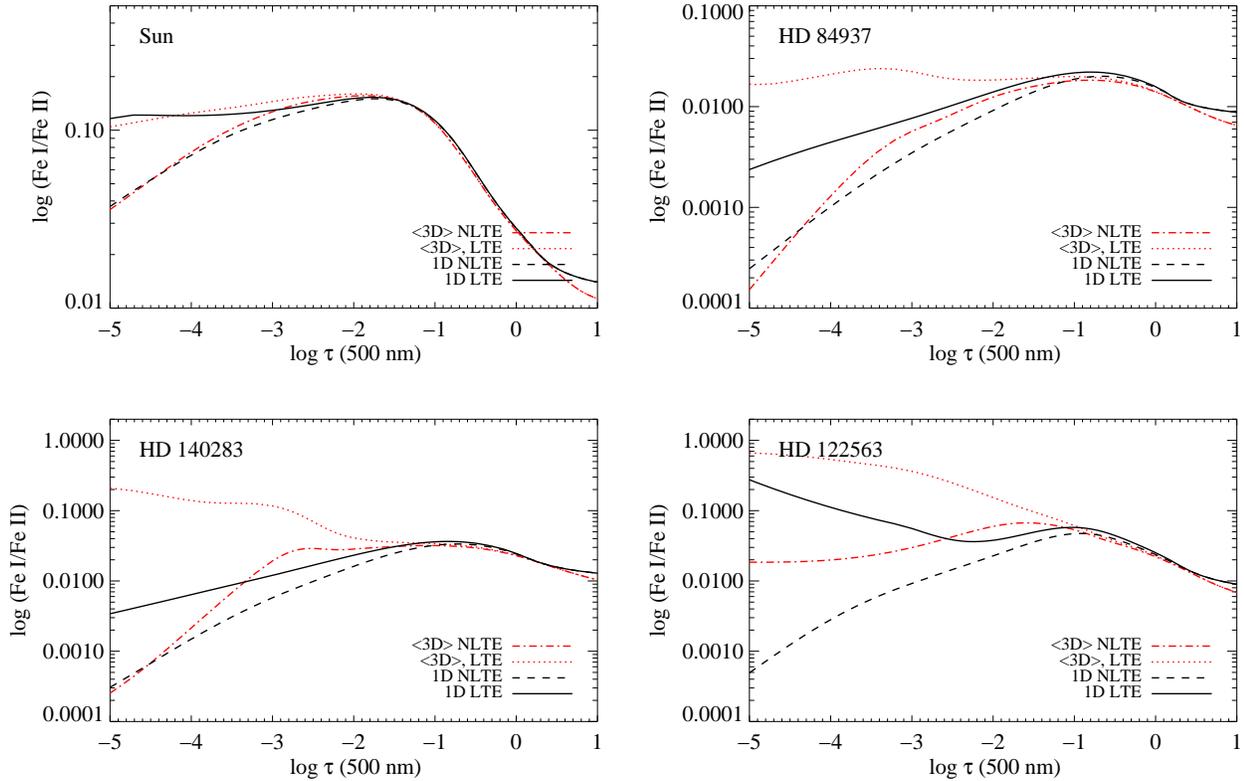}
\caption{Total number densities of \ion{Fe}{i}/\ion{Fe}{ii} for the Sun, HD
140283, HD 84937, and HD 122563.}
\label{total_number}
\end{figure*}

\begin{figure}
\centering
\includegraphics[scale=0.45]{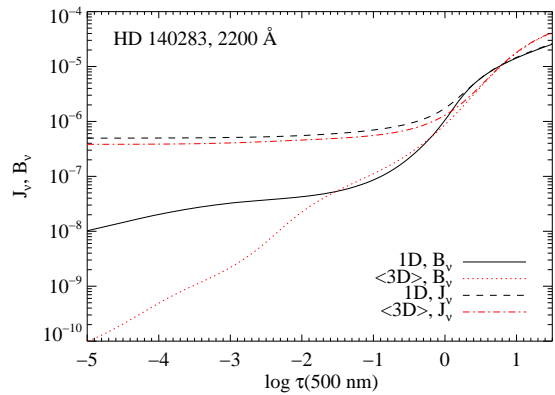}
\caption{Mean radiation field compared to the local Planck function in the UV
continuum computed with the 1D and $\td$ model atmospheres of the metal-poor
sub-giant HD 140283.}
\label{jnu}
\end{figure}

A comparison of the departure coefficients computed with 1D and $\td$ model
atmospheres reveals important differences (Fig. \ref{dep}), which
depend on stellar parameters, although their qualitative behavior is the same.
Generally, metal-poor $\td$ models show more thermalization close to
continuum-forming layers, while in the outer layers \ion{Fe}{i} atoms experience
a larger degree of over-ionization. This can be primarily understood based on
the temperature gradients, which in the $\td$ stratifications are shallower in
the inner layers but steeper in the outer layers compared to 1D models.
We briefly consider the case of the metal-poor subgiant HD 140283, for which the
1D and $\td$ model atmospheres are compared in Fig. \ref{atm_abs}. The $\td$
model leads to smaller deviations from LTE at $-2 \lesssim \opd \lesssim 0$.
That is, for the majority of the underpopulated \ion{Fe}{i} levels, $b_i$(1D)
$<$ $b_i$($\td$). Inspection of the respective T$(\tau)$ relations in Fig.
\ref{atm_abs} shows that the $\td$ model is cooler than the \textsc{marcs} model
at $0 \lesssim \opd \lesssim 1$, where the UV continua form, but it is $\sim
100$ K hotter higher up. Figure \ref{jnu} illustrates mean angle-averaged
intensities at $2200$ \AA, i.e., at the wavelength sampled by the ionization
edges of the important \ion{Fe}{i} levels. These are shown for the 1D and $\td$
models, and are compared with the local Planck functions $B_{\nu}$ at each
optical depth. At $-1.5 \lesssim \opd \lesssim 0$, the $J_{\nu} -
B_{\nu}$ imbalance is smaller in the $\td$ model compared to the 1D model, and,
thus, 
over-ionization is less efficient. In this range of optical depths, densities
in the $\td$ stratification are also slightly larger than in the 1D model (Fig.
\ref{atm_abs}). At $\opd \lesssim -2$, the situation is reverse. The $\td$ model
is nearly $1000$ K cooler and less dense than the 1D model. The $J_{\nu} -
B_{\nu}$ imbalance becomes very large causing significant underpopulation of
the lower \ion{Fe}{i} levels, while uppermost \ion{Fe}{i} levels and the
\ion{Fe}{ii} ground state and excited levels develop appreciable overpopulation
compared to their LTE occupation numbers. In contrast, 1D models predict nearly
thermalized \ion{Fe}{ii} level populations, over the whole optical depth scale.

Figure \ref{total_number} shows the ratios of \ion{Fe}{i}/\ion{Fe}{ii} number
densities for the four cases: \textsc{marcs} and $\td$ model atmospheres, NLTE
and LTE. In the LTE case, about $\sim 5 - 10$ percent of the element is in the
form of \ion{Fe}{i} above $\opd \sim 0$. In the $\td$ models, due to their
steeper T($\tau$) relation, LTE recombination of \ion{Fe}{ii} to \ion{Fe}{i} is
even more efficient, raising \ion{Fe}{i}/\ion{Fe}{ii} to $20 - 50$ percent.
Under NLTE, this effect is suppressed since ionization balance is primarily set
by radiative ionization and recombination. For the metal-poor stars, \ion{Fe}{i}
occupies less then $1$ percent of the total element abundance at $\opd < -2$
notwithstanding much cooler outer layers of the $\td$ models. We note, however,
that the huge differences in the ionization balance of \ion{Fe}{i}/\ion{Fe}{ii}
in the outer layers of the models are not important for our NLTE abundance
determinations. Most of the \ion{Fe}{i} lines observed in spectra of the
selected metal-poor stars form close to the continuum forming regions, $-2 <
\opd < 0$.

Extensive tests demonstrated that the behavior of departure coefficients is
robust against the differences in terms of model atmospheres and NLTE codes used
for the statistical equilibrium calculations. Perhaps, the only systematic
effect is that \textsc{multi} predicts slightly stronger NLTE effects than
\textsc{detail} mainly due to the differences in the background opacity. Also,
the comparison of the LTE and NLTE equivalent widths obtained with
\textsc{multi} and \textsc{detail}/\textsc{siu} show agreement to within $1 - 5$
percent for the lines with $\EW > 10 ~\mA$; NLTE $\EW$'s are somewhat divergent
for the weakest lines, with the relative $(\textsc{multi}-\textsc{detail})$
differences being about $10$ percent. These differences shall be taken as a
formal error, which arises as a consequence of different implementations of the
same basic physics in the various codes. Still, this is not of a big concern in
our work because the calibration of the models and grid calculations are
consistently performed with same code (\textsc{multi}). Furthermore, the
magnitude of the NLTE effects is initially effectively controlled by a free
parameter, the scaling factor $\SH$ for the efficiency of the \ion{H}{i}
collisions.
%
%
\section{Spectroscopic stellar parameters}{\label{sec:solspec}}
%
%
\subsection{Method}{\label{sec:method}}
Spectroscopic stellar parameters, including metallicity, effective temperature,
gravity, and microturbulence, were determined for each model atmosphere
(\textsc{marcs}, \textsc{mafags}, $\td$) in LTE and NLTE using the following
approach.

In a first step, we performed full spectrum synthesis for the reference stars
with the \textsc{siu} line formation code using the 1D \textsc{mafags-os} model
atmospheres computed with parameters described in Sect. \ref{sec:obs}.
The \textsc{mafags-os} models were used at this stage because they include
partial pressures for molecules, which are important contributors to the
background opacity of cooler stars. The profiles of all diagnostic Fe lines
computed in LTE
and NLTE were visually fitted to the observed spectra. The lines were computed
with depth-dependent Voigt profiles taking into account various external
broadening mechanisms. The equivalent widths of the Fe lines were then obtained
from the best-fitting NLTE profiles, excluding contribution of blends. The
$\EW$'s are given in Tables \ref{ew1} and \ref{ew2} of the Appendix and
are accurate to within $1-2$ percent. 

In a second step, the equivalent widths $\EW$'s were applied to determine
spectroscopic parameters using the grids of LTE and NLTE equivalent
widths computed with \textsc{multi} for \textsc{marcs} and $\td$ model
atmospheres.
For each star, we constructed a local model atmosphere grid with two effective
temperature points (the IRFM value and a value $200$\,K lower), two values of
microturbulence ($1$ and $2$ km s$^{-1}$), and a range of plausible
metallicities ($\Delta$[Fe/H] up to $\pm 1.5$ dex) with a step-size of
$0.25$\,dex. Equivalent widths were computed for the grid models in LTE as well
as in NLTE for two values of the \ion{H}{i} collision efficiency parameter, $\SH
=0.1$ and $\SH =1$. In order to save computational time in the construction of
the local model grids in the $\td$ case, the $\td$ models were simply adjusted
by multiplying the temperature and electron pressure at a given optical depth
with the ratios of those quantities obtained from corresponding 1D models. Based
on our experience with scaling 1D and 3D model stellar atmospheres to different
stellar parameters, this procedure appears accurate to first-order level. The
grid was thus constructed for \textsc{marcs} and $\td$ models.

\textsc{mafags-os} model parameters were adjusted by assuming the equivalent
widths to have the same sensitivity to stellar parameters as for \textsc{marcs}
model, which is reasonable considering the similarity between the two codes.

An iterative procedure was then applied to determine spectroscopic
parameters from \ion{Fe}{i} and \ion{Fe}{i} lines. First of all the
microturbulent velocity was determined by flattening the slope of \ion{Fe}{ii}
line abundances with reduced equivalent width, thereby circumventing
differential NLTE effects with line strength of \ion{Fe}{i} lines. Only for the
most metal-poor star G 64-12, the microturbulence is poorly defined due to the
lack of strong \ion{Fe}{ii} lines. We then followed two different approaches,
keeping either $\log g$ or $\Teff$ fixed, and optimizing the other parameter to
establish ionization balance between the \ion{Fe}{i} and \ion{Fe}{ii} lines. In
order to avoid biased results, we have not attempted to adjust the oscillator
strengths of the Fe lines, giving preference to the experimental data from the
laboratory measurements, or their weighted means if few measurements were
available (Sect. \ref{sec:line_par}). Hence, all abundance results are absolute
values. Furthermore, we did not perform a differential stellar abundance
analysis with respect to the Sun. First, that would conflict with the use of the
reference stars as a calibration sample. Second, a differential approach would
introduce yet an another source of error related to the usual problems of
fitting the strong Fe lines in the solar spectrum, in particular the ambiguity
between the effect of van der Waals damping and abundance on a line profile.

The standard technique to infer effective temperatures by flattening the
slope of \ion{Fe}{i} line abundances with excitation potential is particularly
sensitive to the choice of transition probabilities. We found, however, that it
is only weakly sensitive to $\SH$, which is the main parameter we seek to
constrain. Initial attempts to vary all parameters simultaneously, i.e.
microturbulence from line strengths, effective temperature by excitation balance
and surface gravity by ionization balance, did not give a conclusive answer in
terms of the best choice of $\SH$. This apparent weakness is likely related to
the problem of multidimensionality in the parameter space, and, as such, shall
be inherent to the method itself. In addition, we do not exclude remaining
systematic uncertainties in the atomic and atmospheric models. It is possible
that a differential approach between similar stars would be more successful in
establishing temperatures based on excitation balance. We therefore reduced the
dimensionality of the problem by not enforcing strict excitation equilibrium and
focusing only on the ionization balance, as described above. As the results will
show, an adequately flat trend with excitation potential is anyway naturally
achieved for the majority of stars, HD 122563 being a notable exception. 

The analysis described above was applied to each star in the sample
using the \textsc{marcs}, \textsc{mafags}, and $\td$ model atmospheres, thus
yielding four desired quantities: $\Teff$, $\log g$, [Fe/H], and $\Vmic$.
The results are discussed in Sect. \ref{sec:abstars}. Table \ref{tab:fin_param}
gives the mean metallicities determined using the reference $\Teff$ and $\log
g$, and abundances averaged over the measured \ion{Fe}{i} and
\ion{Fe}{ii} lines together with their standard deviations are presented in
Table \ref{tab:ion_abu}.

%
%
\subsection{Observations and reference stellar parameters}{\label{sec:obs}}
Our reference sample consists of six late-type stars (Table
\ref{tab:init_param}), and includes two solar-metallicity stars (the Sun and
Procyon), two metal-poor dwarfs (HD 84937 and G 64-12), a metal-poor subgiant
(HD 140283), and a very bright metal-poor giant (HD 122563). The following
observed spectra were adopted here. For the Sun, we used the KPNO flux spectrum
(Kurucz et al. 1984). The UVES observations of Procyon (HD 61421), HD 84937, HD
140283, and HD 122563 were taken from the UVES-POP survey
\citep{2003Msngr.114...10B}. These spectra have a slit-determined resolution of
$\lambda/\Delta\lambda \sim 80\,000$ and a signal-to-noise ratio $S/N \sim 300$
near $5000$ \AA. The UVES spectrum for G 64-12 was taken from ESO/ST-ECF Science
Archive Facility (67.D-0554(A)). The number of Fe lines suitable for the
spectrum synthesis at this spectral quality, is $40$ (Procyon) to $10$ (G
64-12). For comparison, we also used the Keck/HIRES spectra of G 64-12 with
$\lambda/\Delta\lambda \sim 100\,000$ and $S/N \sim 500$ kindly provided by J.
Melendez (private communication).

A crucial step in our analysis is the choice of the reference $\Teff$ and $\log
g$ values for the selected stars, which are used as a benchmark for testing the
new NLTE model atom and $\td$ model atmospheres. These parameters were taken
from the literature, giving preference to the least model-dependent methods,
such as interferometry, infra-red flux method (IRFM), and parallaxes. The
adopted values and their uncertainties are given in the Table
\ref{tab:init_param}. A brief description of these data is given below.
%
%
\begin{table}
\centering
\caption{Input parameters for the reference stars. Parallaxes $\pi$
  and their uncertainties are also given. See text.}
\label{tab:init_param}
\renewcommand{\footnoterule}{}
\renewcommand{\tabcolsep}{3pt}
\begin{tabular}{l cc lrc rcc}
\hline\noalign{\smallskip}
Star & $T_{\rm eff}$ & $\sigma$ & Ref. & $\log g$ & $\sigma$
& $\pi$ & $\sigma$ & [Fe/H] \\
\noalign{\smallskip}\hline\noalign{\smallskip}
Sun          & $5777$ &     &   & $4.44$  &      &        &       &     \\
Procyon      & $6543$ &  84 & a & $3.98$  & 0.02 & 284.56 & 1.26  &  -0.03 \\
HD 84937     & $6408$ &  66 & b & $4.13$  & 0.09 & 3.83   & 0.78  &  -2.16 \\
HD 140283    & $5777$ &  55 & b & $3.70$  & 0.08 & 17.16  & 0.68  &  -2.38 \\
HD 122563    & $4665$ &  80 & c & $1.64$  & 0.16 & 4.22   & 0.35  &  -2.51 \\
G  64-12     & $6464$ &  61 & b & $4.3^*$&      & 0.57   & 2.83  &  -3.12 \\
\noalign{\smallskip}\hline
\end{tabular}
References: a - \citet{2005ApJ...633..424A}; b - \citet{2010A&A...512A..54C};
c - Casagrande (private communication)\\
$^*$ $\log g$ is derived from the upper limit on the parallax
\end{table}

For the four metal-poor stars, we adopted the IRFM effective temperatures by
\citet{2010A&A...512A..54C}. The mean internal uncertainty of the data is about
$70$ K, which includes the uncertainty on the zero point of the $\Teff$ scale,
reddening and photometric errors. 

The IRFM value for the metal-poor giant HD 122563, $\Teff = 4665 \pm 80$ K, was
kindly provided by L. Casagrande (private communication). This estimate
incorporates a correction due to reddening, $E(B-V) = 0.005$, which was
determined from the interstellar Na D lines detected in the UVES spectrum. Our
equivalent widths for the $5889.95$ and $5895.92$ \AA\ \ion{Na}{i} lines are
$22.3$ and $12.4$ \mA, respectively.

For Procyon, an interferometric estimate of angular diameter is available
\citep*{2005ApJ...633..424A}, which gives $\Teff = 6545 \pm 83$ K and combined
with a very accurate parallax $\pi = 284.56 \pm 1.26$ milli-arcsec, surface
gravity can be estimated, $\log g = 3.99 \pm 0.02$. 

For the other stars, we determined surface gravities from the {\sc Hipparcos}
parallaxes \citep{2007A&A...474..653V} using the masses estimated from the
tracks of \citet{2000ApJ...532..430V} by \citet{2004A&A...413.1045G,
2006A&A...451.1065G} and apparent bolometric magnitudes from
\citet{2010A&A...512A..54C}. The uncertainties were computed by mapping the
uncertainty in mass\footnote{Note that the uncertainty in mass is nominal and
was determined by comparing positions of the stars on the HRD with theoretical
isochrones.} (0.2 M$_\odot$ for HD 122563 and G 64-12, and $0.1$ M$_\odot$ for
HD 84937 and HD 140283), temperature and parallax into the range of
possible gravities ($\Delta \logg \leq 0.16$ dex). The only exception is G
64-12, for which the parallax is too uncertain, $\pi = 0.57 \pm 2.83$
milli-arcsec. From comparison with metal-poor evolutionary tracks for reasonable
masses (Fig. \ref{tracks}), one may also conclude that within the errors of the
given $\Teff$ ($6464$ K) surface gravity of G 64-12 is in the range $\log g =
3.8 \ldots 4.6$ dex. We, thus adopt the value derived from the upper limit on
the parallax, $\log g = 4.3$ and assign a nominal error of $0.3$ dex. In
comparison, \citet*{2004A&A...415..993N,2007A&A...469..319N} and
\citet*{2009A&A...500.1143F} estimate $\log g = 4.26$.
%
%
\subsection{Line selection}{\label{sec:line_par}}
%
%
%
\begin{table}
\begin{minipage}{\linewidth}
\renewcommand{\footnoterule}{}
\tabcolsep2.0mm
\caption{Parameters of the \ion{Fe}{i} lines used for the solar analysis.}
\label{sun2601}
\begin{center}
\begin{tabular}{l|ccccc|cc}
\noalign{\smallskip}\hline\noalign{\smallskip} ~~~$\lambda$ &
Lower & Upper & g$_l$ & g$_u$ & $\Elow$ & $\log gf$ & $\log C_6$ \\
~~$\AA$  &  level & level & & & [eV] & & \\ 
\noalign{\smallskip}\hline\noalign{\smallskip}
4445.47  & \Fe{a}{5}{D}{}{}      &  \Fe{z}{7}{F}{\circ}{} &   5 &   5 & 0.09  &  -5.410 &  -31.8 \\
4494.56  & \Fe{a}{5}{P}{}{}      &  \Fe{x}{5}{D}{\circ}{} &   5 &   7 & 2.20  &  -1.136 &  -31.2 \\
4920.50  & \Fe{z}{7}{F}{\circ}{} &  \Fe{e}{7}{D}{}{}      &  11 &   9 & 2.80  &   0.068 &  -30.5 \\
4994.13  & \Fe{a}{5}{F}{}{}      &  \Fe{z}{5}{F}{\circ}{} &   9 &   7 & 0.92  &  -3.080 &  -31.7 \\
5044.21  & \Fe{z}{7}{F}{\circ}{} &  \Fe{e}{7}{D}{}{}      &   9 &  11 & 2.85  &  -2.017 &  -30.6 \\
5198.71  & \Fe{a}{5}{P}{}{}      &  \Fe{y}{5}{P}{\circ}{} &   3 &   5 & 2.22  &  -2.135 &  -31.3 \\
5216.27  & \Fe{a}{3}{F}{}{}      &  \Fe{z}{3}{F}{\circ}{} &   5 &   5 & 1.61  &  -2.150 &  -31.5 \\
5225.53  & \Fe{a}{5}{D}{}{}      &  \Fe{z}{7}{D}{\circ}{} &   3 &   3 & 0.11  &  -4.789 &  -31.9 \\
5232.94  & \Fe{z}{7}{P}{\circ}{} &  \Fe{e}{7}{D}{}{}      &   9 &  11 & 2.94  &  -0.060 &  -30.6 \\
5236.20  & \Fe{c}{3}{F}{}{}      &  \Fe{t}{3}{D}{\circ}{} &   5 &   3 & 4.19  &  -1.497 &  -31.3 \\
5242.49  & \Fe{a}{1}{I}{}{}      &  \Fe{z}{1}{H}{\circ}{} &  13 &  11 & 3.63  &  -0.967 &  -31.3 \\
5247.05  & \Fe{a}{5}{D}{}{}      &  \Fe{z}{7}{D}{\circ}{} &   5 &   7 & 0.09  &  -4.960 &  -31.9 \\
5250.21  & \Fe{a}{5}{D}{}{}      &  \Fe{z}{7}{D}{\circ}{} &   1 &   3 & 0.12  &  -4.938 &  -31.9 \\
5269.54  & \Fe{a}{5}{F}{}{}      &  \Fe{z}{5}{D}{\circ}{} &  11 &   9 & 0.86  &  -1.321 &  -31.8 \\
5281.79  & \Fe{z}{7}{P}{\circ}{} &  \Fe{e}{7}{D}{}{}      &   5 &   7 & 3.04  &  -0.830 &  -30.5 \\
5379.57  & \Fe{b}{1}{G}{}{}      &  \Fe{z}{1}{H}{\circ}{} &   9 &  11 & 3.70  &  -1.514 &  -31.3 \\
5383.37  & \Fe{z}{5}{G}{\circ}{} &  \Fe{e}{5}{H}{}{}      &  11 &  13 & 4.31  &   0.645 &  -30.4 \\
5434.52  & \Fe{a}{5}{F}{}{}      &  \Fe{z}{5}{D}{\circ}{} &   3 &   1 & 1.01  &  -2.122 &  -31.7 \\
5491.84  & \Fe{c}{3}{F}{}{}      &  \Fe{u}{3}{D}{\circ}{} &   5 &   7 & 4.19  &  -2.190 &  -30.4 \\
5600.22  & \Fe{z}{3}{P}{\circ}{} &  \Fe{g}{5}{D}{}{}      &   3 &   3 & 4.26  &  -1.420 &  -30.4 \\
5661.35  & \Fe{z}{3}{P}{\circ}{} &  \Fe{g}{5}{D}{}{}      &   1 &   3 & 4.28  &  -1.756 &  -31.3 \\
5662.52  & \Fe{y}{5}{F}{\circ}{} &  \Fe{g}{5}{D}{}{}      &  11 &   9 & 4.18  &  -0.573 &  -30.5 \\
5696.09  & \Fe{y}{3}{F}{\circ}{} &  \Fe{e}{5}{H}{}{}      &   9 &   9 & 4.55  &  -1.720 &  -30.2 \\
5701.54  & \Fe{b}{3}{F}{}{}      &  \Fe{y}{3}{D}{\circ}{} &   9 &   7 & 2.56  &  -2.163 &  -31.3 \\
5705.46  & \Fe{y}{5}{F}{\circ}{} &  \Fe{g}{5}{D}{}{}      &   3 &   3 & 4.30  &  -1.360 &  -30.5 \\
5778.45  & \Fe{b}{3}{F}{}{}      &  \Fe{y}{3}{D}{\circ}{} &   7 &   7 & 2.59  &  -3.440 &  -31.3 \\
5855.08  & \Fe{y}{3}{F}{\circ}{} &  \Fe{e}{5}{H}{}{}      &   7 &   9 & 4.61  &  -1.480 &  -30.3 \\
5916.25  & \Fe{a}{3}{H}{}{}      &  \Fe{y}{3}{F}{\circ}{} &   9 &   9 & 2.45  &  -2.994 &  -31.4 \\
5956.69  & \Fe{a}{5}{F}{}{}      &  \Fe{z}{7}{P}{\circ}{} &  11 &   9 & 0.86  &  -4.550 &  -31.8 \\
6065.48  & \Fe{b}{3}{F}{}{}      &  \Fe{y}{3}{F}{\circ}{} &   5 &   5 & 2.61  &  -1.530 &  -31.3 \\
6082.71  & \Fe{a}{5}{P}{}{}      &  \Fe{z}{3}{P}{\circ}{} &   3 &   3 & 2.22  &  -3.570 &  -31.5 \\
6151.62  & \Fe{a}{5}{P}{}{}      &  \Fe{y}{5}{D}{\circ}{} &   7 &   5 & 2.18  &  -3.282 &  -31.6 \\
6173.33  & \Fe{a}{5}{P}{}{}      &  \Fe{y}{5}{D}{\circ}{} &   3 &   1 & 2.22  &  -2.880 &  -31.6 \\
6200.31  & \Fe{b}{3}{F}{}{}      &  \Fe{y}{3}{F}{\circ}{} &   5 &   7 & 2.61  &  -2.416 &  -31.3 \\
6219.28  & \Fe{a}{5}{P}{}{}      &  \Fe{y}{5}{D}{\circ}{} &   5 &   5 & 2.20  &  -2.433 &  -31.6 \\
6240.65  & \Fe{a}{5}{P}{}{}      &  \Fe{z}{3}{P}{\circ}{} &   3 &   5 & 2.22  &  -3.287 &  -31.5 \\
6252.56  & \Fe{a}{3}{H}{}{}      &  \Fe{z}{3}{G}{\circ}{} &  13 &  11 & 2.40  &  -1.687 &  -31.4 \\
6265.13  & \Fe{a}{5}{P}{}{}      &  \Fe{y}{5}{D}{\circ}{} &   7 &   7 & 2.18  &  -2.547 &  -31.6 \\
6297.79  & \Fe{a}{5}{P}{}{}      &  \Fe{y}{5}{D}{\circ}{} &   3 &   5 & 2.22  &  -2.715 &  -31.6 \\
6311.50  & \Fe{b}{3}{P}{}{}      &  \Fe{y}{3}{D}{\circ}{} &   5 &   5 & 2.83  &  -3.141 &  -31.4 \\
6430.85  & \Fe{a}{5}{P}{}{}      &  \Fe{y}{5}{D}{\circ}{} &   7 &   9 & 2.18  &  -2.006 &  -31.6 \\
6498.94  & \Fe{a}{5}{F}{}{}      &  \Fe{z}{7}{F}{\circ}{} &   7 &   7 & 0.96  &  -4.695 &  -31.8 \\
6518.37  & \Fe{b}{3}{P}{}{}      &  \Fe{y}{3}{D}{\circ}{} &   5 &   7 & 2.83  &  -2.298 &  -31.4 \\
6574.23  & \Fe{a}{5}{F}{}{}      &  \Fe{z}{7}{F}{\circ}{} &   5 &   5 & 0.99  &  -5.010 &  -32.4 \\
6593.87  & \Fe{a}{3}{H}{}{}      &  \Fe{z}{5}{G}{\circ}{} &  11 &  11 & 2.43  &  -2.394 &  -31.4 \\
6609.11  & \Fe{b}{3}{F}{}{}      &  \Fe{z}{3}{G}{\circ}{} &   9 &   9 & 2.56  &  -2.682 &  -31.4 \\
6699.14  & \Fe{d}{3}{F}{}{}      &  \Fe{u}{3}{D}{\circ}{} &   9 &   7 & 4.59  &  -2.101 &  -31.5 \\
6726.67  & \Fe{y}{5}{P}{\circ}{} &  \Fe{e}{5}{P}{}{}      &   5 &   3 & 4.61  &  -1.000 &  -30.5 \\
6739.52  & \Fe{a}{3}{F}{}{}      &  \Fe{z}{5}{F}{\circ}{} &   7 &   7 & 1.56  &  -4.794 &  -31.7 \\
6750.15  & \Fe{a}{3}{P}{}{}      &  \Fe{z}{3}{P}{\circ}{} &   3 &   3 & 2.42  &  -2.605 &  -31.4 \\
6793.26  & \Fe{c}{3}{F}{}{}      &  \Fe{w}{5}{G}{\circ}{} &   9 &   9 & 4.08  &  -2.326 &  -30.8 \\
6810.26  & \Fe{y}{5}{P}{\circ}{} &  \Fe{e}{5}{P}{}{}      &   5 &   7 & 4.61  &  -0.986 &  -30.4 \\
6837.01  & \Fe{d}{3}{F}{}{}      &  \Fe{u}{3}{G}{\circ}{} &   9 &   9 & 4.59  &  -1.690 &  -31.6 \\
6854.82  & \Fe{d}{3}{F}{}{}      &  \Fe{3}{H}{P}{}{}      &   9 &  11 & 4.59  &  -1.926 &  -31.6 \\
6945.20  & \Fe{a}{3}{P}{}{}      &  \Fe{z}{3}{P}{\circ}{} &   3 &   5 & 2.42  &  -2.454 &  -31.4 \\
6978.85  & \Fe{a}{3}{P}{}{}      &  \Fe{z}{3}{P}{\circ}{} &   1 &   3 & 2.48  &  -2.480 &  -31.4 \\
7401.68  & \Fe{c}{3}{F}{}{}      &  \Fe{w}{3}{D}{\circ}{} &   5 &   3 & 4.19  &  -1.500 &  -31.2 \\
7912.87  & \Fe{a}{5}{F}{}{}      &  \Fe{z}{7}{D}{\circ}{} &  11 &   9 & 0.86  &  -4.848 &  -31.9 \\
8293.51  & \Fe{a}{3}{D}{}{}      &  \Fe{y}{3}{D}{\circ}{} &   5 &   5 & 3.30  &  -2.203 &  -31.3 \\
\noalign{\smallskip}\hline\noalign{\smallskip}
\end{tabular}
\end{center}
\end{minipage}
\end{table}
%

%
%
\begin{table}
\begin{minipage}{\linewidth}
\renewcommand{\footnoterule}{}
\tabcolsep2.0mm
\caption{Parameters of the \ion{Fe}{ii} lines used for the solar analysis.}
\label{sun2602}
\begin{center}
\begin{tabular}{l|ccccc|cc}
\noalign{\smallskip}\hline\noalign{\smallskip} ~~~$\lambda$ & Lower & Upper & g$_l$ & g$_u$ & $\Elow$ &
$\log gf$ & $\log C_6$ \\
~~$\AA$  & level & level & & & [eV] & & \\
\noalign{\smallskip}\hline\noalign{\smallskip}
4491.40  &   \Fe{b}{4}{F}{}{}  &  \Fe{z}{4}{F}{\circ}{} &   4 &   4 & 2.856 &  -2.71 &  -32.1 \\
4508.29  &   \Fe{b}{4}{F}{}{}  &  \Fe{z}{4}{D}{\circ}{} &   4 &   2 & 2.856 &  -2.44 &  -32.0 \\
4576.34  &   \Fe{b}{4}{F}{}{}  &  \Fe{z}{4}{D}{\circ}{} &   6 &   6 & 2.844 &  -2.95 &  -32.0 \\
4582.84  &   \Fe{b}{4}{F}{}{}  &  \Fe{z}{4}{F}{\circ}{} &   6 &   8 & 2.844 &  -3.18 &  -32.1 \\
4583.84  &   \Fe{b}{4}{F}{}{}  &  \Fe{z}{4}{F}{\circ}{} &  10 &   8 & 2.807 &  -1.93 &  -32.0 \\
4620.52  &   \Fe{b}{4}{F}{}{}  &  \Fe{z}{4}{D}{\circ}{} &   8 &   8 & 2.828 &  -3.21 &  -32.0 \\
4923.93  &   \Fe{a}{6}{S}{}{}  &  \Fe{z}{6}{P}{\circ}{} &   6 &   4 & 2.891 &  -1.26 &  -32.1 \\
5018.44  &   \Fe{a}{6}{S}{}{}  &  \Fe{z}{6}{P}{\circ}{} &   6 &   6 & 2.891 &  -1.10 &  -32.1 \\
5169.03  &   \Fe{a}{6}{S}{}{}  &  \Fe{z}{6}{P}{\circ}{} &   6 &   8 & 2.891 &  -1.00 &  -32.1 \\
5197.58  &   \Fe{a}{4}{G}{}{}  &  \Fe{z}{4}{F}{\circ}{} &   6 &   4 & 3.230 &  -2.22 &  -32.1 \\
5234.62  &   \Fe{a}{4}{G}{}{}  &  \Fe{z}{4}{F}{\circ}{} &   8 &   6 & 3.221 &  -2.18 &  -32.1 \\
5264.81  &   \Fe{a}{4}{G}{}{}  &  \Fe{z}{4}{D}{\circ}{} &   6 &   4 & 3.230 &  -3.13 &  -32.0 \\
5284.11  &   \Fe{a}{6}{S}{}{}  &  \Fe{z}{6}{F}{\circ}{} &   6 &   8 & 2.891 &  -3.11 &  -32.1 \\
5325.55  &   \Fe{a}{4}{G}{}{}  &  \Fe{z}{4}{F}{\circ}{} &   8 &   8 & 3.221 &  -3.16 &  -32.1 \\
5414.07  &   \Fe{a}{4}{G}{}{}  &  \Fe{z}{4}{D}{\circ}{} &   8 &   8 & 3.221 &  -3.58 &  -32.0 \\
5425.26  &   \Fe{a}{4}{G}{}{}  &  \Fe{z}{4}{F}{\circ}{} &  10 &  10 & 3.199 &  -3.22 &  -32.1 \\
6239.95  &   \Fe{b}{4}{D}{}{}  &  \Fe{z}{4}{P}{\circ}{} &   2 &   4 & 3.889 &  -3.41 &  -32.0 \\
6247.56  &   \Fe{b}{4}{D}{}{}  &  \Fe{z}{4}{P}{\circ}{} &   6 &   4 & 3.892 &  -2.30 &  -32.0 \\
6369.46  &   \Fe{a}{6}{S}{}{}  &  \Fe{z}{6}{D}{\circ}{} &   6 &   4 & 2.891 &  -4.11 &  -32.1 \\
6432.68  &   \Fe{a}{6}{S}{}{}  &  \Fe{z}{6}{D}{\circ}{} &   6 &   6 & 2.891 &  -3.57 &  -32.1 \\
6456.38  &   \Fe{b}{4}{D}{}{}  &  \Fe{z}{4}{P}{\circ}{} &   8 &   6 & 3.903 &  -2.05 &  -32.0 \\
6516.08  &   \Fe{a}{6}{S}{}{}  &  \Fe{z}{6}{D}{\circ}{} &   6 &   8 & 2.891 &  -3.31 &  -32.1 \\
7222.39  &   \Fe{b}{4}{D}{}{}  &  \Fe{z}{4}{D}{\circ}{} &   4 &   2 & 3.889 &  -3.26 &  -32.0 \\
7224.49  &   \Fe{b}{4}{D}{}{}  &  \Fe{z}{4}{D}{\circ}{} &   2 &   2 & 3.889 &  -3.20 &  -32.0 \\
7515.83  &   \Fe{b}{4}{D}{}{}  &  \Fe{z}{4}{D}{\circ}{} &   8 &   6 & 3.903 &  -3.39 &  -32.0 \\
7711.72  &   \Fe{b}{4}{D}{}{}  &  \Fe{z}{4}{D}{\circ}{} &   8 &   8 & 3.903 &  -2.50 &  -32.0 \\
\noalign{\smallskip}\hline\noalign{\smallskip}
\end{tabular}
\end{center}
\end{minipage}
\end{table}

%
%
\begin{table}
\begin{minipage}{\linewidth}
\renewcommand{\footnoterule}{}
\tabcolsep2.0mm
\caption{Parameters of the Fe I lines used for the abundance analysis of the metal-poor stars.
$gf$-values are taken from NIST.}
\label{stars26}
\begin{center}
\begin{tabular}{l|ccccc|cc}
\noalign{\smallskip}\hline\noalign{\smallskip} ~~~$\lambda$ & Lower &
Upper & g$_l$ & g$_u$ & $\Elow$ & $\log gf$ & $\log C_6$ \\
~~$\AA$  &  level & level & & & [eV] & & \\
\noalign{\smallskip}\hline\noalign{\smallskip}
3581.19  &   \Fe{a}{5}{F}{}{}      &  \Fe{z}{5}{G}{\circ}{} &  11 &  13 & 0.86  &   0.406  &  -31.6  \\
3618.77  &   \Fe{a}{5}{F}{}{}      &  \Fe{z}{5}{G}{\circ}{} &   5 &   7 & 0.99  &  -0.003  &  -31.5  \\
3719.93  &   \Fe{a}{5}{D}{}{}      &  \Fe{z}{5}{F}{\circ}{} &   9 &  11 & 0.00  &  -0.432  &  -31.8  \\
3737.13  &   \Fe{a}{5}{D}{}{}      &  \Fe{z}{5}{F}{\circ}{} &   7 &   9 & 0.05  &  -0.574  &  -31.6  \\
3745.56  &   \Fe{a}{5}{D}{}{}      &  \Fe{z}{5}{F}{\circ}{} &   5 &   7 & 0.09  &  -0.771  &  -31.8  \\
3758.23  &   \Fe{a}{5}{F}{}{}      &  \Fe{y}{5}{F}{\circ}{} &   7 &   7 & 0.96  &  -0.027  &  -31.6  \\
3820.43  &   \Fe{a}{5}{F}{}{}      &  \Fe{y}{5}{D}{\circ}{} &  11 &   9 & 0.86  &   0.119  &  -31.6  \\
4045.81  &   \Fe{a}{3}{F}{}{}      &  \Fe{y}{3}{F}{\circ}{} &   9 &   9 & 1.49  &   0.280  &  -31.5  \\
4235.94  &   \Fe{z}{7}{D}{\circ}{} &  \Fe{e}{7}{D}{}{}      &   9 &   9 & 2.43  &  -0.341  &  -30.5  \\
4250.79  &   \Fe{a}{3}{F}{}{}      &  \Fe{z}{3}{G}{\circ}{} &   7 &   7 & 1.56  &  -0.714  &  -31.5  \\
4415.12  &   \Fe{a}{3}{F}{}{}      &  \Fe{z}{5}{G}{\circ}{} &   5 &   7 & 1.61  &  -0.615  &  -31.5  \\
5586.76  &   \Fe{z}{5}{F}{\circ}{} &  \Fe{e}{5}{D}{}{}      &   9 &   7 & 3.37  &  -0.144  &  -30.4  \\
\noalign{\smallskip}\hline\noalign{\smallskip}
\end{tabular}
\end{center}
\end{minipage}
\end{table}

The Fe lines for the abundance calculations were selected on a star by star
basis, i.e., carefully inspecting the observed stellar spectra. We rejected
lines affected by blends, strong damping wings or are located in the
spectral windows where continuum placement is uncertain. Our solar line list
includes $59$ lines of \ion{Fe}{i} and $24$ lines of \ion{Fe}{ii} in the
wavelength range $4400 - 8500$ \AA. In spectra of the metal-poor stars most of
these lines are very weak. Thus, $12$ strong lines of \ion{Fe}{i} were added to
the analysis of the metal-poor stars. All line parameters are given in Tables
\ref{sun2601}, \ref{sun2602}, and \ref{stars26}.

The $gf$-values adopted in this work are weighted averages of different
experimental values. An elaborate discussion of their accuracy can be found in
\citet{2001A&A...366..981G}. The largest weights are typically assigned to the
$\log gf$ values measured at Oxford (\citealt{1979MNRAS.186..633B};
\citealt*{1979MNRAS.186..657B};
\citealt*{1979MNRAS.186..673B};
\citealt{1980MNRAS.191..445B}; \citealt*{1982MNRAS.199...33B}; 
\citealt{1982MNRAS.199...43B}; \citealt*{1982MNRAS.201..595B};
\citealt{1986MNRAS.220..549B}).
According to the NIST database, the uncertainties of the data are $3 - 10\%$.
Smaller weights are assigned to the $gf$-values of the Hannover group
(\citealt*{1991A&A...248..315B}, \citealt*{1994A&A...282.1014B}, uncertainties
typically $10 - 25\%$) and \citet[][uncertainties $25 -50
\%$]{1991JOSAB...8.1185O}. The $gf$-values for the \ion{Fe}{ii} transitions are
taken from \citet{2009A&A...497..611M}, who renormalized the branching fractions
from \citet{1998A&A...340..300R} and other theoretical sources to the
experimental lifetimes (\citealt[e.g., that of ][]{1999A&A...342..610S,
2004A&A...414.1169S}). To avoid biased results, we excluded any of their 'solar'
values, which were obtained by a 1D LTE spectroscopic analysis of the solar flux
spectrum. That is, for the \ion{Fe}{ii} transitions $\lambda\lambda\ 5284.1,
6239.95, 6247.56, 6456.38\ \AA$ we kept the NIST-recommended values.

\begin{figure*}
\centering
\includegraphics[scale=0.5]{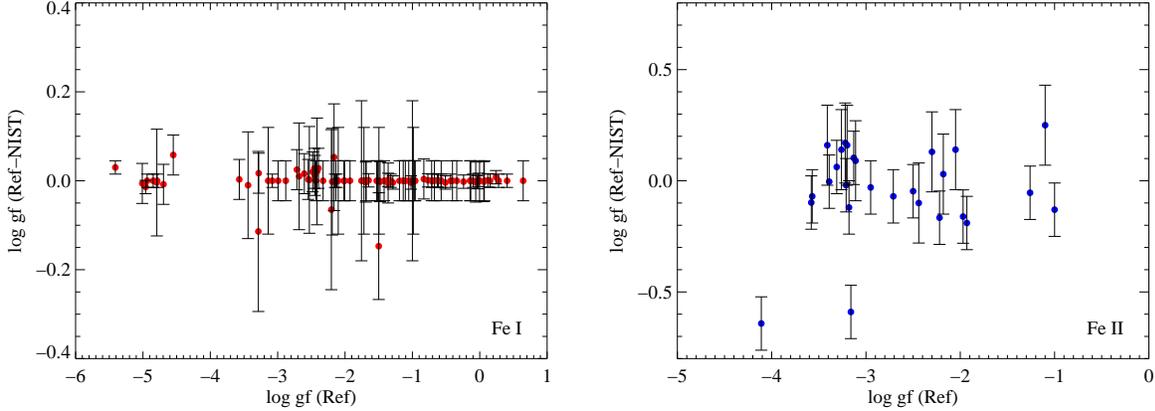}
\caption{Comparison of transition probabilities adopted in this work (Ref) with
that from the NIST database.} 
\label{gf}
\end{figure*}

Our adopted transition probabilites for the \ion{Fe}{i} and \ion{Fe}{ii} lines
are compared to the NIST-recommended values in Fig. \ref{gf}. There are small
differences for the weak \ion{Fe}{ii} transitions, which reflect the
discrepancies between the NIST \ion{Fe}{ii} data based essentially on the
\citet{1998A&A...340..300R} study and the adopted values of
\citet{2009A&A...497..611M}. Note that \citet{1999A&A...347..348G} in the
re-analysis of the solar Fe abundance discuss the possibility of underestimated
$\log gf$'s from \citet{1998A&A...340..300R} for the optical lines, although
they do not find the same problem for the \ion{Fe}{ii} UV lines.

Line broadening due to elastic collisions with \ion{H}{i} atoms was computed
from the tables of \citet*{2000yCat..41420467B} and \citet{2005A&A...435..373B}.
For the four \ion{Fe}{i} lines ($4445.47$, $5600.22$, $5661.35$, $8293.51$ \AA),
broadening cross-sections $\sigma$ and their velocity dependence $\alpha$ were
kindly provided by P. Barklem (2011, private communication). For seven lines,
the \citet{1955QB461.U55......} values increased by a factor of $1.5$ were
adopted. In Tables \ref{sun2601}, \ref{sun2602}, and and \ref{stars26} these
values are given in terms of commonly-used van der Waals damping constants $\log
C_6$.
%
%
\subsection{Sun and Procyon}{\label{sec:sun_procyon}}
Owing to the superior quality of the solar spectrum, the initial spectrum
synthesis for the Sun was performed taking into account solar rotation with
$V_{\rm rot , \odot} = 1.8$ \kms\, and a radial-tangential macroturbulence
velocity $\Vmac$, which was adjusted for each line separately. The typical
values of $\Vmac$ required to match the observed spectrum are $2.5 \ldots 4$
\kms\ for the profile fitting with the 1D models, and $1.6 \ldots 3 $ \kms\ for
that with the $\td$ model. We note that in full 3D line formation calculations
\citet{2000A&A...359..729A} found that no macroturbulence was necessary to fit
the line profiles due to the Doppler shifts from the convective motions, which
are not explicitly taken into account in the $\td$ models used here.

\begin{figure*}
\hbox{\includegraphics[scale=0.35]{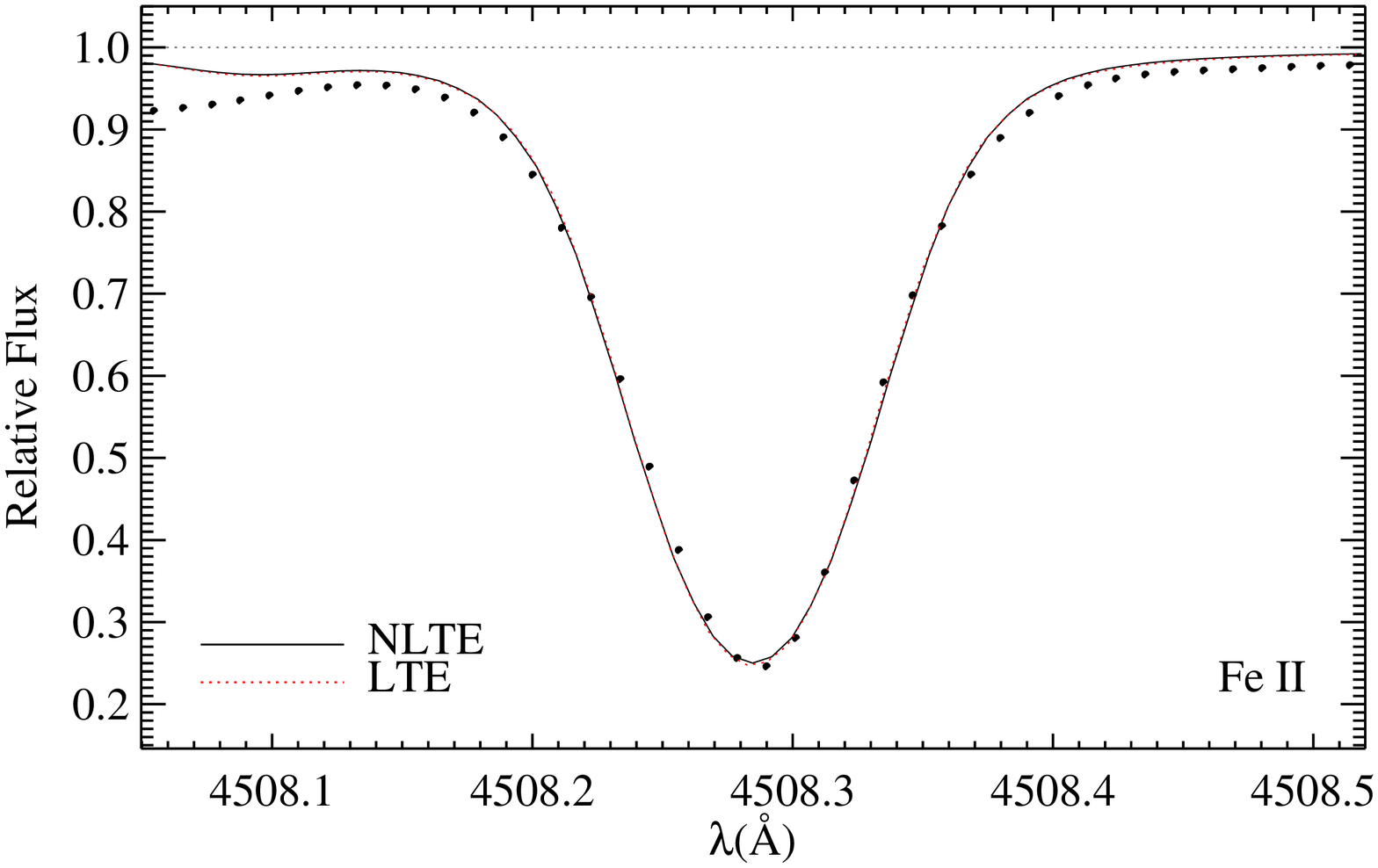}\hfill\hspace{-5mm}
\includegraphics[scale=0.35]{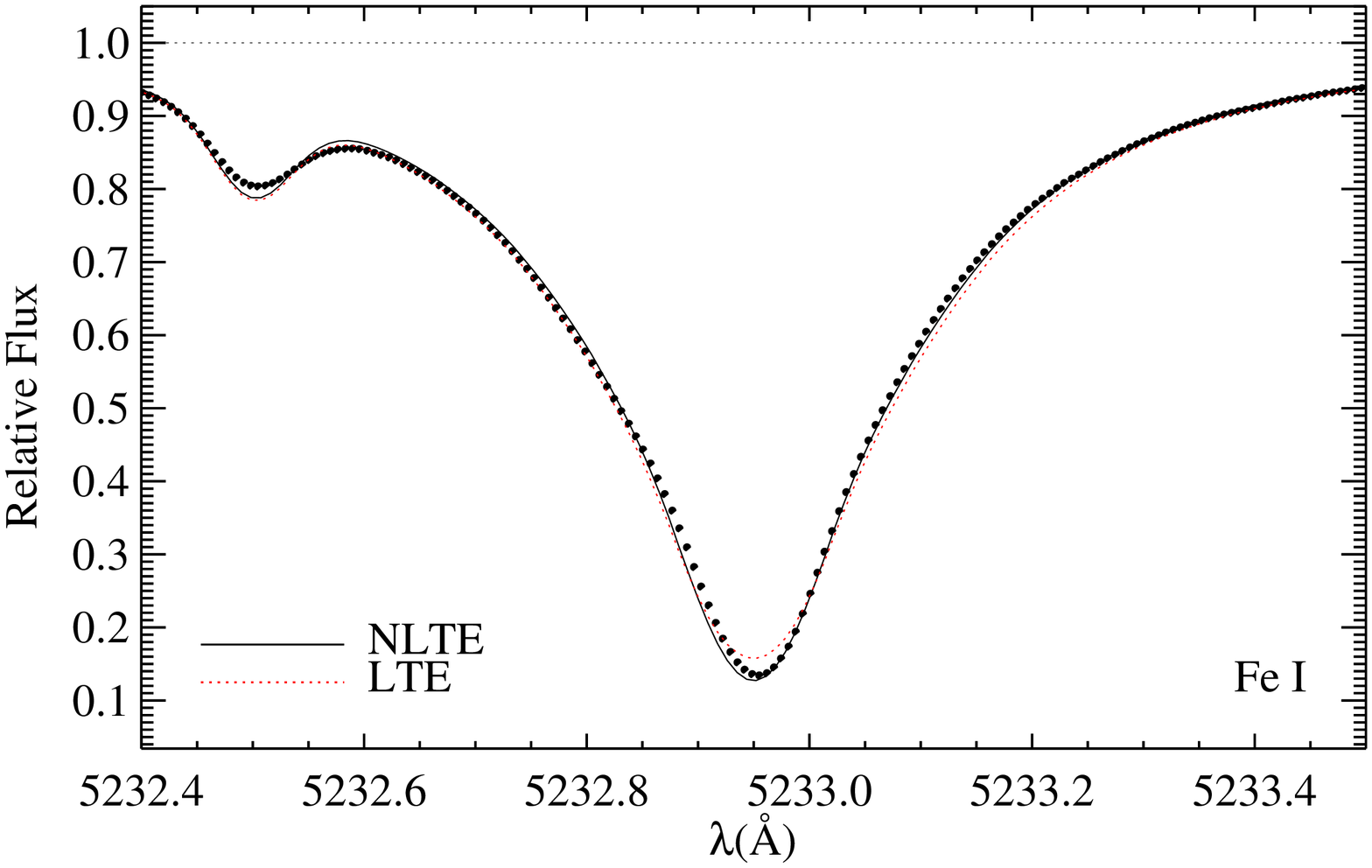}}\vspace{-3mm}
\hbox{\includegraphics[scale=0.35]{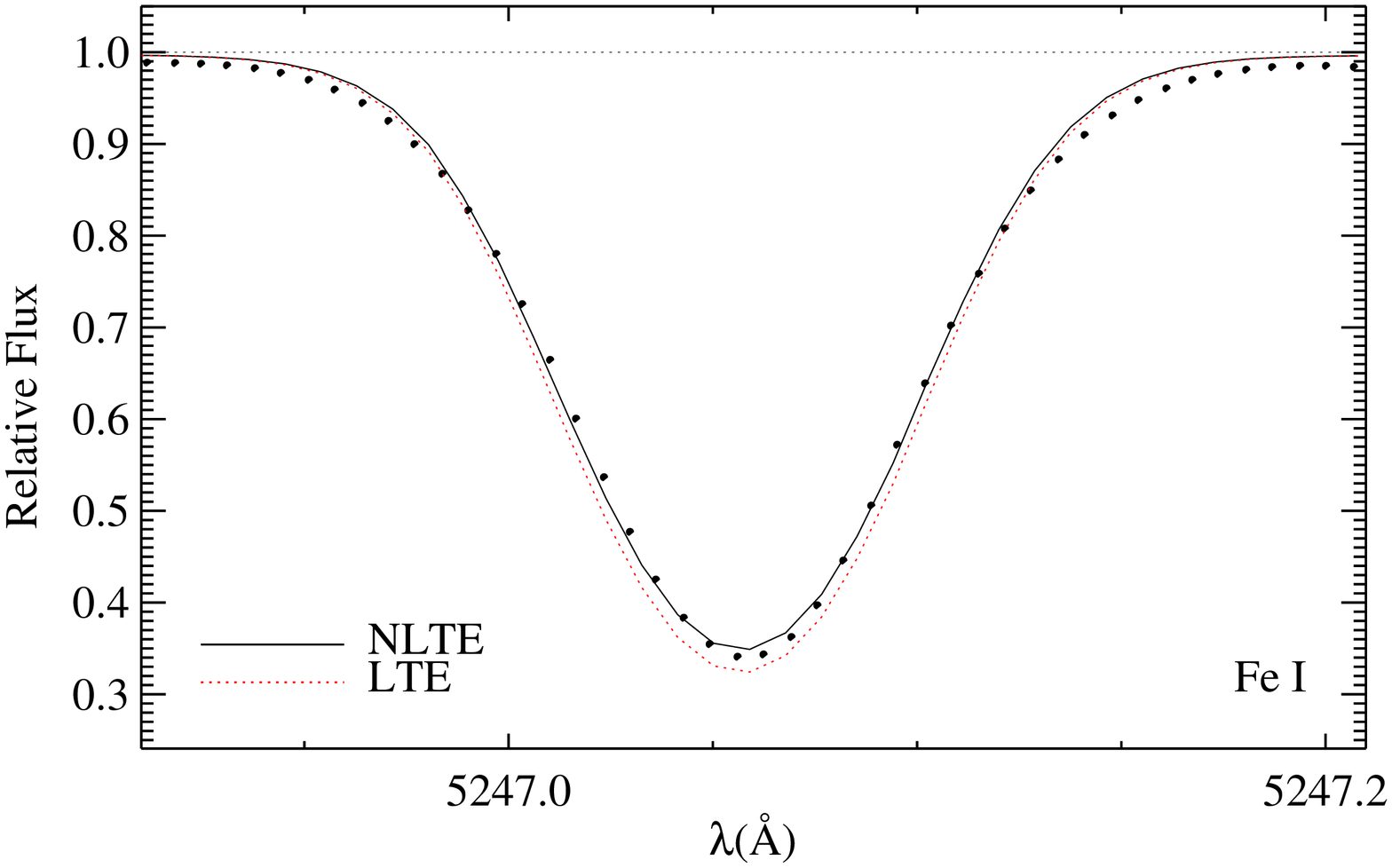}\hfill\hspace{-5mm}
\includegraphics[scale=0.35]{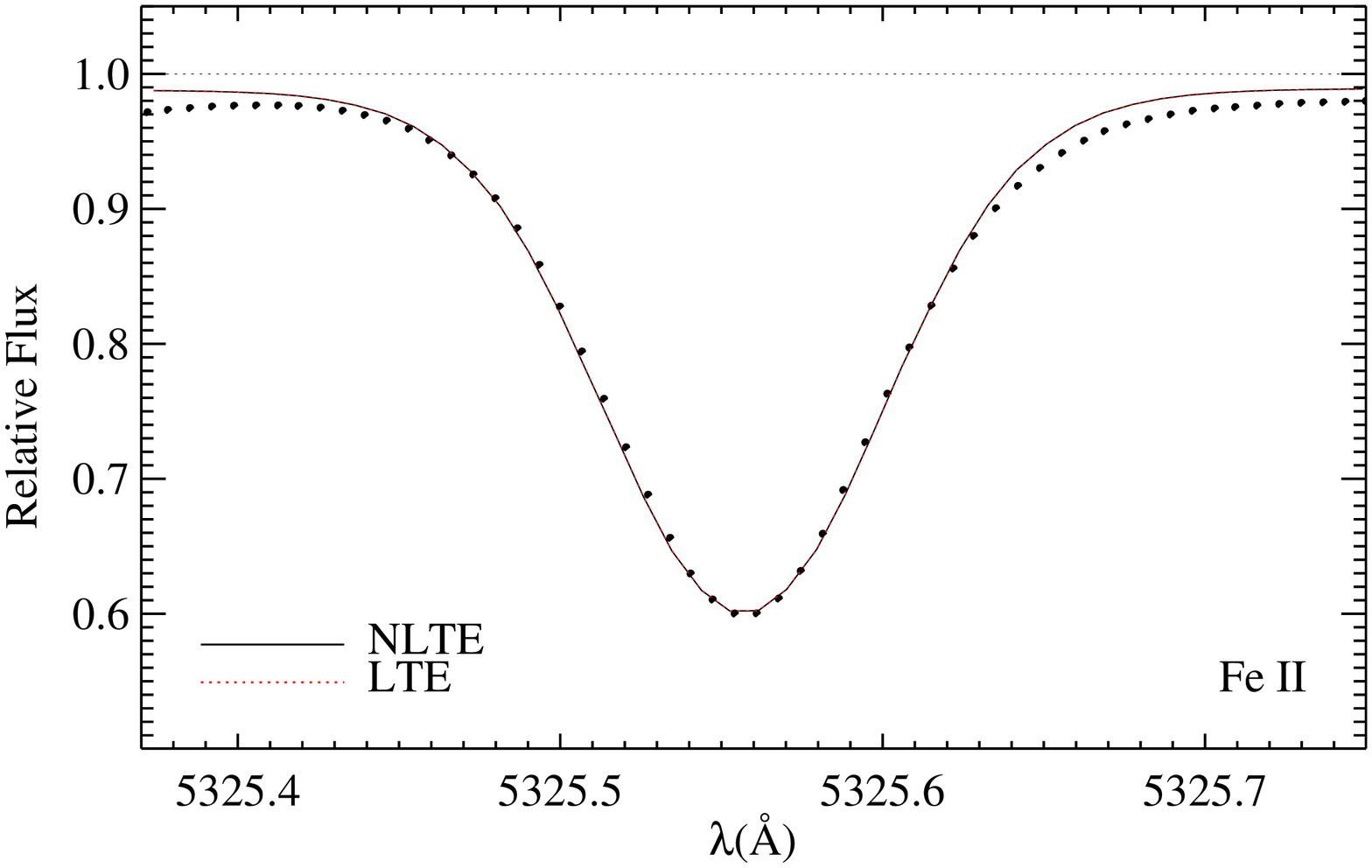}}\vspace{-3mm}
\hbox{\includegraphics[scale=0.35]{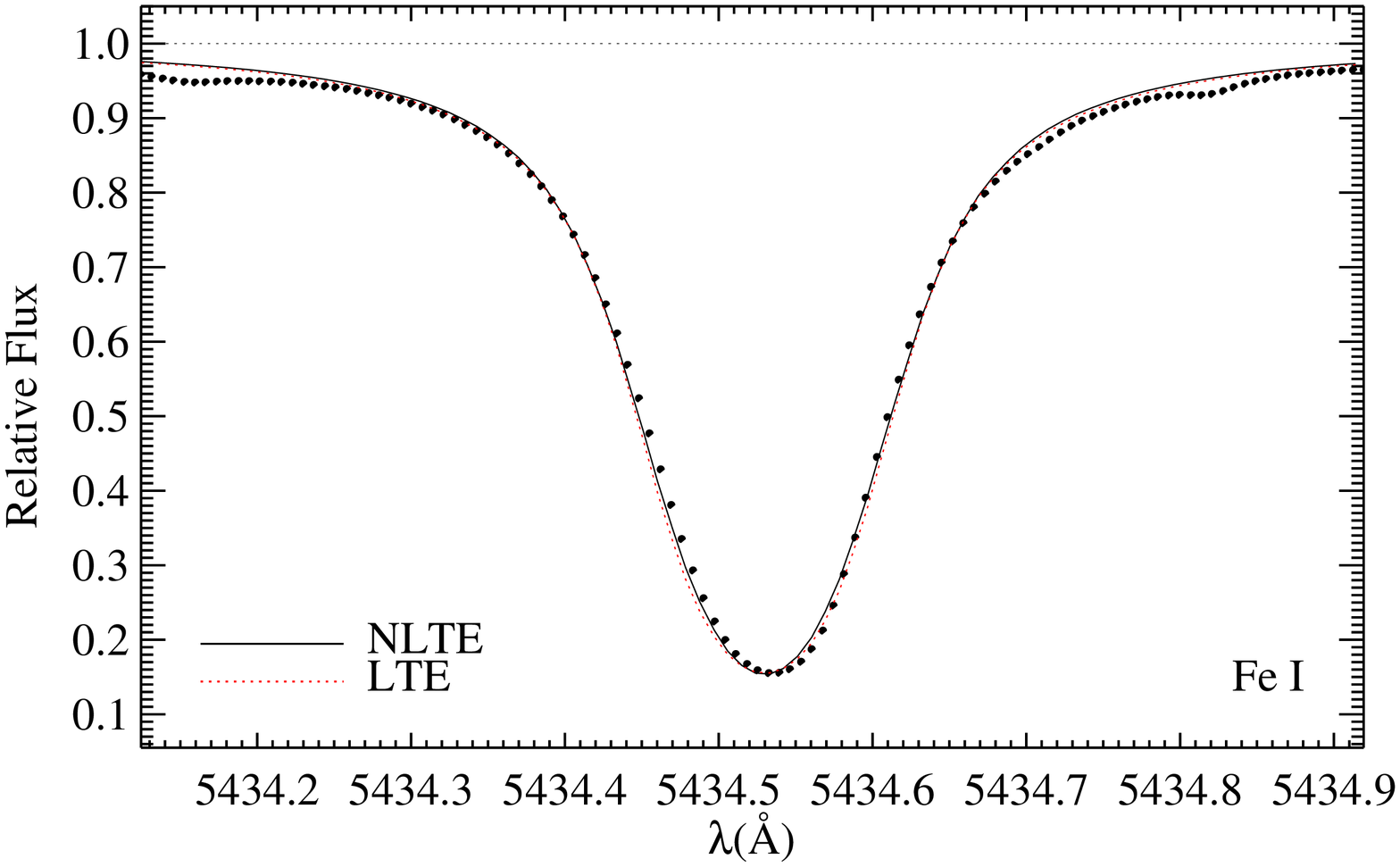}\hfill\hspace{-5mm}
\includegraphics[scale=0.35]{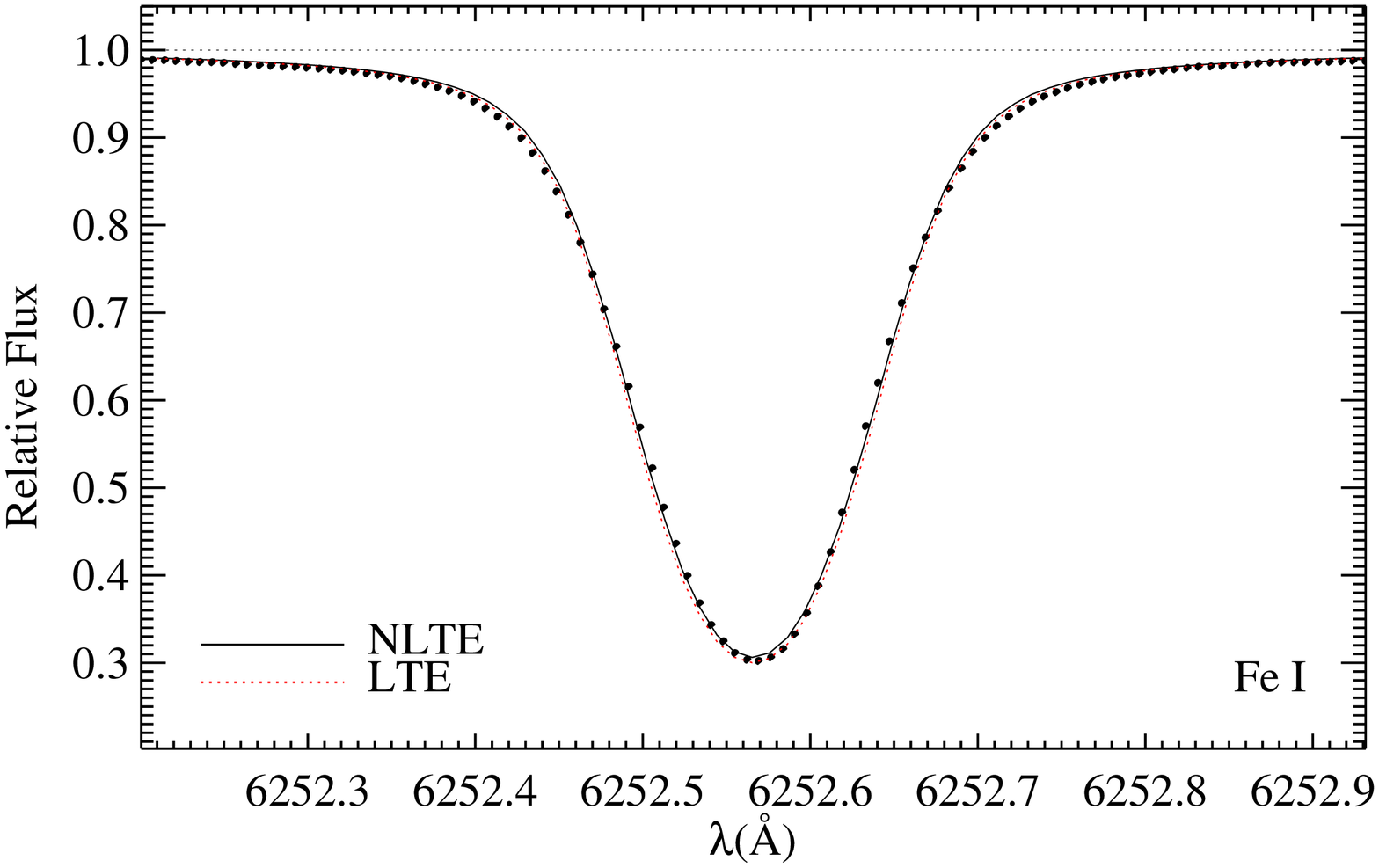}}\vspace{-3mm}
\caption{Synthetic (black trace) and observed (dots) profiles of selected
\ion{Fe}{i} and \ion{Fe}{ii} lines in the solar flux spectrum. The results are
shown for the \textsc{mafags-os} solar model in NLTE (black trace) and LTE
(dotted trace).}
\label{profiles}
\end{figure*}

Selected \ion{Fe}{i} and \ion{Fe}{ii} solar line profiles computed under LTE and
NLTE with the $\td$ models are compared with the observations in Fig.
\ref{profiles}. In general, NLTE \ion{Fe}{i} line profiles are weaker compared
to LTE, which is driven by the NLTE effect on the line opacity. For the
strongest \ion{Fe}{i} lines, forming very far out in the atmospheres, deviation
of the line source function from the Planck function, $S_{\rm ij} <
B_{\nu}(\Te)$, leads to line core darkening. However, their wings, which
dominate the total line strength, are formed at the depths where $b_i < 1$ due
to over-ionization, and the NLTE absorption coefficient is smaller. Therefore
the
NLTE equivalent widths are smaller compared to LTE for a given abundance. As
discussed in Sect. \ref{sec:statec}, NLTE effects on the \ion{Fe}{ii} lines are
negligible. We also inspected the influence of inelastic \ion{H}{i} collisions
on the line profiles finding that the profile shapes are rather insensitive to
the $\SH$ value. Any Fe line could be fitted by slightly adjusting the abundance
or the damping parameters within the error bars (typically $10$ percent).

\begin{figure*}
{\rotatebox{90}
{\includegraphics[scale=0.7]{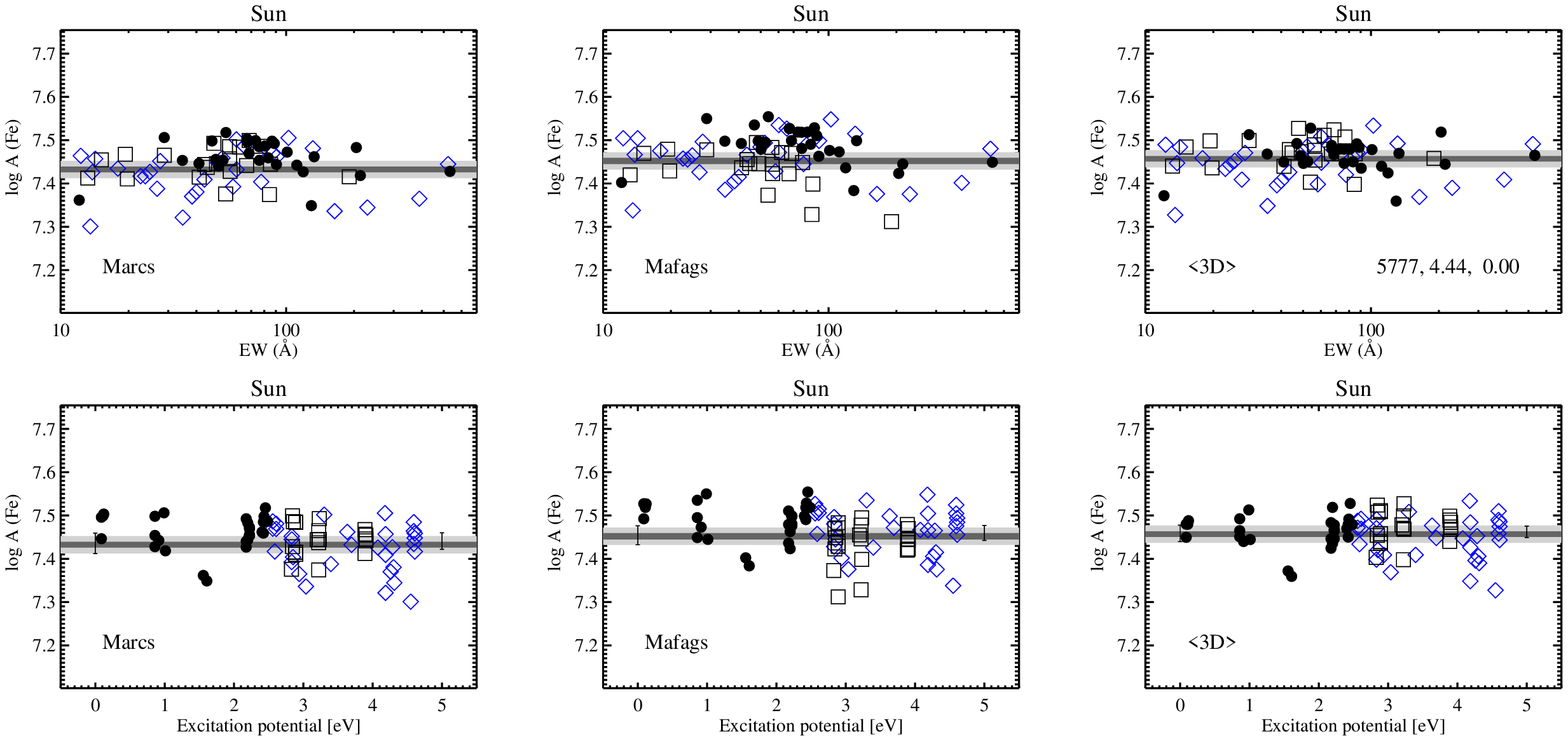}}}
{\rotatebox{90}
{\includegraphics[scale=0.7]{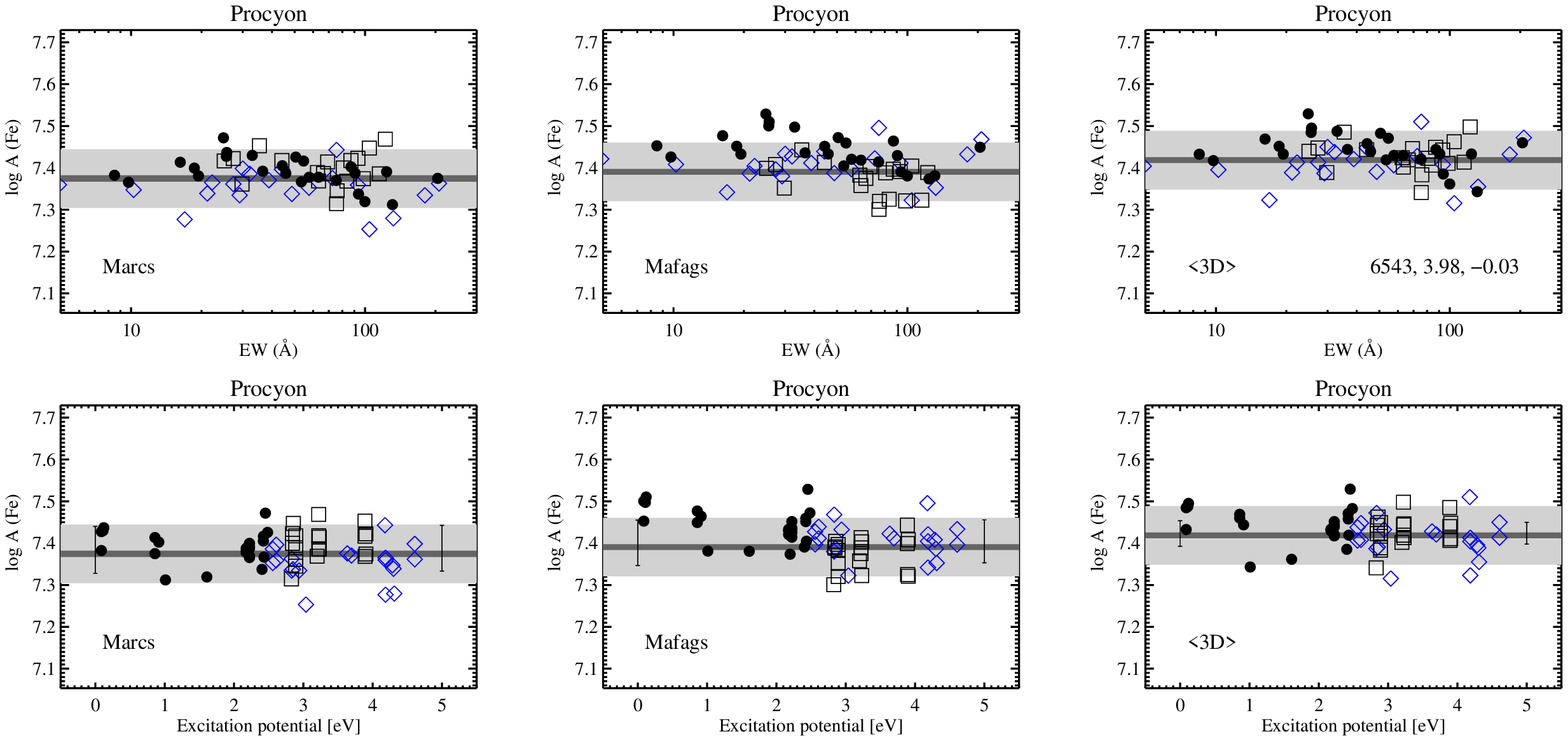}}}
\caption{NLTE abundances determined from the low-excitation \ion{Fe}{i}
(filled circles), high-excited \ion{Fe}{i} (blue diamonds), and \ion{Fe}{ii}
(squares) lines using \textsc{marcs-os}, \textsc{mafags-os}, and $\td$ model
atmospheres for the Sun (top) and Procyon (bottom). Small vertical bars on the
left and right-hand side of each plot show the size of NLTE abundance
corrections for
the low-excited ($< 2.5$ eV) and high-excited \ion{Fe}{i} lines, respectively.
They are leveled out at the mean NLTE abundance obtained from the \ion{Fe}{i}
and \ion{Fe}{ii} lines with $\SH = 1$. The bar's upper and lower ends correspond
to the mean NLTE abundance obtained with $\SH = 0.1$ and the LTE abundance,
respectively. The dark and light shaded regions indicate the standard and
total errors of the mean (see text).}
\label{st12}
\end{figure*}

The mean Fe abundances computed with the \textsc{mafags-os}, \textsc{marcs},
and the $\td$ model atmospheres are given in the Table \ref{tab:fin_param} (see
also Fig. \ref{st12}). The total errors of the mean are only shown for the NLTE
results obtained with the $\td$ models. The total error was computed as:

$$\sigma_{\rm tot} = \left((\sigma_{\rm tot, Fe I}^2 + \sigma_{\rm tot, Fe
II}^2)/2 \right)^{1/2}$$
where,
$$\sigma_{\rm tot,~Fe} = (\sigma_{\rm Fe}^2 + \sigma_{\log gf,
\rm{Fe}}^2 + \sigma_{\rm \log g, Fe}^2 + \sigma_{\Teff, \rm{Fe}}^2)^{1/2}$$
The total error includes the following uncertainties: observational errors
given by the standard deviation, errors stemming from the reference values of
the surface gravity and effective temperatures (Table \ref{tab:init_param}), and
in the $\log gf$ values. For the latter, we adopted the uncertainties given by
NIST and assumed they are uncorrelated, since the $gf$-values come from
different sources in the literature\footnote{Note that it is not possible to
assess systematic errors in $\log gf$'s.}. That is, the total uncertainty
$\sigma_{\log gf}$ is decreased by a factor $\sqrt N$, where $N$ is the number
of lines, and it takes into account the relative number of the \ion{Fe}{i} and
\ion{Fe}{ii} lines analyzed per star.

We find that the excitation and ionization equilibrium of Fe in the Sun and
Procyon is well satisfied under NLTE with both 1D and $\td$ models (Fig.
\ref{st12}, Table \ref{tab:ion_abu}). The statistical
uncertainty of the mean abundance, which corresponds to the standard error of
the line sample, is not larger than $0.01$ dex. Some lines with $\EW > 100\ \mA$
seem to indicate slightly lower abundances, by $\sim 0.03$ dex, compared to the
rest of the sample both in 1D and $\td$ modelling. We do not assign much weight
to these lines. First, it is almost impossible to discern the influence of
abundance, damping, and weak
blends in the line wings. Second, their asymmetric profiles
are clearly shaped by convective velocity fields, which are not accounted for
in our models. As a consequence, it is hardly possible to assign a unique
'best-fitting' abundance to such lines at all, that, in addition to a systematic
error caused by the neglect of 3D effects, introduces subjectivity in abundance
estimates.

The mean NLTE abundances of Fe in the Sun and Procyon, determined using the
$\td$ model atmospheres, are $7.46 \pm 0.02$ respectively $7.43 \pm 0.07$ dex.
1D models (\textsc{marcs} or \textsc{mafags-os}) yield slightly lower values,
which are, however, all consistent within the total error. For the Sun, the
latter only reflects the statistical uncertainty of the $gf$-values, as
described above. We also find that adopting LTE or scaling down the
cross-sections for inelastic collisions with \ion{H}{i} by an order of
magnitude, $\SH =0.1$ does not yield any significant effect on the mean
abundance. The difference between the two extreme cases $\Delta$(NLTE-LTE) is
$\approx 0.02$ dex. Thus, based on the analysis of the solar-metallicity stars
only it is not possible to single out the optimum value of $\SH$. It turns out,
however, that this problem can be solved using the metal-poor stars with
reliable stellar parameters (Sect. \ref{sec:abstars}).

Comparing our results with the published values of the Fe abundance in the Sun
and Procyon, we find full agreement with all data obtained with similar
techniques and model atmospheres. We list only few most recent examples.
\citet{2009A&A...497..611M} using a $\td$ solar model very similar to that
adopted in this work and LTE approach obtain $\logeFetwos\ = 7.45$ dex. The NLTE
1D estimate by \citet{2001A&A...380..645G} obtained using the
\textsc{mafags-odf} (note the difference with the \textsc{mafags-os} version
used here) models is $\logeFesun\ = 7.48 ... 7.51$ dex. In the follow-up NLTE
study with the \textsc{mafags-os} models, \citet{2011A&A...528A..87M} find
$\logeFesun\ = 7.41 \ldots 7.56$ depending on the source of $gf$-values used for
the \ion{Fe}{ii} lines. For Procyon, \citet{2011A&A...528A..87M} claim
discrepant abundances from the NLTE analysis of the \ion{Fe}{i} and \ion{Fe}{ii}
lines, $\logeFeone\ = 7.46 \pm 0.07$ respectively $\logeFetwo\ = 7.52 \pm 0.05$
dex (their Table 3, solution $\SH = 0.1$) suggesting that either increasing
$\Teff$ by $80$ K or decreasing $\log g$ by $0.15$ dex may solve the problem.
Their results for Procyon with $\SH = 1$, which is consistent with our study,
are discrepant, $\logeFeone\ = 7.41 \pm 0.07$ respectively
$\logeFetwo\ = 7.53 \pm 0.05$ dex, which is most likely due to the different
sources of $\log gf$ values for the \ion{Fe}{ii} lines.

We emphasize that further, more detailed, comparison to these and other studies
is not meaningful because of various methodical differences, such as atomic
data, line selection, and microturbulence, to name just a few. 
%
%
\subsection{Analysis of metal-poor stars}\label{sec:abstars}
%
%
\begin{table*}
\centering
\caption{Metallicities \textbf{[Fe/H]} and microturbulence $\Vmic$ (km s$^{-1}$)
  determined relative to the Sun for the reference
  stars. The $\Vmic$ for the \textsc{mafags} models are identical to that
  obtained with the \textsc{marcs} models and are not shown. NLTE results refer to $\SH = 1$. The references to the  $\Teff$ and $\logg$ values are given in the Table 1. See text.}
\label{tab:fin_param}
\begin{tabular}{lcc@{\hspace{1.1cm}}|cccc@{\hspace{1.1cm}}cc@{\hspace{1.1cm}}ccccc}
\hline
\noalign{\smallskip}
 Star & \multicolumn{2}{c}{fixed} &  \multicolumn{2}{c}{derived $\rightarrow$} & & & & & & & &\\
     & $T_{\rm eff}$ & $\log g$ & 
 \multicolumn{4}{c}{\textsc{marcs}} & \multicolumn{2}{c}{\textsc{mafags}} &
 \multicolumn{4}{c}{\textsc{$\td$}} & $\sigma_{\rm tot}$ \\
  &  &  & NLTE & $\xi$ & LTE & $\xi$ & NLTE & LTE & NLTE & $\xi$ &
 LTE & $\xi$ & NLTE, $\td$ \\
\noalign{\smallskip}
\hline
\noalign{\smallskip}
Sun          & $5777$ & $4.44$ & ~7.44 & 1.02 & ~7.43 & 1.02 & ~7.46 & ~7.45 & ~7.46 & 1.08 & ~7.45 & 1.09 & 0.02 \\
Procyon      & $6543$ & $3.98$ & -0.06 & 1.93 & -0.08 & 1.94 & -0.05 & -0.07 & -0.03 & 2.05 & -0.04 & 2.06 & 0.07 \\
HD 84937     & $6408$ & $4.13$ & -2.07 & 1.42 & -2.10 & 1.43 & -2.05 & -2.09 & -2.03 & 1.38 & -2.12 & 1.35 & 0.08 \\
HD 140283    & $5777$ & $3.70$ & -2.38 & 1.26 & -2.42 & 1.27 & -2.40 & -2.43 & -2.40 & 1.18 & -2.53 & 1.17 & 0.07 \\
HD 122563    & $4665$ & $1.64$ & -2.55 & 1.60 & -2.60 & 1.61 & -2.48 & -2.52 & -2.57 & 1.66 & -2.66 & 1.67 & 0.24 \\
G  64-12     & $6464$ & $4.30$ & -3.12 & 1.75 & -3.24 & 1.75 & -3.16 & -3.23 & -3.10 & 1.75 & -3.37 & 1.75 & 0.17 \\
\noalign{\smallskip}
\hline
\end{tabular}
\end{table*}

%
\begin{table*}
\centering
\caption{Spectroscopic $\Teff$, $\logg$, $\Vmic$ (km s$^{-1}$), and [Fe/H] values. The parameters, 
  which were kept fixed in the calculations, are given in the column 2.}
\label{tab:opt_param}
\renewcommand{\footnoterule}{}
\renewcommand{\tabcolsep}{3.5pt}
\begin{tabular}{lc ccc|ccc|ccc|ccc|ccc|ccc}
\hline\noalign{\smallskip}
Star & fixed  & \multicolumn{2}{c}{derived $\rightarrow$} & & & & & & & & & & & & & & & \\
     & $\log g$/$T_{\rm eff}$ & 
       \multicolumn{3}{c}{1D, LTE} & \multicolumn{3}{c}{1D, $\SH=1$} &
       \multicolumn{3}{c}{1D, $\SH=0.1$} & \multicolumn{3}{c}{$\td$, LTE} &
       \multicolumn{3}{c}{$\td$, $\SH=1$} & 
       \multicolumn{3}{c}{$\td$, $\SH=0.1$} \\
\noalign{\smallskip}\hline\noalign{\smallskip}
       Sun &   4.44 &   5794 &  1.0 &   7.44 &   5780 &  1.0 &   7.45 &   5753 &  1.0 &   7.46 &   5805 &  1.1 &   7.47 &    5793 $\pm$  35 &  1.1 &    7.47 $\pm$  0.03 &   5775 &  1.0 &   7.48 \\
   Procyon &   3.98 &   6631 &  1.9 &  -0.04 &   6578 &  1.9 &  -0.05 &   6487 &  1.9 &  -0.05 &   6576 &  2.0 &  -0.04 &    6546 $\pm$  43 &  2.0 &   -0.04 $\pm$  0.04 &   6499 &  2.0 &  -0.05 \\
   HD84937 &   4.13 &   6405 &  1.4 &  -2.11 &   6328 &  1.4 &  -2.13 &   6115 &  1.4 &  -2.16 &   6483 &  1.3 &  -2.08 &    6314 $\pm$  69 &  1.4 &   -2.10 $\pm$  0.07 &   6134 &  1.5 &  -2.12 \\
  HD140283 &   3.70 &   5772 &  1.2 &  -2.43 &   5703 &  1.2 &  -2.45 &   5510 &  1.2 &  -2.49 &   6078 &  1.2 &  -2.36 &    5820 $\pm$  65 &  1.2 &   -2.38 $\pm$  0.07 &   5586 &  1.3 &  -2.41 \\
  HD122563 &   1.64 &   4754 &  1.6 &  -2.51 &   4705 &  1.6 &  -2.51 &   4609 &  1.5 &  -2.49 &   4838 &  1.7 &  -2.53 &    4755 $\pm$  59 &  1.7 &   -2.50 $\pm$  0.14 &   4669 &  1.7 &  -2.48 \\
    G64-12 &   4.30 &   6592 &  1.8 &  -3.15 &   6471 &  1.8 &  -3.17 &   6006 &  1.8 &  -3.34 &   6757 &  1.8 &  -3.18 &    6439 $\pm$ 152 &  1.8 &   -3.17 $\pm$  0.17 &   6258 &  1.8 &  -3.16 \\
 & &  &  &  &  &  &  &  &  &  &  &  &  & & & & & & \\
       Sun &   5777 &   4.40 &  1.0 &   7.44 &   4.43 &  1.0 &   7.45 &   4.48 &  1.0 &   7.47 &   4.38 &  1.0 &   7.46 &    4.40 $\pm$  0.06 &  1.0 &    7.47 $\pm$  0.03 &   4.44 &  1.0 &   7.48 \\
   Procyon &   6543 &   3.81 &  1.9 &  -0.11 &   3.93 &  1.9 &  -0.07 &   4.08 &  1.9 &  -0.04 &   3.92 &  2.0 &  -0.05 &    3.99 $\pm$  0.14 &  2.0 &   -0.04 $\pm$  0.08 &   4.06 &  2.0 &  -0.02 \\
   HD84937 &   6408 &   4.14 &  1.4 &  -2.11 &   4.26 &  1.4 &  -2.07 &   4.57 &  1.4 &  -1.98 &   4.00 &  1.3 &  -2.12 &    4.28 $\pm$  0.16 &  1.4 &   -2.04 $\pm$  0.08 &   4.53 &  1.5 &  -1.95 \\
  HD140283 &   5777 &   3.72 &  1.2 &  -2.42 &   3.84 &  1.2 &  -2.39 &   4.21 &  1.2 &  -2.26 &   3.11 &  1.1 &  -2.59 &    3.63 $\pm$  0.14 &  1.1 &   -2.41 $\pm$  0.08 &   4.05 &  1.3 &  -2.26 \\
  HD122563 &   4665 &   1.22 &  1.6 &  -2.65 &   1.48 &  1.6 &  -2.57 &   1.83 &  1.6 &  -2.45 &   0.91 &  1.7 &  -2.74 &    1.31 $\pm$  0.26 &  1.7 &   -2.59 $\pm$  0.19 &   1.61 &  1.7 &  -2.48 \\
    G64-12 &   6464 &   4.06 &  1.8 &  -3.26 &   4.29 &  1.8 &  -3.18 &   4.90 &  1.8 &  -3.05 &   3.75 &  1.8 &  -3.37 &    4.34 $\pm$  0.20 &  1.8 &   -3.16 $\pm$  0.10 &   4.56 &  1.8 &  -3.04 \\
\noalign{\smallskip}\hline
\end{tabular}
\end{table*}

We start with a discussion of how NLTE affects the Fe abundances obtained with
1D and $\td$ models, delineating the key differences. In Sect. \ref{sec:teff},
we then illustrate how these differences propagate in the determination of
effective temperatures and surface gravities.

%
\begin{table*}
\centering
\caption{LTE and NLTE abundances determined from the Fe I and Fe II lines using the reference values of $\Teff$ and $\log g$ (Tables 1 and 5). 
NLTE results refer to $\SH = 1$. Standard deviations $\sigma$ of the Fe I respectively Fe II mean abundances are also indicated.}
\label{tab:ion_abu}
\begin{tabular}{l cccccc@{\hspace{0.8cm}}cccc@{\hspace{0.8cm}}cccccc}
\hline
\noalign{\smallskip}
 Star 
  & \multicolumn{6}{c}{\textsc{marcs}} 
  & \multicolumn{4}{c}{\textsc{mafags}}
  & \multicolumn{6}{c}{\textsc{$\td$}} \\
  & \multicolumn{2}{c}{LTE} & \multicolumn{4}{c}{NLTE} 
  & \multicolumn{2}{c}{LTE} & \multicolumn{2}{c}{NLTE} 
  & \multicolumn{2}{c}{LTE} & \multicolumn{4}{c}{NLTE} \\  
   &  Fe I & Fe II & Fe I & $\sigma$ &  Fe II &  $\sigma$ & Fe I & Fe II &  Fe I & Fe II &  Fe I & Fe II &  Fe I & $\sigma$ & Fe II & $\sigma$ \\
\noalign{\smallskip}
\hline
\noalign{\smallskip}
Sun             &  7.43 & 7.45 & 7.44 & 0.05 & 7.44 & 0.04 &  7.46 & 7.44 & 7.48 & 7.43 &   7.44 & 7.47 & 7.45 & 0.04 & 7.47 & 0.04 \\
Procyon      &  7.33 & 7.40 & 7.37 & 0.04 & 7.40 & 0.04 &  7.39 & 7.38 & 7.43 & 7.37 &   7.40 & 7.43 & 7.43 & 0.04 & 7.43 & 0.04 \\
HD 84937   &  5.32 & 5.33 & 5.39 & 0.07 & 5.33 & 0.04 &  5.36 & 5.38 & 5.42 & 5.38 &   5.30 & 5.38 & 5.45 & 0.06 & 5.38 & 0.03 \\
HD 140283 &  5.02 & 5.01 & 5.07 & 0.09 & 5.01 & 0.04 &  5.03 & 5.01 & 5.09 & 5.01 &   4.86 & 5.09 & 5.05 & 0.05 & 5.09 & 0.04 \\
HD 122563 &  4.78 & 4.95 & 4.87 & 0.12 & 4.95 & 0.05 &  4.93 & 4.95 & 5.00 & 4.95 &   4.71 & 5.01 & 4.85 & 0.08 & 5.01 & 0.05 \\
G  64-12     &  4.25 & 4.23 & 4.36 & 0.05 & 4.24 & 0.09 &  4.23 & 4.21 & 4.34 & 4.22 &   4.20 & 4.28 & 4.39 & 0.05 & 4.30 & 0.08 \\
\noalign{\smallskip}
\hline
\end{tabular}
\end{table*}

\subsubsection{Metallicities}{\label{sec:meta}}

The most important test of our new models is whether they are able to recover
ionization equilibrium of \ion{Fe}{i} and \ion{Fe}{ii} for the stars with
parameters determined by independent techniques, given a unique $\SH$. As
described in Sect. \ref{sec:obs}, a small set of metal-poor stars in different
evolutionary stages was selected for this purpose with $\Teff$ known from the
interferometry and/or IRFM, and gravities well constrained by parallaxes.
Metallicities and microturbulence parameters were then determined keeping the
$\Teff$ and $\logg$ fixed, and varying the value of $\SH$ in the NLTE
calculations.

\begin{figure*}
{\rotatebox{90}
{\includegraphics[scale=0.7]{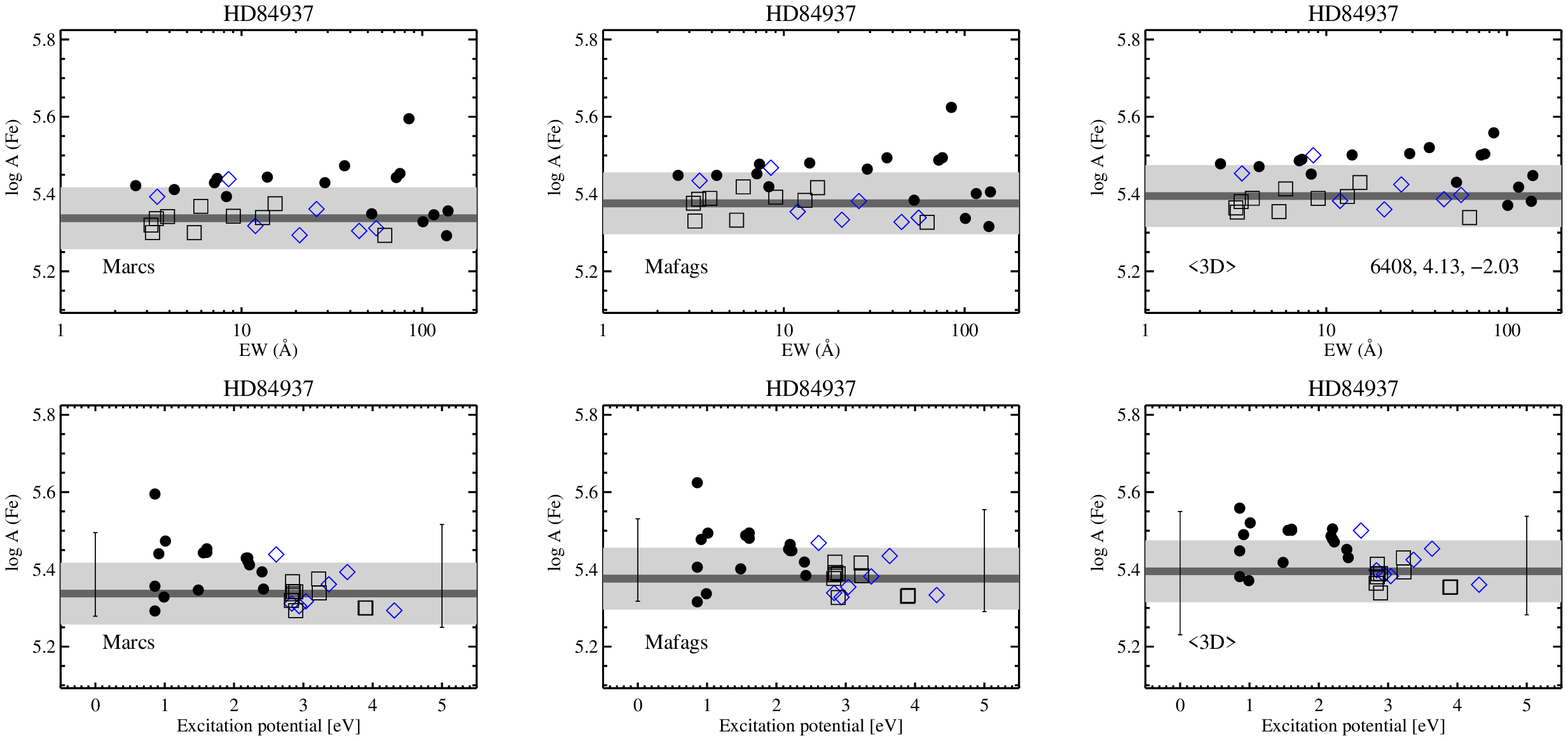}}}
{\rotatebox{90}
{\includegraphics[scale=0.7]{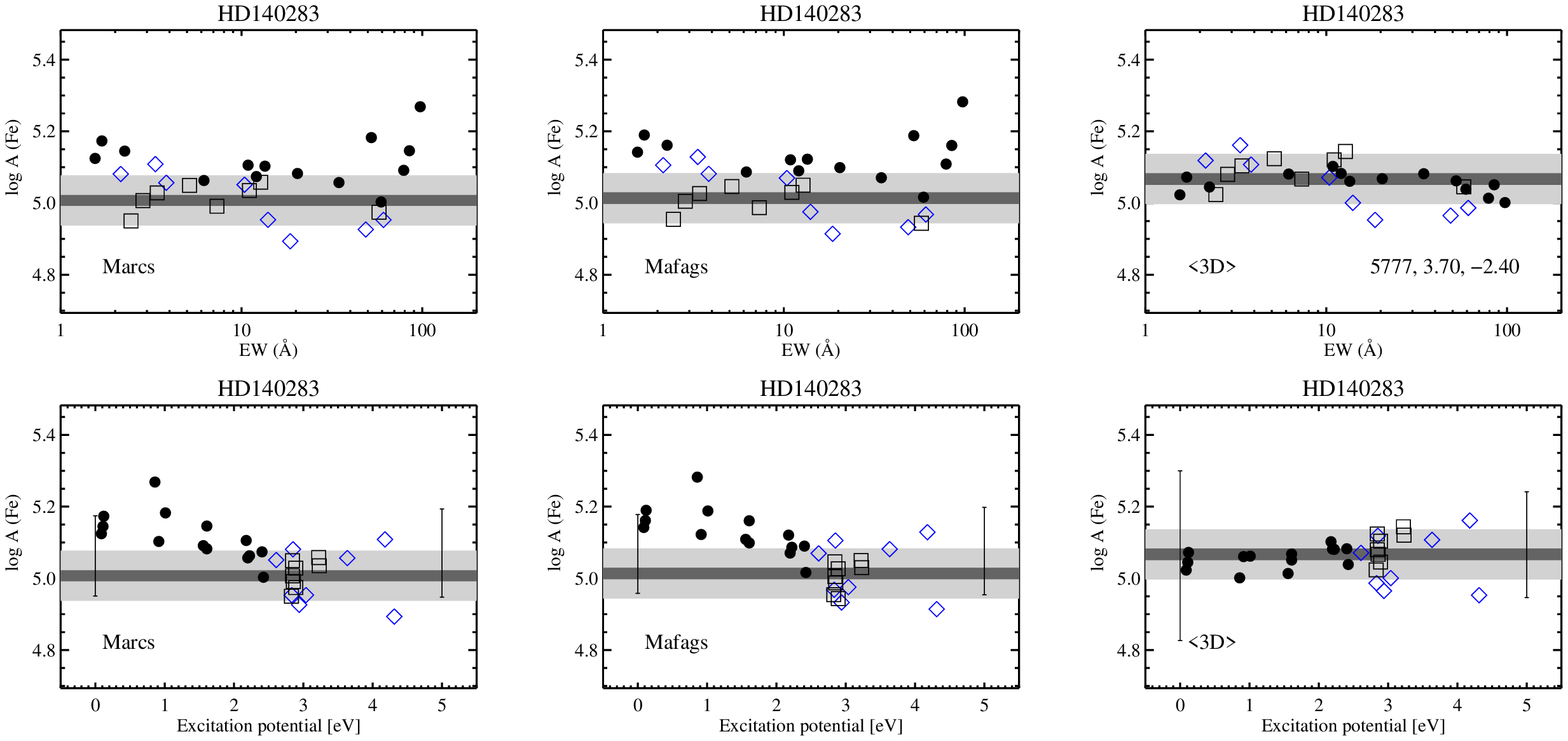}}}
\caption{Abundances determined from the Fe I and Fe II lines using stellar
parameters from Table \ref{tab:fin_param} for HD 84937 (top) and HD 140283
(bottom).}
\label{st34}
\end{figure*}

\begin{figure*}
{\rotatebox{90}
{\includegraphics[scale=0.7]{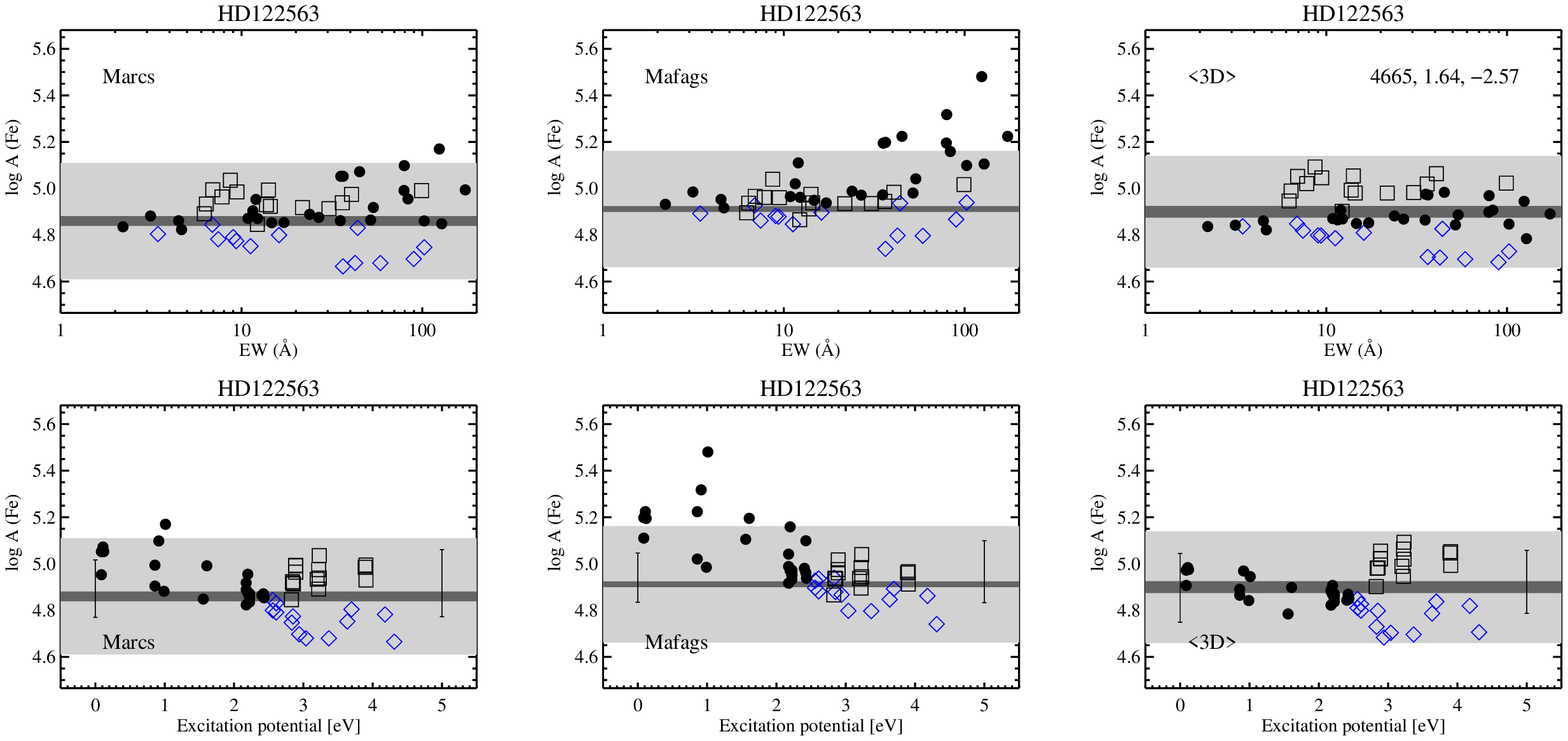}}}
{\rotatebox{90}
{\includegraphics[scale=0.7]{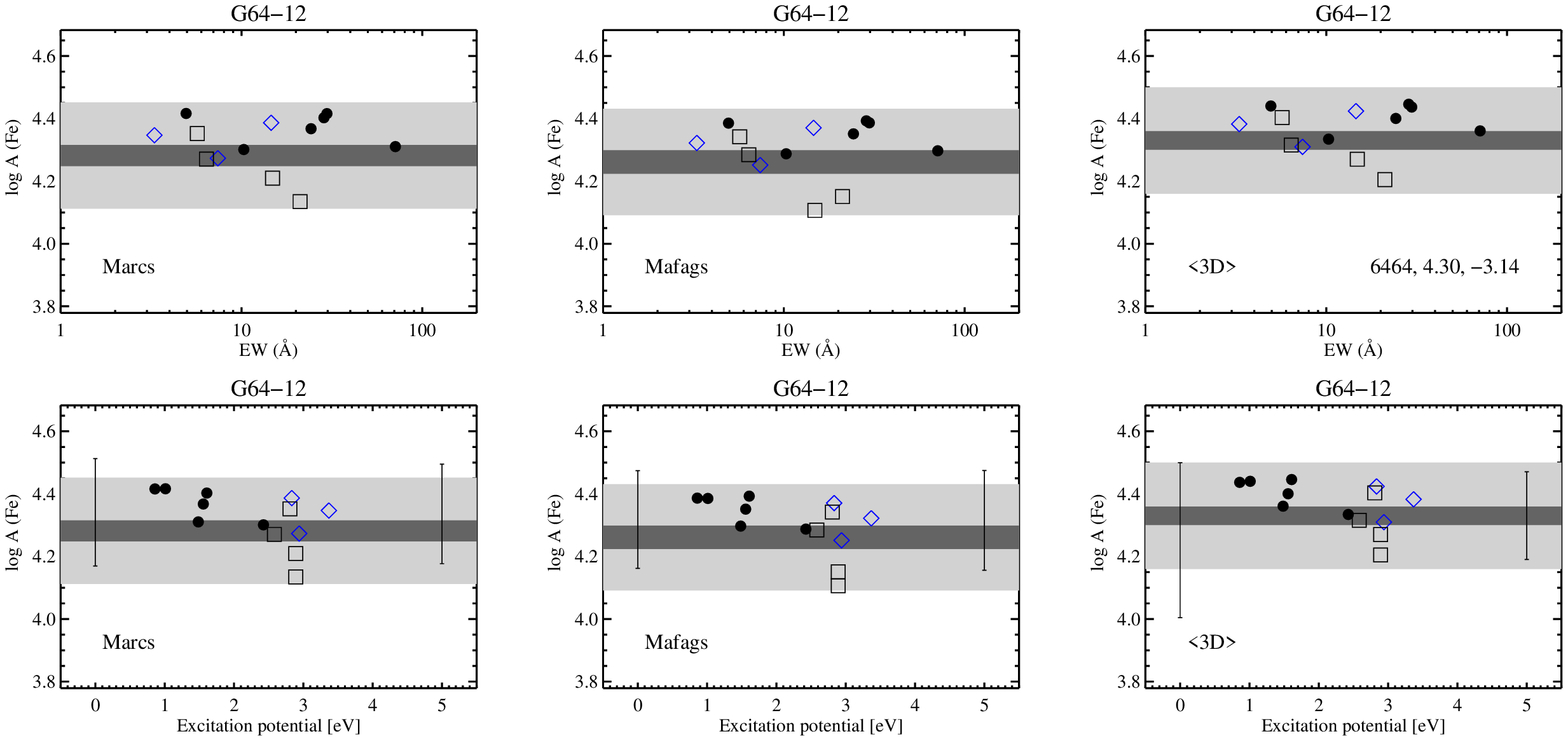}}}
\caption{Abundances determined from the Fe I and Fe II lines using stellar
parameters from Table \ref{tab:fin_param} for HD 122563 (top) and G 64-12
(bottom). Note that the total errors are large as they also reflect the
uncertainties in the input stellar parameters.}
\label{st56}
\end{figure*}

We find that the solution providing optimum ionization balance of Fe for all
metal-poor stars is achieved with $\SH = 1$, i.e, unscaled Drawin's \ion{H}{i}
collisions (see also discussion in Sect. \ref{sec:teff}). The Fe abundances
determined in this way, along with the optimized $\td$ NLTE microturbulence
$\Vmic$, are given in the Table \ref{tab:fin_param}. These are mean abundances
from the \ion{Fe}{i} and \ion{Fe}{ii} lines, and the errors were computed as
described in Sect. \ref{sec:sun_procyon}. For all stars, but the Sun and G 64-12
with its very uncertain parallax, the errors are dominated by the uncertainties
of the reference $\Teff$. The abundances are shown in Figs. \ref{st12},
\ref{st34}, and \ref{st56} as a function of line equivalent widths and lower
level excitation potentials (in eV). The error bars in the figures indicate the
size of NLTE abundance corrections for the low-excitation ($< 2.5$ eV) and
high-excitation \ion{Fe}{i} lines. They are levelled out at the mean NLTE
abundance obtained from the \ion{Fe}{i} and \ion{Fe}{ii} lines with $\SH = 1$.
The bar's upper and lower ends correspond to the mean NLTE abundance obtained
with $\SH = 0.1$ and the LTE abundance, respectively. The abundances
averaged over the measured \ion{Fe}{i} respectively \ion{Fe}{ii} lines are
presented in Table \ref{tab:ion_abu} along with their
one sigma errors. Note that these are not total propagated errors as described
in Sec. 4.4, but standard deviations of the mean abundance for each ionization
stage. These quantities mainly serve to show the difference in the line scatter
between the 1D (\textsc{marcs}) and $\td$ models.
\begin{figure*}
\includegraphics[scale=0.6]{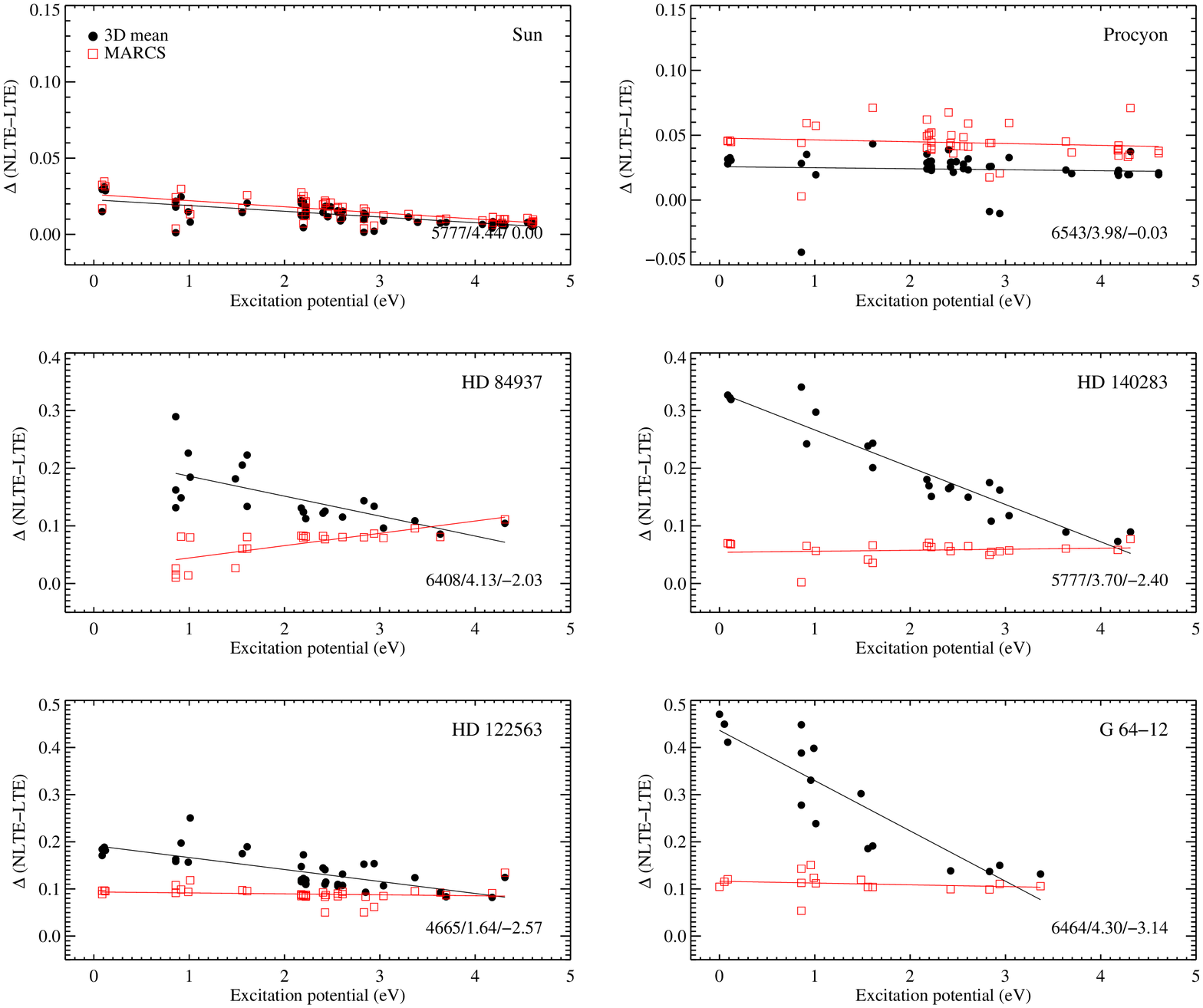}
\caption{The NLTE abundance corrections to the \ion{Fe}{i} lines
obtained with \textsc{Marcs} and $\td$ model atmospheres for the same reference
$\Teff$ and $\log g$ (see text). The metallicities indicated are that obtained
from the NLTE spectroscopic analysis with the $\td$ model atmospheres.}
\label{delta}
\end{figure*}

The major differences between 1D and $\td$, as well as LTE and NLTE, are
apparent from inspection of the Fe abundances in the Table \ref{tab:fin_param}
and the NLTE abundance corrections\footnote{NLTE abundance correction,
$\Delta_{\rm Fe} = \log \rm{A (Fe)}_{\rm NLTE} - \log \rm{A (Fe)}_{\rm LTE}$, is
the difference between the NLTE and LTE abundances required to fit a line with a
given equivalent width} shown in Fig. \ref{delta}. For the Sun and Procyon, the
NLTE effects are rather subtle. The difference between the LTE and NLTE
abundances determined from the \ion{Fe}{i} lines is even slightly smaller in the
$\td$ case compared to 1D (Fig. \ref{delta}, top panels). The basic reason is
that the temperature gradient in the solar-metallicity \textsc{marcs} models is
slightly steeper than in the $\td$ models (Fig. \ref{atm_abs}). At $\opd > 0$,
where the UV-blue continuum is formed, $\left( d T/d \tau \right)_{\rm 1D} >
\left( d T/d \tau \right)_{\rm 3D}$, whereas in the outer layers the gradients
are very similar. Over-ionization in \ion{Fe}{i} by radiation field emerging
from the hot deep layers is, thus, somewhat stronger in the \textsc{marcs}
models.

Over-ionization is more important at low metallicities (see Sect.
\ref{sec:statec} of this paper, and a more extensive discussion in Paper II).
This causes non-negligible differences between the NLTE and LTE abundances
inferred from the \ion{Fe}{i} lines. The differences are in the range $0.05 -
0.15$ dex for 1D models and weak \ion{Fe}{i} lines, but increase to $0.2 - 0.5$
dex for the strong low-excitation \ion{Fe}{i} lines with the $\td$ models (Fig.
\ref{delta}). The larger NLTE effects of metal-poor $\td$ models compared to 1D
models are due to the larger decoupling between the radiation temperature and
the local temperature, owing to the dramatically cooler outer layers of the
$\td$ models compared to 1D (see Sect. \ref{sec:statec} and Fig. \ref{jnu}).
NLTE corrections to the \ion{Fe}{ii} lines are negative and are not larger than
$-0.03$ dex. As a result, for the metal-poor stars, the difference between the
mean LTE and NLTE Fe abundances is, at least, twice as large in the $\td$ case
compared to that obtained with \textsc{marcs} and \textsc{mafags-os} models. On
the other side, the same phenomenon, which leads to amplified NLTE effects in
the
$\td$ calculations, i.e., very steep $\left( d T/d \tau \right)$ above $\opd
\approx 0$, is also responsible for the LTE strengthening of the \ion{Fe}{i}
lines in $\td$ compared to 1D. The \ion{Fe}{ii} lines are only slightly weaker
in $\td$ calculations. Our LTE 1D metallicities are, thus, systematically larger
than the $\td$ results, in agreement with other studies
\citep*[e.g., ][]{2007A&A...469..687C}. We note, however, that direct comparison
of our results with that of, e.g., \citet{2007A&A...469..687C}, is not
meaningful since they performed LTE line formation calculations with full 3D
hydrodynamic model atmospheres.

The fact that the $\td$ models also predict somewhat lower LTE Fe abundances,
yet larger NLTE abundance corrections than 1D models, has the important
consequence that the mean NLTE metallicities in 1D and $\td$ turn out to be in
agreement for the whole range of stellar parameters ($\Delta
\log\varepsilon_{\rm Fe} \leq 0.1$).

However, even though the mean metallicities are similar, individual \ion{Fe}{i}
line abundances are still somewhat discrepant, and show systematic trends with
line equivalent width and excitation potential of the lower level of a
transition, $E_{\rm low}$ (Figs. \ref{st34}, \ref{st56}). NLTE abundances
inferred from the low-excitation lines of \ion{Fe}{i} with 1D models
are too large compared to the high-excitation lines and lines of \ion{Fe}{ii}.
A very similar picture is obtained with the LTE approximation, and it reverses
sign in LTE $\td$. A combination of $\td$ and NLTE alleviates this discrepancy,
so that \ion{Fe}{i} and \ion{Fe}{ii} abundances become more consistent.
The improvement comes from the different sensitivity of the NLTE effects to
$E_{\rm low}$ in the 1D and $\td$ case. This is clearly seen in Fig.
\ref{delta}. A least-square fit to the NLTE abundance corrections is also shown,
which is a simple albeit very illustrative measure of the mean NLTE effect on
the excitation balance of \ion{Fe}{i}. In the 1D case, the \ion{Fe}{i} lines
have rather similar NLTE corrections irrespective of their excitation potential
and equivalent width. In the $\td$ case, however, the low-excitation
transitions, $E_{\rm low} \leq 2$ eV, tend to experience significantly larger
departures from LTE compared to the higher-excitation transitions. The reason
is that, owing to the dramatically cooler surfaces of the metal-poor $\td$
models, low-excitation \ion{Fe}{i} lines become stronger and more sensitive to
the physical conditions in the outer atmospheric layers, where the influence of
non-local radiation field is extreme, not only decreasing the line opacity but
also pushing the line source functions to super-thermal values. Thus, the NLTE
$\td$ profiles of these lines are weaker compared to LTE. The magnitude of
this effect depends on stellar parameters (see also the discussion in Sect.
\ref{sec:statec}). As a consequence, one would also expect that excitation
balance achieved under LTE with the $\td$ models will strongly overestimate
$\Teff$, and the error shall increase with $\Teff$ and decreasing metallicity as
indicated by the increasing slope of $\Delta ({\rm NLTE-LTE})$ vs $E_{\rm low}$
for more metal-poor and hotter stars.

A corollary is that accurate metallicities for late-type stars can be obtained
with standard 1D model atmospheres if NLTE effects in \ion{Fe}{i} are taken into
account. However, this is true only if the following condition is satisfied:
a sufficiently large number of \ion{Fe}{i} and \ion{Fe}{ii} lines of different
types are included in an analysis, so that individual line-to-line abundance
discrepancies cancel out. This approach, however, does not eliminate residual
trends of abundance with line equivalent width and excitation potential.
Although these trends can be, at least in part, corrected for by adjusting
microturbulence, a better approach is to restrict an analysis with 1D
hydrostatic models to high-excitation \ion{Fe}{i} lines only, as also advocated
in the literature \citep[e.g., ][]{1983PASP...95..101G}.
%
%
\subsubsection{Effective temperatures and surface gravities}{\label{sec:teff}}
\begin{figure*}
\includegraphics[scale=0.45]{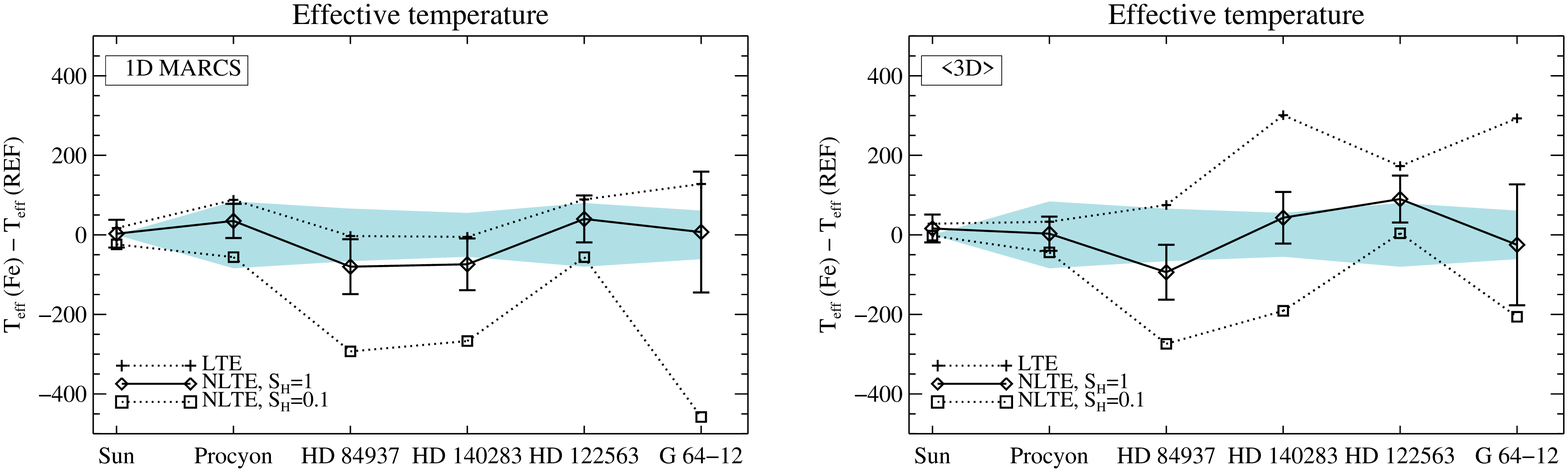}
\includegraphics[scale=0.45]{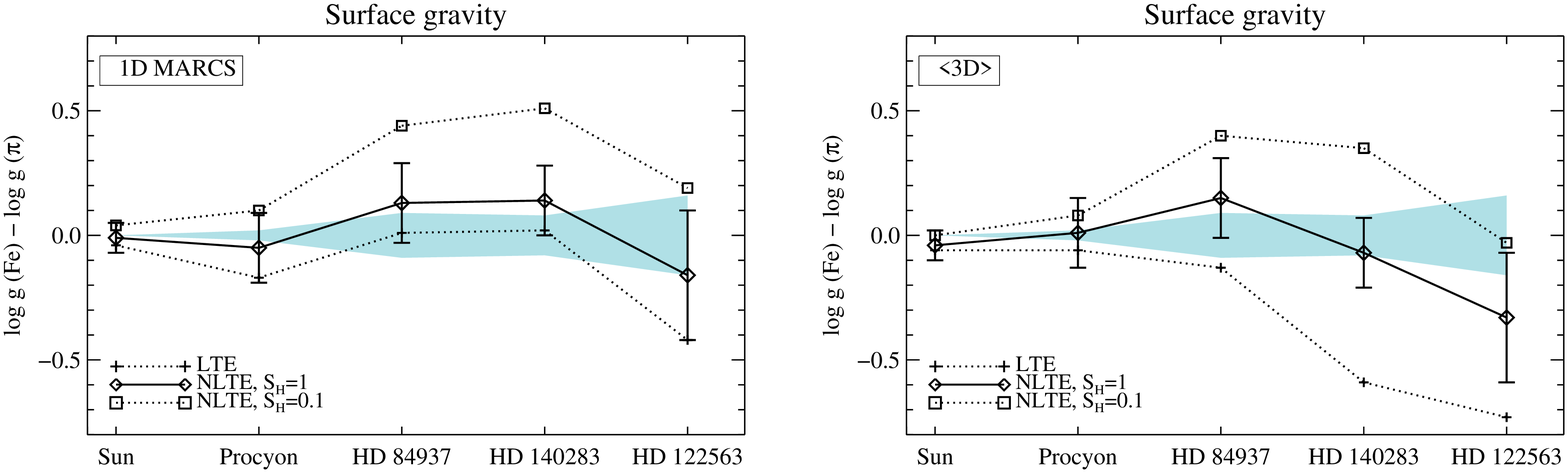}
\caption{Comparison of spectroscopic stellar parameters, $\Teff$ and $\log g$, 
with the data obtained by independent methods, IRFM and parallaxes. G 64-12 is
not shown in the bottom figure, because its astrometric gravity is too
uncertain.}
\label{teff_logg}
\end{figure*}
The spectroscopic effective temperatures and surface gravities determined
according to the procedure described in Sect. \ref{sec:method} are given in the
Table \ref{tab:opt_param}.

Fig. \ref{teff_logg} (top panel) plots the effective temperatures obtained for
the reference stars adopting a fixed surface gravity inferred from stellar
parallaxes. The results are shown for the three cases: LTE, NLTE with the
standard model atom ($\SH = 1$), and NLTE with Drawin's \ion{H}{i}
collision cross-sections scaled by a factor of $\SH = 0.1$. The reference
temperature scale, T$_{\rm eff, REF}$, refers to the IRFM results (see Sect.
\ref{sec:obs}) for all stars but Procyon with the interferometric angular
diameters; their uncertainties are shown with blue filled areas. The
uncertainties of our spectroscopic $\Teff$'s were estimated by mapping the
errors in atomic transition probabilities, observational errors, and errors in
the input surface gravity into a range in temperatures. In order not to
overload the figures, these errors are shown for the NLTE $\SH = 1$ solution
only. We emphasize that the $\Teff$'s were obtained by enforcing ionization
balance, but not requiring null slopes with excitation potential for the reasons
discussed above.

The overall best result in terms of consistency between the spectroscopic
effective temperatures, T$_{\rm eff, Fe}$, and effective temperatures obtained
by independent methods, T$_{\rm eff, REF}$, is achieved under NLTE using the
reference model atom with $\SH = 1$, i.e. standard Drawin's \ion{H}{i}
collisions. The 1D and $\td$ models perform similarly, predicting only minor
offsets of T$_{\rm eff, Fe}$ from the reference IRFM $\Teff$ scale, with an rms
error of the order of $\sim 50$ K. Assuming LTE or decreasing the efficiency of
inelastic \ion{H}{i} collisions in the NLTE statistical equilibrium calculations
by an order of magnitude, $\SH = 0.1$, leads to larger discrepancies. The
latter result is very important. At low metallicity NLTE effects in \ion{Fe}{i}
are extremely sensitive to the magnitude of collisional thermalization by
\ion{H}{i}. Our analysis, thus, not only confirms that it is possible to
constrain the absolute magnitude of atomic collision cross-sections using
metal-poor stars, but also demonstrates that a single scaling factor (unity in
our case) to the classical Drawin's formula is a reasonable approximation. What
concerns the LTE $\Teff$ values, they are still within the errors of T$_{\rm
eff, REF}$ in the 1D case. We caution, however, that the NLTE effects for our
stars are small in 1D, and this statement may not apply for stellar parameters
where the NLTE effects are more significant, i.e. for stars with higher $\Teff$
and lower $\log g$. The discrepancy between the LTE spectroscopic and the IRFM
scale increases to $200$ K for the metal-poor stars when $\td$ models are
employed, which is consistent with the discussion in the previous section.

Our second approach, in which the $\Teff$ values are kept fixed and surface
gravities are obtained by forcing Fe ionization equilibria, returns similar
results. Fig. \ref{teff_logg} (bottom panel) compares the spectroscopic $\log
g$'s with the values determined from parallaxes, the shaded regions indicate the
uncertainty of the latter (see Sect. \ref{sec:obs}) and the error bars of the
former account for the uncertainties in $\log gf$, IRFM effective
temperatures, and observational uncertainties. As seen from this figure, only
the NLTE approach with $\SH = 1$ gives $\log g$ values on average consistent
with that inferred from parallaxes. In $\td$, LTE gravities are systematically
too low. In 1D, the discrepancy is smaller, but appears to increase with
$\Teff$. The assumption of $\SH = 0.1$ brings the \ion{Fe}{i} abundances so
high, that much larger gravities are generally needed to match \ion{Fe}{i} with
\ion{Fe}{ii}.

\subsubsection{Comparison with other studies}

Korn et al. (2003) performed a study similar to ours aiming to constrain
the efficiency of inelastic \ion{H}{i} collisions from the NLTE 1D Fe
calculations for a small sample of late-type stars. Using the \textsc{mafags-os}
model atmospheres and the Drawin's formulae for \ion{H}{i} collision
cross-sections, they determined $\SH = 3$ as a scaling factor to the latter
yielding satisfactory ionization balance of \ion{Fe}{i} and \ion{Fe}{ii}. For
the stars in common, HD 140283, HD 84937, and Procyon, their results are
consistent with our values for $\SH = 1$. This study was, however, later
superseded by that of Mashonkina et al. (2011), in which the Fe model atom of
\citet[][b]{2001A&A...380..645G}, also used by Korn et al. (2003), was updated
by more recent atomic data. This work, in fact, showed that a lower efficiency
of inelastic \ion{H}{i} collisions ($\SH = 0.1$) is needed to satisfy ionization
balance of Fe. Their results for HD 84937 and HD 122563 are in agreement with
our values within the error bars: $-2.00 \pm 0.07$ (Fe I) respectively $-2.08
\pm 0.04$ (Fe II) for HD 84937, and $-2.61 \pm 0.09$ (Fe I) respectively $-2.56
\pm 0.07$ (Fe II) for HD 122563. They also find that the NLTE \ion{Fe}{i}-based
abundances are correlated with line excitation potential, with a slope of
$0.013$ dex eV$^{-1}$ for HD 84937 and $-0.054$ dex eV$^{-1}$ for HD 122563.
From Fig. \ref{st34}, we can not confirm their slope for HD 84937 with 1D
models, however, we would get exactly same results, adopting Mashonkina et al.
linelists. Their positive slope for HD 84937 would likely had vanished
had they included few more subordinate near-UV \ion{Fe}{i} lines with very
accurate $gf$-values in their analysis. For HD 122563, however, we also find a
small discrepancy between the high-excited \ion{Fe}{i} and \ion{Fe}{ii} lines,
in addition to residual slopes of the NLTE Fe abundances with $\Elow$:
$-0.11$~dex eV$^{-1}$ for the \textsc{mafags-os}, $-0.09$~dex eV$^{-1}$ for
\textsc{marcs}, and $-0.05$ dex eV$^{-1}$ for the $\td$ model. One of the
important differences between our and their study, which also explains our more
extreme slopes in 1D, is the way we handle microturbulence. We use \ion{Fe}{ii}
lines, whereas Mashonkina et al. (2011) rely on the \ion{Fe}{i} lines.
Nevertheless, the fact that our results obtained with 1D model atmospheres
support the findings of Mashonkina et al. (2011) is reassuring.
%
%
\subsubsection{HD 122563}{\label{sec:hd122563}}

HD 122563 is one of the most well-studied metal-poor giants. Nevertheless, its
parameters are badly constrained. For example, Fulbright (2000) gives $\Teff =
4425$ K, $\log g = 0.6$,  and [Fe/H] $=-2.6$, whereas according to Mashonkina et
al. (2011) the star is better described by a model with $\Teff = 4600$ K, $\log
g = 1.6$, and [Fe/H] $\approx -2.58$.

At $\SH = 1$, no combination of parameters can achieve a fully satisfactory
spectroscopic solution for this star, consistently achieving adequate excitation
and ionization balance. This is evident from the residual offset between the
high-excitation \ion{Fe}{i} and \ion{Fe}{ii} lines in Fig. \ref{st56}. In NLTE
with $\SH = 1$, the spectroscopic $\log g$ is $0.3$ dex lower than the value
derived from the Hipparcos parallax. The discrepancy can be minimized by
adopting less efficient hydrogen collisions, which, however, contradicts the
spectroscopic results of the other metal-poor stars.

Based on HD122563 alone, the evidence for a lower $\SH$ is thus limited.
Similar investigation of larger samples of metal-poor giants and dwarfs with
well-defined parameters may help to better clarify the situation.
%
%
\subsection{Comparison with evolutionary tracks}{\label{sec:evolution}}

\begin{figure*}
\centering
\hbox{
\resizebox{\columnwidth}{!}{\includegraphics[scale=1.0]{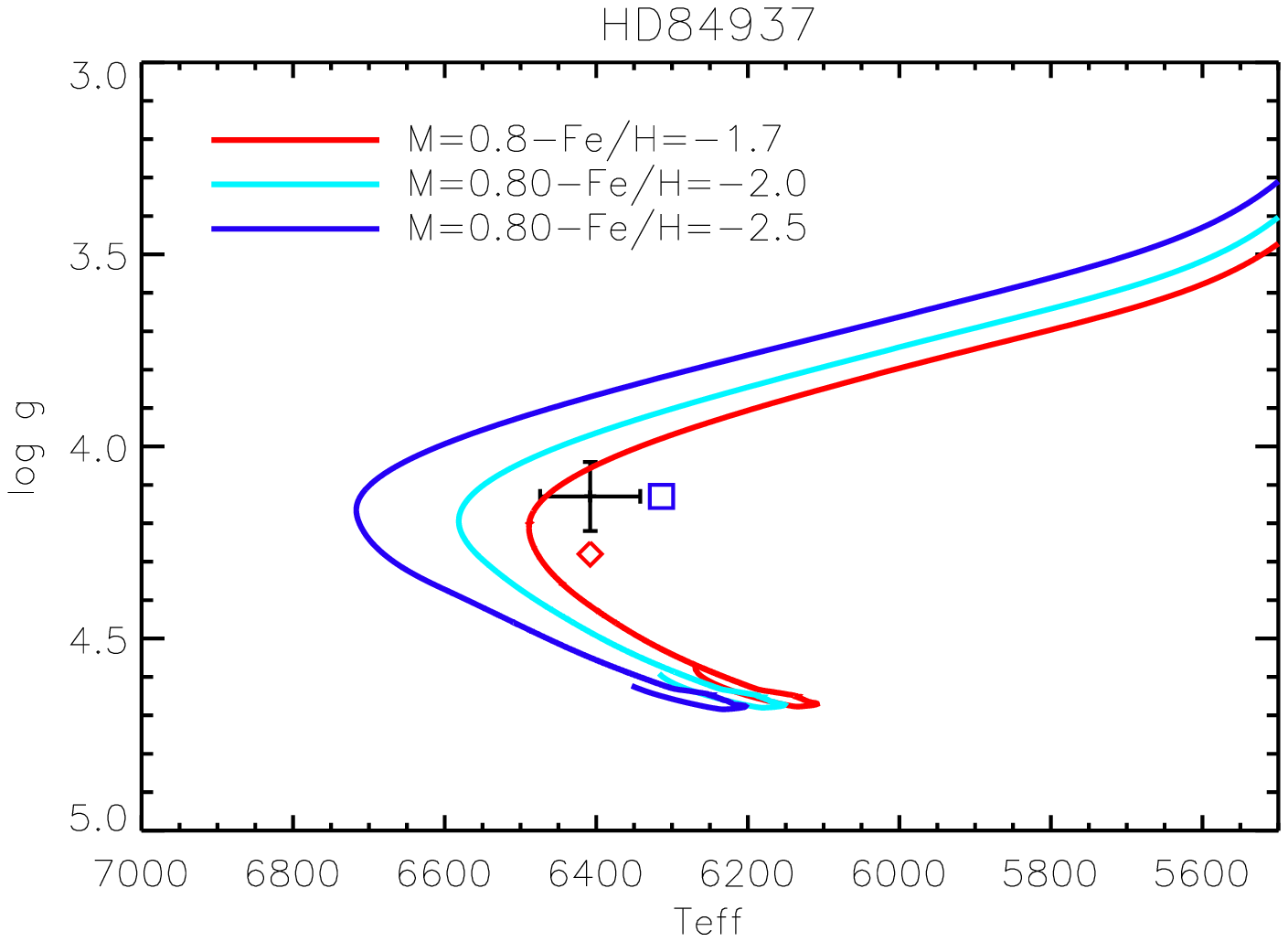}}
\resizebox{\columnwidth}{!}{\includegraphics[scale=1.0]{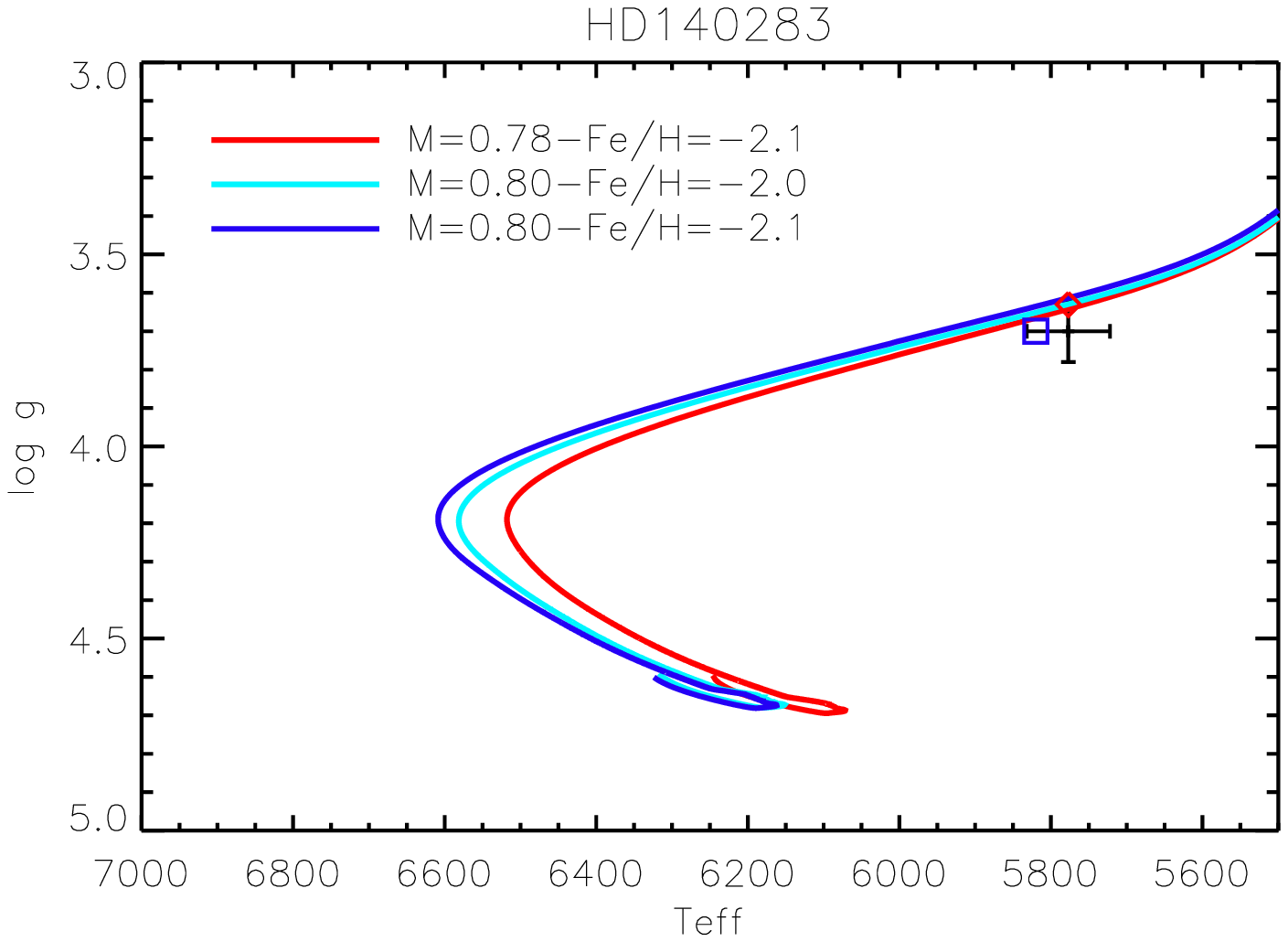}}}
\hbox{
\resizebox{\columnwidth}{!}{\includegraphics[scale=1.0]{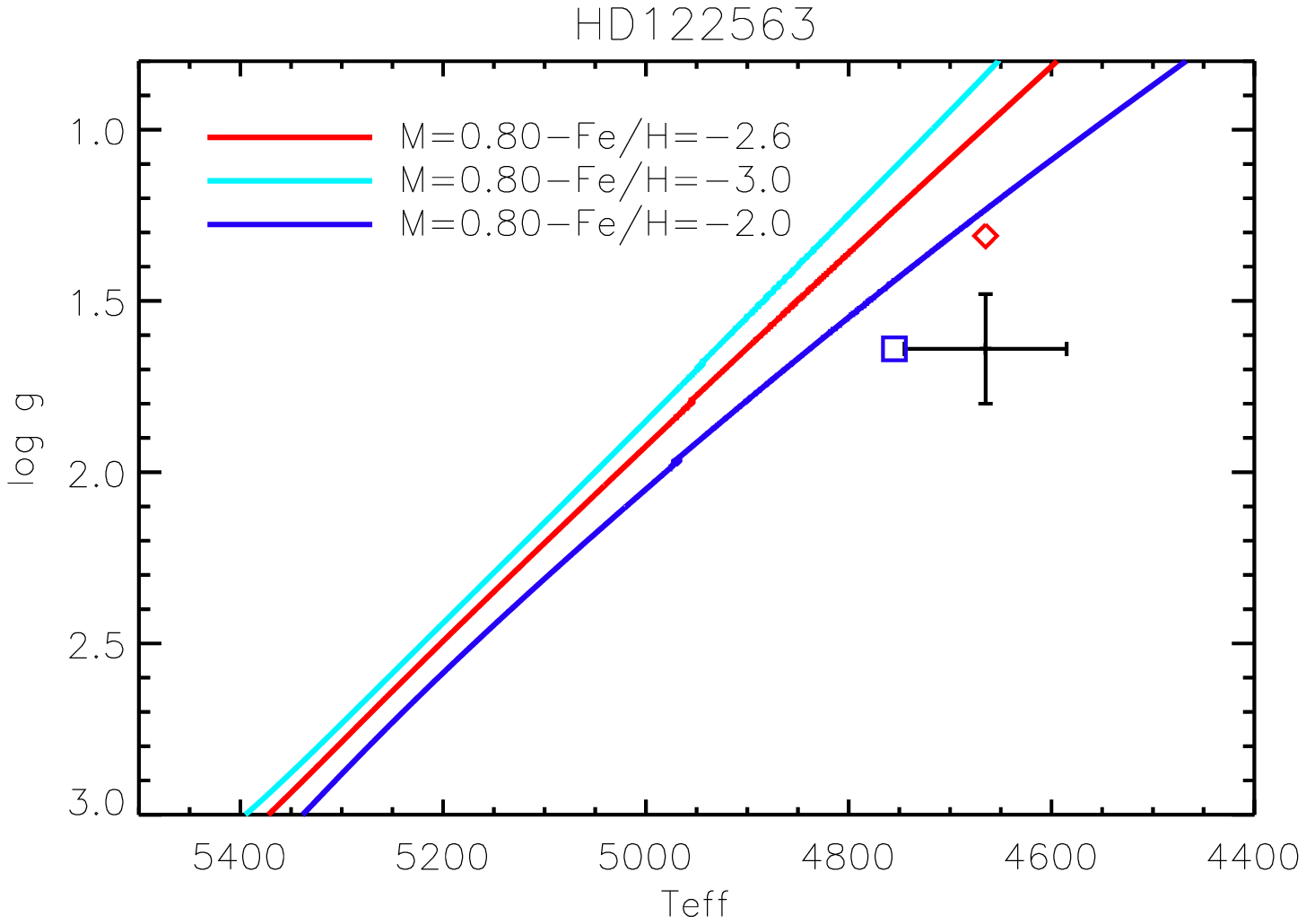}}
\resizebox{\columnwidth}{!}{\includegraphics[scale=1.0]{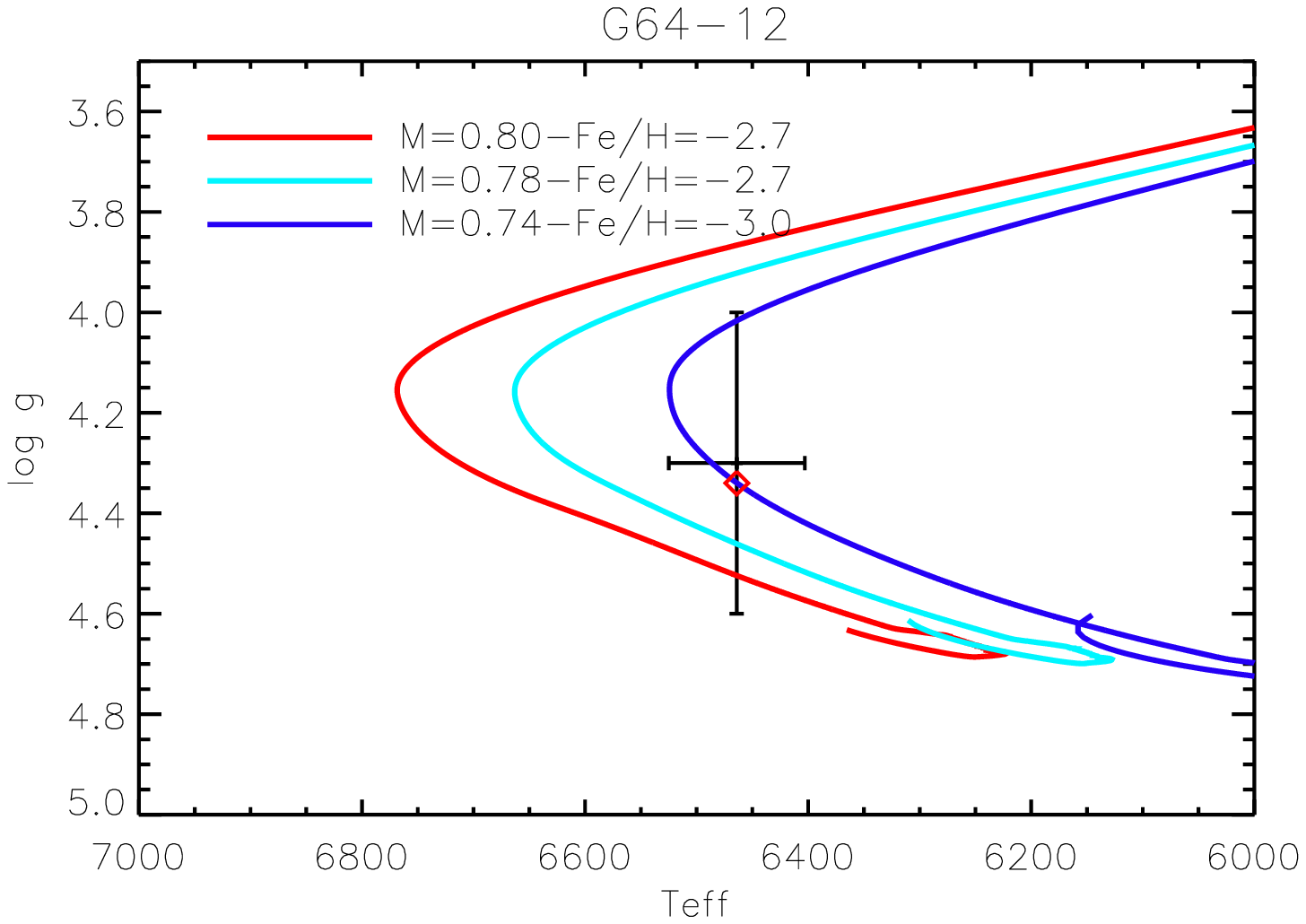}}
}
\caption{The position of the four metal-poor stars in the $\Teff$-$\log g$
plane. The black cross with error bars is the result with the IRFM $\Teff$ and
astrometric surface gravity, and the blue square and red diamond correspond to
the spectroscopic NLTE $\td$ $\Teff$ and $\log g$ from this work,
respectively. \textsc{garstec} evolutionary tracks are also shown for
comparison. For G 64-12 only spectroscopic $\log g$/IRFM $\Teff$ solution is
shown.}
\label{tracks}
\end{figure*}

In an attempt to assess the trustworthiness of the spectroscopic results versus
those obtained from independent measurements, we compared them to evolutionary
tracks. The position of the four metal-poor stars in the $\Teff$-$\log g$ plane
is shown in Fig. \ref{tracks}. The black cross with error bars indicated is the
result with the IRFM $\Teff$ and astrometric surface gravity, and the blue
square and red diamond correspond to our spectroscopic $\td$ NLTE $\Teff$ and
$\log g$, respectively. For comparison, evolutionary tracks with close
parameters computed with the \textsc{garstec} code \citep{2008Ap&SS.316...99W}
are shown. These were kindly made available to us by A. Serenelli (2011, private
communication). In order not to overload the figures, we show only the NLTE
$\td$ spectroscopic results, while briefly discussing other solutions from the
Table \ref{tab:opt_param} below.

For all metal-poor stars, a combination of astrometric gravity and spectroscopic
$\Teff$/[Fe/H], obtained with NLTE $\SH = 1$ or LTE with 1D and $\td$ models,
appears to be most consistent with stellar evolution calculations. The error
bars of our data are omitted in the plots for clarity (they are given in the
Table \ref{tab:opt_param}). 1D LTE $\Teff$'s are somewhat larger than the 1D
NLTE results, that makes the stars a bit more massive. It is interesting that
even with $\td$ LTE effective temperatures, which are $\sim 50 - 300$ K larger
compared to NLTE results, the stars still match the tracks with adequate mass
and metallicity, which, in fact, also implies more realistic ages.

A combination of IRFM $\Teff$ and the spectroscopic $\td$ LTE $\logg$/[Fe/H]
makes the agreement with stellar evolution less satisfactory. In particular, HD
140283 is now in the 'forbidden' zone of the HRD. Situation is even worse for
the NLTE $\SH = 0.1$ results, obtained with 1D and with $\td$ model atmospheres.
In this case, all four metal-poor stars are some $\sim 2 - 3 \sigma$ in $\Teff$
and $\logg$ away from any reasonable track.

HD 122563 is a somewhat outstanding object. Assuming the tracks to be realistic,
it also becomes clear that the IRFM $\Teff$ and/or luminosity of the star
inferred from its relatively accurate parallax are too low. Other attempts, such
as variation of the mass or decreasing the $\alpha$-enhancement, either do not
have any effect or make the discrepancy with masses and ages even larger. In
this respect, it is encouraging that our spectroscopic $\td$ NLTE values of
$\Teff$ and $\log g$ bring the star into a much better agreement with stellar
evolution, although these value appear to be shifted away from the presumably
more reliable IRFM temperature and astrometric gravity.

%
%
\section{Summary and conclusions}{\label{sec:conclusions}}
One important problem in spectroscopy of late-type stars is to quantify
systematic errrors in the determination of $\Teff$, $\log g$, and [Fe/H] from
the \ion{Fe}{i}/\ion{Fe}{ii} excitation-ionization equilibrium due to the
assumptions of LTE and hydrostatic equilibrium in the models.

We have constructed a new extensive NLTE Fe model atom using the most up-to-date
theoretical and experimental atomic data available so far. The model has been
applied to the analysis of the Sun and a number of well-studied late-type stars
with parameters determined by other independent methods. We have used standard
1D LTE \textsc{marcs} and \textsc{mafags-os} model atmospheres, as well as
average stratifications computed from full time-dependent, 3D, hydrodynamical
simulations of stellar surface convection. In addition, we compared the
commonly-used NLTE line formation program packages \textsc{detail}/\textsc{siu}
and \textsc{multi}. Despite obvious differences in their numerical scheme and
input physical data, we find that the final results are consistent both in terms
of the NLTE statistical equilibria and absolute LTE and NLTE abundances.

Our LTE and NLTE results for the 1D models are similar to most of the previous 
findings in the literature. Statistical equilibrium of Fe, which is a minority
ion at the typical conditions of these cool and dense atmospheres, favors lower
number densities of \ion{Fe}{i} compared to LTE. The number densities of
\ion{Fe}{ii} are hardly affected by NLTE. In general, this leads to a weakening
of \ion{Fe}{i} lines compared to LTE, which, in turn, requires larger Fe
abundance to fit a given observed spectral line. The magnitude of departures
from LTE depends on stellar parameters. With 1D hydrostatic model atmospheres
the NLTE corrections on \ion{Fe}{i} lines do not exceed $0.1$ dex for stars with
[Fe/H] $> -3.5$, and they are negligible for the \ion{Fe}{ii} lines. The
situation changes dramatically for the mean 3D ($\td$) models with their cooler
surfaces, and, thus, steeper T$(\tau)$ relations. The NLTE abundance corrections
can be as large as $\sim 0.5$ dex for the resonance \ion{Fe}{i} lines formed in
very metal-poor atmospheres. In contrast to LTE, NLTE strengths of the 
\ion{Fe}{i} lines predicted by the $\td$ and 1D models are rather similar,
because the line formation is largely dictated by the radiation field forming
around the optical surface, $\opd \sim 0$, where the thermal structures of $\td$
and 1D models are similar. This suggests that the full 3D NLTE results, once
they become feasible, will be even closer to our $\td$ NLTE calculations.

The solar Fe abundance obtained under NLTE using the $\td$ model atmospheres is
$7.46 \pm 0.06$ dex, and the uncertainty is fully dominated by the errors of the
experimental transition probabilities. Adopting LTE or 1D hydrostatic model
atmospheres lowers the solar Fe abundance by $\sim 0.02$ dex. For Procyon we
infer slightly sub-solar metallicity, [Fe/H] $= -0.03$ dex.

We find that the \ion{Fe}{i}/\ion{Fe}{ii} ionization balance can be well
established for all reference stars with the 1D and $\td$ model atmospheres
and the NLTE model atom with Drawin's \ion{H}{i} collision cross-sections.
Strong resonance and subordinate \ion{Fe}{i} lines are very sensitive to the
atmospheric structure, thus, classical 1D models fail to provide consistent
excitation balance. A better agreement among \ion{Fe}{i} lines spanning a range
of excitation potential is obtained with the mean 3D models, although the
optimum solution, which fully eliminates the correlation of the abundance and
line excitation potential, may necessitate self-consistent NLTE modelling with
full 3D hydrodynamical model atmospheres. The assumption of LTE in combination
with $\td$ models leads to large errors in $\Teff$ and $\log g$ inferred from
the Fe ionization balance, yet we find satisfactory results for certain
combinations of stellar parameters using standard hydrostatic model atmospheres.

The mean NLTE metallicities determined with 1D and $\td$ models are in agreement
in the range of stellar parameters investigated here, although there are marked
residual slopes of abundance with line excitation potential and strength. We
can, thus, conclude that accurate, albeit not precise, metallicities for
late-type stars can be obtained with classical hydrostatic 1D model atmospheres
if NLTE effects in \ion{Fe}{i} are taken into account and a sufficiently large
number of \ion{Fe}{i} and \ion{Fe}{ii} lines with different excitation
potentials is used, so that individual line-to-line abundance discrepancies
cancel out. An alternative solution, which avoids additional biases introduced
by adjusting microturbulence parameter, is to use high-excitation \ion{Fe}{i}
lines only. Thus, the results obtained with classical 1D models depend on the
choice of the line list and, in particular, on the balance of the number of
lines of different types. A combination of $\td$ and NLTE alleviates this
problem.

Our NLTE effective temperatures and gravities are consistent with the parameters
determined by other less model-dependent methods, in particular with IRFM
$\Teff$'s and $\log g$'s inferred from parallaxes. The results for the metal-
poor dwarfs and subgiant are also consistent with stellar evolution predictions.
For HD 122563, the results are inconclusive. In addition to a residual slope of
\ion{Fe}{i}-based abundances with line excitation potential, the NLTE $\td$
spectroscopic gravity is roughly $0.3$ dex lower than the astrometric result. On
the one side, the latter is somewhat uncertain due to the difficulty of
determining the mass of this metal-poor giant. Furthermore, comparisons with
evolutionary tracks favor higher $\Teff$ and lower $\log g$ which would be more
consistent with our spectroscopic estimates. On the other side, the
discrepancies might be indicative of remaining systematic uncertainties in the
stellar atmosphere models, such as neglect of convective inhomogeneities and
atmospheric extension, which  become increasingly important with increasing
$\Teff$ and decreasing $\log g$. The importance of the latter is also amplified
at higher metallicities.

\section*{Acknowledgments} This work is based on observations collected at the
European Southern Observatory, Chile, 67.D-0086A and ESO DDT Program ID
266.D-5655 (PI Christlieb). We made extensive use of the NIST and Kurucz's
atomic databases. We acknowledge valuable discussions with Lyudmila Mashonkina,
and are indebted to Thomas Gehren for providing the codes for calculations
performed in this work. We thank Jorge Melendez for providing the observed
spectra for some of the studied stars. We are grateful to Manuel Bautista for
providing the photoionization cross-sections for Fe and to Aldo Serenelli for
providing the \textsc{garstec} evolutionary tracks.

\bibliographystyle{mn2e} 
\bibliography{references}

\appendix
%
%
\begin{table*}
\caption{Equivalent widths of the \ion{Fe}{i} lines (in units of \mA).}
\label{ew1}
\begin{tabular}{lcccccc}
\hline
 Line (\AA) &  Sun  & Procyon & HD 84937 & HD 140283 & HD 122563 & G 64-12 \\
\hline
 3581.19 &         &            &    136.30  &            &            &     90.00 \\
 3618.77 &         &            &    100.95  &            &            &     61.80 \\
 3719.93 &         &            &            &            &            &     89.00 \\
 3737.13 &         &            &            &            &            &     80.60 \\
 3745.56 &         &            &            &            &            &     71.35 \\
 3758.23 &         &            &            &            &            &     64.00 \\
 3820.43 &         &            &    138.90  &            &            &     70.40 \\
 4045.81 &         &            &    115.95  &            &            &     71.02 \\
 4235.94 &         &            &     52.55  &     59.23  &    102.58  &     10.30 \\
 4250.79 &         &            &     71.90  &     78.99  &    128.14  &     24.26 \\
 4415.12 &         &            &     75.30  &     84.90  &            &     28.59 \\
 4445.47 &  41.05  &      8.49  &            &            &     12.02  &           \\
 4494.56 & 204.79  &    123.69  &     28.94  &     34.62  &     83.36  &           \\
 4920.50 & 522.83  &    207.10  &     55.68  &     61.03  &    102.54  &     14.60 \\
 4994.13 & 111.49  &     87.12  &      7.33  &     13.49  &     79.57  &           \\
 5044.21 &  77.48  &     48.67  &            &      2.15  &      9.35  &           \\
 5198.71 & 101.08  &     75.32  &      4.25  &      6.20  &     35.22  &           \\
 5216.27 & 129.41  &     99.89  &     13.90  &     20.41  &     79.26  &           \\
 5225.53 &  73.14  &     32.95  &            &      2.26  &     45.02  &           \\
 5232.94 & 389.77  &    180.23  &     44.80  &     48.73  &     89.72  &      7.40 \\
 5236.20 &  34.77  &     16.93  &            &            &            &           \\
 5242.49 &  90.82  &     72.53  &      3.42  &      3.84  &     11.22  &           \\
 5247.05 &  67.15  &     25.53  &            &      1.55  &     36.52  &           \\
 5250.21 &  66.95  &     25.64  &            &      1.69  &     35.45  &           \\
 5269.54 & 533.08  &    204.43  &     84.46  &     97.47  &    172.99  &     29.70 \\
 5281.79 & 164.04  &    104.41  &     11.93  &     14.03  &     42.52  &           \\
 5379.57 &  60.26  &     38.82  &            &            &      3.45  &           \\
 5383.37 & 229.36  &    132.15  &     20.96  &     18.62  &     36.45  &           \\
 5434.52 & 213.36  &    130.93  &     37.18  &     52.31  &    124.32  &      4.94 \\
 5491.84 &  14.23  &      4.91  &            &            &            &           \\
 5586.76 &         &            &     26.01  &            &     58.67  &      3.30 \\
 5661.35 &  24.87  &     10.25  &            &            &            &           \\
 5662.52 & 102.45  &     75.49  &            &      3.34  &      7.44  &           \\
 5701.54 &  86.97  &     57.59  &            &            &     16.14  &           \\
 5705.46 &  39.97  &     21.17  &            &            &            &           \\
 5916.25 &  54.06  &     24.82  &            &            &            &           \\
 5956.69 &  52.47  &     16.22  &            &            &     11.56  &           \\
 6065.48 & 131.32  &     93.35  &      8.48  &     10.40  &     43.92  &           \\
 6082.71 &  34.67  &      9.75  &            &            &      2.21  &           \\
 6151.62 &  50.17  &     19.41  &            &            &      4.66  &           \\
 6173.33 &  68.47  &     36.51  &            &            &     10.86  &           \\
 6200.31 &  73.66  &     44.14  &            &            &      9.00  &           \\
 6219.28 &  90.55  &     63.29  &      2.60  &            &     26.80  &           \\
 6240.65 &  48.75  &     18.63  &            &            &      4.49  &           \\
 6252.56 & 133.16  &     93.79  &      8.26  &     12.09  &     51.82  &           \\
 6265.13 &  88.59  &     57.68  &            &            &     23.83  &           \\
 6297.79 &  76.07  &     45.84  &            &            &     14.69  &           \\
 6430.85 & 119.06  &     90.63  &      7.09  &     10.88  &     53.67  &           \\
 6518.37 &  58.16  &     29.10  &            &            &            &           \\
 6574.23 &  28.80  &            &            &            &      3.14  &           \\
 6593.87 &  86.59  &     53.43  &            &            &     17.17  &           \\
 6609.11 &  65.27  &     32.12  &            &            &      6.90  &           \\
 6699.14 &   8.96  &            &            &            &            &           \\
 6726.67 &  49.91  &     27.36  &            &            &            &           \\
 6739.52 &  12.08  &            &            &            &            &           \\
 6750.15 &  75.95  &     44.28  &            &            &     12.29  &           \\
 6793.26 &  13.81  &            &            &            &            &           \\
 6810.26 &  52.15  &     30.02  &            &            &            &           \\
 6837.01 &  17.92  &            &            &            &            &           \\
 6854.82 &  12.28  &            &            &            &            &           \\
 6945.20 &  83.35  &     54.60  &            &            &            &           \\
 6978.85 &  80.56  &     50.56  &            &            &            &           \\
 7401.68 &  43.41  &     22.22  &            &            &            &           \\
 7912.87 &  46.89  &            &            &            &            &           \\
 8293.51 &  60.14  &            &            &            &
 &           \\
\hline
\end{tabular}
\end{table*}
\begin{table*}
\caption{Equivalent widths of the \ion{Fe}{ii} lines (in units of \mA).}
\label{ew2}
\begin{tabular}{lcccccc}
\hline
 Line (\AA) &  Sun  & Procyon & HD 84937 & HD 140283 & HD 122563 & G 64-12 \\
\hline
 4233.17 &         &            &            &            &            &      6.40 \\
 4491.40 &  76.77  &    104.32  &      9.02  &      7.32  &     30.49  &           \\
 4508.29 &         &            &            &            &            &           \\
 4576.34 &  68.55  &     93.19  &      5.96  &      5.16  &     21.74  &           \\
 4582.84 &  56.51  &     80.64  &      3.38  &      2.85  &     14.46  &           \\
 4583.84 &         &            &            &            &            &      5.70 \\
 4620.52 &  53.91  &     75.55  &      3.16  &      2.45  &     12.24  &           \\
 4923.93 & 190.40  &            &     62.07  &     57.67  &     99.18  &     14.85 \\
 5169.03 &         &            &            &            &            &     21.10 \\
 5197.58 &  85.22  &    115.09  &     13.08  &     11.07  &     36.14  &           \\
 5234.62 &  84.37  &    122.16  &     15.35  &     12.77  &     40.63  &           \\
 5264.81 &  47.81  &     69.23  &            &            &      8.68  &           \\
 5284.11 &  62.01  &     87.00  &      3.90  &      3.42  &            &           \\
 5325.55 &  44.71  &     66.42  &            &            &      6.23  &           \\
 5414.07 &  28.83  &     44.12  &            &            &            &           \\
 5425.26 &  43.50  &     63.26  &            &            &      6.39  &           \\
 6239.95 &  13.18  &     25.00  &            &            &            &           \\
 6247.56 &  56.21  &     83.62  &      3.22  &            &      9.44  &           \\
 6369.46 &  19.75  &     29.78  &            &            &            &           \\
 6432.68 &  41.16  &     62.63  &            &            &      7.80  &           \\
 6456.38 &  66.79  &     98.26  &      5.48  &            &     13.74  &           \\
 6516.08 &  56.09  &     75.98  &            &            &     14.12  &           \\
 7222.39 &  19.35  &     35.36  &            &            &            &           \\
 7515.83 &  15.16  &     27.30  &            &            &            &           \\
 7711.72 &  49.09  &            &            &            &      6.94  &           \\
\hline
\end{tabular}
\end{table*}

\bsp

\label{lastpage} \end{document}